\documentclass{article}[11pt]

\usepackage{amsmath, amsthm, amssymb}
\usepackage[english]{babel}
\usepackage{dsfont}
\usepackage{mathdots}
\usepackage{natbib}
\usepackage{graphicx}
\usepackage{enumerate}
\usepackage{color}
\usepackage{float}
\usepackage{placeins}
\usepackage{url}
\usepackage{comment}
\usepackage[disable]{todonotes}
\usepackage{rotating}
\usepackage{multirow}
\usepackage{diagbox}
\usepackage[a4paper, left=2.5cm, right=2.5cm, top=3cm, bottom=3cm]{geometry}

\newtheorem{thm}{{Theorem}}
\newtheorem{lem}[thm]{{Lemma}}
\newtheorem{rmk}{{Remark}}
\newtheorem{asp}{{Assumption}}

\newcommand{\argmin}{\mbox{argmin}}
\newcommand{\var}{\mbox{Var}}
\newcommand{\cov}{\mbox{Cov}}

\newcommand{\kp}{k_A}
\newcommand{\kpp}{k_\Gamma}
\newcommand{\cp}{k_{2}(p)}

\newcommand{\cpu}{k_u}
\newcommand{\cpB}{k_B}
\newcommand{\cpD}{k_D}

\def\eps{\varepsilon}
\newcommand{\gp}{g_n}
\newcommand{\gpeps}{\widetilde{g}_n}

\newcommand{\grest}{g_{\mathcal{R}}}
\newcommand{\grestr}{g_{\mathcal{R},r}}

\newcommand{\R}{\mathds{R}}

\newcommand{\N}{\mathds{N}}
\newcommand{\Z}{\mathds{Z}}
\newcommand{\E}{\mathds{L}}
\newcommand{\Rot}{R}
\newcommand{\ind}{\mathds{1}}
\newcommand{\Lemmaelf}{Lemma~11} 
\newcommand{\Lemmaneun}{Lemma~12} 
\newcommand{\Gammah}{\hat{ {\Gamma}}^{(st)}(0)}

\newcommand{\Gammas}{{\Gamma}^{(st)}(0)}
\newcommand{\Sigmah}{\hat \Sigma_\eps}
\newcommand{\THR}{\operatorname{THR}_{\lambda}}
\newcommand{\THRarg}[1]{\operatorname{THR}_{#1}}
\newcommand{\veco}{\operatorname{vec}}
\newcommand{\vech}{\operatorname{vech}}
\newcommand{\DESP}{\emph{\!Boot}}
\newcommand{\OLS}{\emph{\!Gaussian}}

\newcommand{\A}{\mathds{A}}
\newcommand{\Sigmaeps}{\Sigma_\eps}
\newcommand{\fracd}[2]{#1/#2}
\renewcommand{\hat}{\widehat}
\renewcommand{\tilde}{\widetilde}

\allowdisplaybreaks
\begin{document}

\begin{center}
{\bf  \huge Structural Inference in Sparse High-Dimensional Vector Autoregressions} 
\end{center} 
\begin{center}
{\bf J. Krampe$^{1}$, E. Paparoditis$^{2}$, and C. Trenkler$^{1}$} 
\end{center}  
\noindent $^1$ University of Mannheim; j.krampe@uni-mannheim.de, trenkler@uni-mannheim.de \\ $^2$ University of Cyprus; stathisp@ucy.ac.cy

\begin{abstract}
We consider statistical  inference for impulse responses in  sparse, structural high-dimensional vector autoregressive (SVAR) systems. 
We introduce consistent estimators of impulse responses in the high-dimensional setting and suggest  valid inference procedures for the same parameters. Statistical inference  in our setting is much more involved  since standard procedures, like the delta-method, do not apply. By using  local projection equations, we first construct a de-sparsified version of regularized estimators of the moving average parameters  associated with the VAR system. We  then obtain estimators of the structural impulse responses by combining the aforementioned de-sparsified estimators with a non-regularized estimator of the contemporaneous impact matrix, also taking into account  the high-dimensionality of the system. We show that the distribution of the derived estimators of  structural impulse responses has a Gaussian limit. We also present a valid bootstrap procedure to estimate this distribution. Applications of the inference procedure in the construction of  confidence intervals for  impulse responses as well as in tests for forecast error variance decomposition are presented.  Our procedure is illustrated by means of simulations.

\end{abstract}

\vspace{0.2cm}

\section{Introduction}
Structural analysis based on impulse responses and forecast error variance decompositions (FEVDs) is an important part of macroceconomic and financial time series analysis. Over the last two decades structural model approaches have become popular that consider a large number of variables, e.g., factor-augmented vector autoregressive (VAR) models \citep{bernanke2005}, structural dynamic factor models \citep{forni2009,stock2005implications}, large Bayesian VARs \citep{banbura2010SIR}, and global VARs \citep{chudik2016theory}. Such large-scale set-ups offer a number of advantages in comparison to small-scale models, e.g., low-dimensional VARs. In particular, they facilitate the measurement of the economic shocks of interest due to a lower degree of information deficiency and they permit to analyse the responses of a larger set of variables to the relevant shocks \citep[Ch.~16]{bernanke2005,stock2016dynamic,kilian2017structural}. Moreover, large-scale models allow  to study interactions within networks of economically relevant sizes, see e.g., \cite{demirer2018estimating}, \cite{barigozzi2017network}, \cite{barigozzi2019nets}. We contribute to this literature on structural analysis by providing asymptotically valid inference approaches for structural impulse response analysis and forecast error variance decompositions in sparse high-dimensional VARs. In order to facilitate inference we suggest an asympotically valid bootstrap approach in addition to the large sample Gaussian approximation.

It is debated whether macroeconomic and financial data are best represented by dense or sparse models \citep{giannone2018illusion,fava2020illusion}. In some applications one may assume sparsity as done e.g., by \cite{demirer2018estimating} for high-dimensional VAR modelling of bank stock return volatilities. In such a set-up one can directly apply our inference procedures to the time series data under consideration. In many other cases a dense structure is more plausible. Then, one may consider a factor model with an idiosyncratic component that is assumed to have a sparse VAR representation. This leads to the combined 'factor plus sparse VAR' approach of \cite{barigozzi2017network} who have studied financial network links based on common and idiosyncratic volatility components. Similarly, \cite{barigozzi2019nets} pre-adjust for a common factor before applying their VAR-based network estimation approach (NETS) to a panel of stock return volatilities. In such applications our structural inference framework remains valid if the factors are assumed to be given. Then, one could, e.g., test whether (idiosyncratic) network links derived from variance decompositions can be regarded as relevant and thereby provide additional statistical information on the type of analysis conducted in \cite{barigozzi2017network} and \cite{barigozzi2019nets}. Finally, our inference results open up the possibility of considering structural factor models in which the economic shocks of interest affect both the common component and the idiosyncratic component that has a sparse VAR structure.

Although high-dimensional modelling approaches provide a number of appealing advantages they also induce numerous challenges for inference. This is no different in our framework. We consider a stable, high-dimensional  structural vector autoregressive (SVAR) system of order $d$. To keep model complexity tractable we impose approximate sparsity assumptions. We consider de-sparsified versions of $\ell_1$-penalized estimators as a vehicle for statistical inference in order to ensure that our estimators have  Gaussian limiting distributions. However, for obtaining these limit results we cannot simply rely on related existing methods and results for reduced form VARs as briefly sketched in the following.


The structural impulse responses as well as the variance ratios obtained from forecast error variance decompositions are nonlinear functions of a high-dimensional parameter vector that comprises the VAR slope parameters and the distinct elements of the variance matrix of the reduced form innovations. However, the Gaussian limiting results for de-sparsified estimators of the VAR slope parameters, obtained e.g.~by \cite{krampe2018bootstrap}, just hold true for a finite subset of the parameters. Therefore, the usual delta method approach for obtaining the limiting distribution of the estimators of the structural parameters does not work in contrast to the case of low-dimensional VARs. Similarly, the high-dimensional set-up will render existing bootstrap approaches invalid in our framework. E.g., a simple model-based bootstrap with i.i.d.~resampling from the estimated residuals will fail since the sample variance matrix of the residual vectors is not a consistent estimator of the variance matrix of the innovations in the high-dimensional case. The parametric bootstrap suggested by \cite{krampe2018bootstrap} for inference in reduced form VARs does not work either since this approach cannot asymptotically imitate the fourth-order moments of the innovations that affect the limit distribution of the estimators of the structural parameters of interest. 

We address the aforementioned challenges in the following way. First, we propose consistent $\ell_1$-regularized estimators of the structural impulse responses that serve as input for the de-sparsified estimation approach. The consistency property can be obtained from existing results based on rather weak assumptions and since no sparsity constraints need to be imposed on the contemporaneous impact matrix, common identification restrictions -- like short-run, long-run, or sign-restrictions -- can be employed; see e.g., \cite{ramey2016shocks} for an extensive survey on various identification methods used in structural VARs. Second, for implementing inference we need to strengthen the approximate sparsity assumptions which now also involve the contemporaneous impact matrix. As a consequence, we focus on short-run identifying restrictions in order to avoid any conflict with necessary sparsity constraints. Third, we construct a new direct de-sparsified estimator of the entries in the moving average (MA) parameter matrices implied by the VAR process. This estimator is based on a local projection approach that avoids to deal with problems due to the nonlinearity of the relationship between the VAR and MA parameters. Eventually, the estimator of the MA parameters is appropriately combined with a non-regularized estimator of the contemporaneous impact matrix in order to obtain an estimator of the structural impulse response of interest. We proceed by showing that the combined estimator has a Gaussian limiting distribution for a finite set of structural parameters. The bootstrap procedure we propose relies on separating the innovations of the SVAR into a subset  representing the structural shocks of interest and a second subset referring to additional shocks or measurement errors. Under independence of the two subsets of SVAR innovations, we show that the limit distribution of the structural impulse response estimators is consistently estimated by our bootstrap procedure.

Finally, we suggest to form confidence intervals for structural impulse responses based on quantiles obtained from the asymptotic or the bootstrap  distribution of the de-sparsified estimators but anchored at the regularized estimators. This suggestions is motivated by the lower degree of variability of the regularized estimators in finite-samples. The latter feature also makes the regularized estimators to be the preferred choice for point estimation. The forecast error variance decompositions can be easily obtained from the regularized estimators of the structural impulse responses and asymptotically valid tests for hypotheses on the decompositions follow from our de-sparsified estimation approach. Therefore, our presentation focuses on the impulse response framework.

Our paper links to several strands of the literature on high-dimensional time series analysis. A number of papers propose approaches for statistical inference on structural impulse responses in dense high-dimensional frameworks. This includes, among others, \cite{yamamoto2019,baililu2016,stock2016dynamic} regarding factor-augmented VARs and structural dynamic factor models, \cite{dees2007exploring} 
in terms of global vector autoregressive models, and \cite{banbura2010SIR,canova2013panel} for large-scale (panel) Bayesian VARs. To the best of our knowledge, no structural inference methods exist for sparse high-dimensional systems. Our paper attempts to fill this gap. 



By now, a quite large literature has emerged that deals with the problem of fitting sparse high-dimensional VAR models using $\ell_1$-penalized estimators; see among others \cite{song2011large}, \cite{han2015direct}, \cite{kock2015oracle}, and \cite{basu2015}. Moreover, high-dimensional VARs and predictive regression models estimated by lasso-type estimators have proven to be useful for forecasting; also in comparison to other estimation and model approaches like ridge and factor models. This applies even if the data generating process has a factor structure  \citep{callot2014forecasting,kascha2015forecasting,medeiros2016forecasting,smeekes2018forecast}. 

Recently, \cite{chaudhry2017}, \cite{neykov2018}, \cite{zheng2019testing}, and \cite{krampe2018bootstrap} have considered inference on the autoregressive slope parameters of high-dimensional reduced form VAR systems which can also be used to derive (bootstrap-based) tests, for instance tests for Granger-causality. These papers rely on de-sparsified estimators that have been initially introduced and investigated in the i.i.d.~regression \citep[e.g.][]{zhang2014,deGeer2014}. 

The rest of the paper is organized as follows. Section 2 introduces the model framework. We describe in Section 3 how to obtain regularized estimators of the structural impulse responses. In Section 4 we suggest a de-sparsified estimator of the structural impulse responses based on the new direct de-sparsified estimator of the reduced form moving average parameters. Moreover, we derive the limiting distribution of the de-sparsified estimator of the structural impulse responses. We also  show how this limiting distribution can be replicated by an asymptotically valid bootstrap scheme. Section 5 deals with structural inference suggesting different ways of obtaining confidence intervals for impulse responses and proposing tests for hypotheses on forecast error variance decompositions. We conduct numerical investigations on impulse response inference in Section 6. Section 7 concludes our findings while technical proofs and additional lemmas are deferred to an Appendix.

Throughout the paper the following  notation is used. For 
a vector $x\in \R^p$,  $\|x\|_0 = \sum_{j=1}^p \ind(x_j \not = 0)$, where $\ind(\cdot)$ is the indicator function, $ \| x \|_1 = \sum_{j=1}^p |x_j|$, $\| x \|_2^2 = \sum_{j=1}^p |x_j|^2$ and $\|x\|_\infty=\max_j |x_j|$.  Furthermore, for a $r\times s$ matrix $B=(b_{i,j})_{i=1,\ldots,r, j=1,\ldots,s}$,  $\|B\|_1=\max_{1\leq j\leq s}\sum_{i=1}^r|b_{i,j}|=\max_j \| B e_j\|_1$, 
$\|B\|_\infty=\max_{1\leq i\leq r}\sum_{j=1}^s|b_{i,j}|=\max_{i} \| e_i^\top B\|_1$ and $\|B\|_{\max}=\max_{i,j} |e_i^\top B e_j|$, where $e_j=(0,\ldots,0,1,0, $ $ \ldots, 0)^\top$ denotes  a unit vector  of appropriate dimension with the one appearing in the $j$th position. Denote the largest absolute eigenvalue of a square matrix $B$ by $\rho(B)$ and let   $\|B\|_2^2=\rho(BB^\top)$. The $p$-dimensional identity matrix is denoted by $I_p$ and  for two matrices $A$ and $B$,  their Kronecker product is denoted by $A\otimes B$; see among others Appendix A.11 in \cite{luetkepohl2007new}. We define   $\E=(e_1\otimes I_p)$ and we denote by 
$I_{d;-\mathcal{J}}\in \R^{d\times (d-|\mathcal{J}|)}$ a $d$-dimensional identity matrix after deleting  all  columns $j \in \mathcal{J}$. 
Here,  $\mathcal{J}$ is a subset of $\{1,2, \ldots, d\}$ and $I_{d;\mathcal{J}}=I_{d;-\mathcal{J}^C}$, where $\mathcal{J}^C$ denotes the complement of the set $\mathcal{J}$. 
For a vector-valued times series $\{X_t\}$, we write $\{X_{t;\mathcal{J}}:=I_{d;\mathcal{J}}^\top X_t\}$ for the  sub-vector containing the  components $j \in \mathcal{J}$ only.

\section{Model Framework}
Let $\{Y_t, t \in \Z\}$ be a stochastic process, where $Y_t$ is a vector of $p$ endogenous variables and $Y_1, Y_2, \ldots, Y_n$ is a time series stemming from $ \{Y_t\}$.  We consider a high-dimensional case such that $p$ is allowed to grow with the sample size $n$. Assume that  $Y_t=\mu+X_t$, where $\mu$ is the  intercept and $\{X_t, t\in \Z\}$ is a $p$-dimensional zero-mean stochastic process. We focus on the process $\{X_t\}$ only since   mean-adjusting the time series does not affect   the theoretical results derived in this paper. We assume that $X_t$ is generated according to the following 
 structural, sparse vector autoregressive model  of order $d$, SVAR($d$),
\begin{align} \label{eq.VAR}
    X_t= \sum_{j=1}^d A_j X_{t-j} + B u_t+D w_t=\sum_{j=1}^d A_j X_{t-j} + \eps_t.
\end{align}
Here, $A_1,\dots,A_d$ are sparse matrices, $ u_t\in \R^{\cpu}$, $B \in \R^{p\times \cpu}, D \in \R^{p\times (p-\cpu)}$, and $\{(u_t^\top,w_t^\top)^\top,t \in \Z \}$ are independent and identically distributed random variables with $E (u_t^\top,w_t^\top)^\top=0$ and $ \var((u_t^\top,w_t^\top)^\top) = I_p$, in short, $(u_t^\top,w_t^\top)^\top \sim i.i.d. (0, I_p)$.  Denote by $\Sigma_\eps =E (\eps_1 \eps_1^\top)$ the variance matrix of $\eps_t=B u_t+D w_t$. Then, $ \Sigma_\eps=BB^\top+ D
D^\top= \Sigma_u + \Sigma_w$, with an obvious notation for $ \Sigma_u$ and $ \Sigma_w$.  We call 
$\{\eps_t\}$ the (reduced form) innovation process. 
The random variables $u_t$ represent  the shocks of interest to  the SVAR system while  $w_t$ are considered as additional shocks or measurement errors. 

Now recall that if  the system is stable, that is if  $\det(\mathcal{A}(z))\not =0$ for all $z\leq 1$, where $\mathcal{A}(z)=I_p-\sum_{j=1}^d A_j z^j$, then $X_t$ also  possesses the  representation
\begin{align} \label{eq.VMA}
X_t=\sum_{j=0}^\infty \Psi_j (B u_{t-j}+D w_{t-j})=:\sum_{j=0}^\infty \Theta_j u_{t-j}+\sum_{j=0}^\infty \Psi_j D  w_{t-j},
\end{align}
with $\Psi_0=I_p$ and $ \Theta_j=\Psi_j B$. Expression (\ref{eq.VMA}) is commonly called  the moving average representation of the SVAR$(d)$ process.
The following relationship between the coefficient matrices $\{A_j,j=1\dots,d\}$ and $\{\Psi_h, h=0,1,\dots,\}$ holds true:  
\begin{align} \label{eq.MACOEF}
    \Psi_h=\sum_{s=1}^{\min(d,h)} \Psi_{h-s} A_s & \Longleftrightarrow 0 = \sum_{s=0}^{\min(d,h)} \Psi_{h-s} A_s, \text{ where } A_0=-I_p, k=1,2,\dots.
\end{align}
Consider now a $r$th shock to the system, that is an  increase by one standard deviation of the $r$th component  of the random variable $u_t$, denoted by $u_{t;r}$.
Then, $e_j^\top \Theta_h e_r=:\Theta_{h;jr}$ quantifies the effect of this increase  on the variable $j$ after   $h$ time points, i.e., on  $X_{t+h;j}$. Given the stretch of  observations  $X_1,\dots,X_n$, our  goal is to estimate the parameters $\Theta_{h;jr}$ for some values  $ h=0,\dots,H$ 
and to do inference for the same set of  parameters.
The  time horizon $H$ is treated  as fixed. 

In this high dimensional  set-up, the  dimension of the system, the sparsity of the parameter matrices as well as the  dimension $\cpu$ of the vector $ u_t$,  may grow with sample size and may depend among other things on the distribution of the innovations as well as on  the application of interest. In this context, if only consistent estimation of the impulse responses is of interest, then we allow for row-wise, approximately sparse matrices $A_j$, $j=1,2, \ldots, d$,  and impose no sparsity restrictions on the innovation's variance matrix $\Sigma_w$ or $B$, respectively. The precise assumptions needed in this case are stated  in Assumption~\ref{ass.Est} of Section~\ref{sec.estimation}. However, for appropriate inference on the impulse responses, the sparsity setting turns out to be  more restrictive. In particular,  row- and column-wise approximate sparsity  restrictions are imposed for the slope matrices $A_j$, $j=1,2, \ldots, d$, and for the matrices  $B$ and $\Sigma_w$. The  additional conditions  required for inference are stated  in Assumption~\ref{ass1} of Section~\ref{section.inferece}.

In the following, we will also rely on the stacked  SVAR(1) representation of a SVAR($d$) system, i.e., on the representation $ W_t = \mathds{A} W_{t-1} +\mathds{U}_t$, where 
$$
\mathds{A}= \begin{pmatrix}
A_1& A_2 & \dots & A_d\\
I_p & 0 & \dots & 0\\
0 & \ddots & \ddots & \vdots\\
0 & \dots & I_p & 0 
\end{pmatrix}\in \R^{dp\times dp}
\text{ and } 
\mathds{U}_t=\E \eps_t=
\begin{pmatrix}
\eps_t \\
0\\
\vdots\\
0
\end{pmatrix}.
$$
Note that using this representation, expression  \eqref{eq.MACOEF} also  can   be written as $\Psi_h=\E^\top \A^h \E$.

\section{Regularized Estimators of Impulse Response Functions} \label{sec.estimation}
In this section we derive an estimator of the impulse responses  $\Theta_{h;jr}=e_j^\top\Psi_h Be_r$ by combining a regularized estimator of the moving average matrix $\Psi_h$ with a non-regularized estimator of the contemporaneous matrix  $B$ of the shocks $u_t$. The estimation of impulse responses is  similar to the finite dimensional SVAR case with  the major difference being that a regularized estimator  of the VAR slope parameters $ A_j$ is used. Before presenting our  results 
regarding  the estimation error,
we first state the assumptions needed.  For this we adapt the concept  of approximately sparse  matrices discussed in \cite{bickel2008} and  define the following  class $\mathcal{U}(k,\zeta) $ of row-wise approximately sparse matrices,
\begin{align*}
    \mathcal{U}(k,\zeta)=\{ &A=(a_{ij})_{i=1,\dots,p_1,j=1,\dots,p_2} \in \R^{p_1\times p_2}: \max_i \sum_{j=1}^{p_2} |a_{ij}|^\zeta\leq k, \|A\|_2\leq M<\infty\}.
\end{align*}
Notice that for $ \zeta=0$  we are in the case of exact sparsity, where  $\sum_{j=1}^{p_2} |a_{ij}|^\zeta$ counts the number of nonzero coefficients in the $i$th row of the matrix $A$.  Allowing for  $ \zeta$ to vary in the interval $[0,1)$ relaxes the exact sparsity assumption to a more flexible  setting,  called  approximate sparsity. This setting is sometimes labelled as weak sparsity in the literature. 

We next state our assumptions on the SVAR system. 
In these assumptions as well as later on in this paper the abbreviations $\gp$ and $ \gpeps$  appear. These abbreviations refer to functions $ \gp=g(p,d,q,n) $  and $\gpeps=g(p,q,n)$, respectively,   where  $g$ is increasing in  the dimension $p$ and its particular form depends  on the number of  finite moments $q$ of the innovations $\varepsilon_t$. These functions can be understood as the price paid by the regularization methods used  and for allowing the  dimension of the SVAR system to grow. 
\begin{asp} \label{ass.Est} {~}
\begin{enumerate}[(i)]
\item  $(A_1,\dots,A_p)\in \mathcal{U}(\kp,\zeta)$, for some $ \zeta \in [0,1)$. 
 \label{ass.Est.1}
\item \label{ass.Est.2} 
There exists a $\varphi\in(0,1)$ such that $\rho(\A)\leq \varphi$ and  for any $m \in \N$, \[\|\mathds{A}^m\|_2 =O(\varphi^m)
\mbox{ and}\ \ 
\|\mathds{A}^m\|_\infty =O(\kp \varphi^m).\]  

\item \label{ass.Est.3} \ $\hat{A}_s^{(re)}$, $s=1,\dots,d$, is  a regularized estimator of $ A_s$, $s=1,2, \ldots, d$,  with corresponding stacked form $\hat \A^{(re)}$ and which   
satisfies $$\|\hat \A^{(re)}-\A\|_\infty=O_P\Big(\kp \big(\frac{\gp}{n}\Big)^{(1-\zeta)/2}\Big).$$ 
\item \label{ass.Est.4} The sample covariance $\sum_{t=1}^n\eps_t\eps_t^\top/n$, satisfies for all $U,V\in \R^{p\times p}$ with $ \|U\|_2=1=\|V\|_2$,
$$
\|1/n\sum_{t=1}^n U(\eps_t \eps_t^\top- \Sigmaeps)V^\top  \|_{\max}=O_P(\sqrt{\gpeps/n})
$$ 
\item There exists some subset $\mathcal{I} \subset \{1,\dots,p\}$ with $|\mathcal{I}|=\cpu$ such that for all $j \in \mathcal{I}$, it holds true that  $e_j^\top \eps_{t}=e_j^\top B u_t$ and the random vector $(e_j^\top \eps_t)_{j \in \mathcal{I}}$ satisfies  $\var((e_j^\top \eps_{t})_{j\in \mathcal{I}})>0$ and $\|\var((e_j^\top \eps_{t})_{j\in \mathcal{I}})^{-1}\|_\infty=O(\cpu)$.
\label{ass.Est.5}
\item For all $j=1,\dots,p$, it holds true that  $E(e_j^\top\eps_t)^q\leq C<\infty$ for some $q\geq 8$. 
\label{ass.Est.6}
\end{enumerate}
\end{asp}

 Assumption \ref{ass.Est}\eqref{ass.Est.1} specifies  the row-wise approximate sparsity of the estimators of the SVAR slope parameters and Assumption \ref{ass.Est}\eqref{ass.Est.2} specifies the stability conditions of the SVAR system.\footnote{The sparsity assumption can be modified  to group sparsity restrictions. It is  important to note  that under such group sparsity restrictions,  a consistent estimation in the sense of part \eqref{ass.Est.3} can be established.} The required rate in estimating  the SVAR slope parameters is specified in part \eqref{ass.Est.3} of Assumption \ref{ass.Est}. This assumption is flexible in the sense that the  regularization method chosen determines the particular form of the function $\gp$. Desired rates are $\gp=\log(dp)+(ndp)^{2/q}$ in the case where  only $q$ moments are finite, and $\gp=\log(dp)$ in the case of  sub-Gaussian innovations.\footnote{Note that in the i.i.d.~Gaussian regression case with exact sparsity $\kp$,  \cite{bellec2018slope} obtain the optimal rate $\log(p/\kp)$ under some conditions for the lasso and the slope estimator.} Candidates for the regularized estimators $ \hat{A}_s^{(re)}$, which fulfill the desired  rate condition in the  sparse SVAR setting considered, are the (adaptive) lasso\footnote{See, among others, Proposition 4.1 in \cite{basu2015} for the vectorized VAR with exact sparsity and Gaussian innovations, Proposition 3.3 therein for a row-wise estimation, Theorem 1 in \cite{kock2015oracle} for the SVAR with exact sparsity and Gaussian innovations, and Section 4 in \cite{kock2015oracle} for the adaptive lasso with exact sparsity and Gaussian innovations. Note that in \cite{kock2015oracle} additional logarithmic terms of the sample size and the dimension  occur. For the Lasso with exact sparsity and sub-Gaussian or sub-Weibull innovations see  Corollary 4 or Corollary 9, respectively, in \cite{Wong2020}. For the Lasso with approximate sparsity and sub-Gaussian or sub-Weibull innovations see Theorem 1 in \cite{masini2019regularized}. Note that given an error bound to $\ell_2$-norm it is straightforward to derive an error bound with respect to the  $\ell_1$-norm, see, among others, the proof of Proposition 4.1 in \cite{basu2015}.} and the Dantzig Selector\footnote{See, among others, Theorem 1 in \cite{han2015direct} for Gaussian innovations and Corollary 1 in \cite{wu2016performance} for sub-Gaussian innovations. The result in \cite{wu2016performance} are derived without assuming a specific sparsity setting. See Corollary 2 in \cite{krampe2020Est} for a thresholded Dantzig selector under approximate sparsity. Note that in the error bounds derived for the Dantzig selector additionally occurs the term $\|\Gammas^{-1}\|_1$.}. Based on finite sample results in simulations, see, e.g.,  \cite{krampe2020Est}, we recommend the use of the adaptive lasso for $\widehat{A}^{(re)}_s$,  which is built up row-wise. More specifically, the corresponding  estimator for the $i$th row of $(A_1,\dots,A_d)$, denoted by $\hat \beta_i^{(re)}$, is obtained as 
\begin{equation} \label{eq.adlasso} 
\hat \beta_i^{(re)} = \argmin_{c=(c_1, \ldots, c_{dp})^\top\in\R^{dp}}\frac{1}{n-d}\sum_{t=d+1}^n\big(X_{t;i} -c^\top (X_{t_1}^\top,\dots,X_{t-d}^\top)^\top\big)^2 + \lambda_A \sum_{s=1}^{dp} \frac{|c_s|}{1/\sqrt{n}+|\widehat{c}_{i,s}|},
\end{equation}
where $ \widehat{c}_i=(\widehat{c}_{i,s}, s=1,2, \ldots, dp)$ are the lasso estimators  of $ \beta_i$ obtained  as $\widehat{c}_i=\argmin_{c\in\R^{dp}}(n-d)^{-1}\sum_{t=d+1}^n\big(X_{t;i} -c^\top (X_{t-1}^\top,\dots,X_{t-d}^\top)^\top\big)^2 + \lambda_A \|c\|_1$ and $\lambda_A$ is a regularization parameter. We notice  here  that a consistent estimation of the SVAR slope parameter also can be obtained without the i.i.d. assumption for the innovations $\varepsilon_t$; see \cite{masini2019regularized} and \cite{Wong2020}. \cite{masini2019regularized} consider the case where these  innovations are a martingale difference process,  which covers, among other things, also the case of conditional heteroskedasticity.

Assumption \ref{ass.Est}\eqref{ass.Est.4} specifies the entry-wise consistency needed for the sample covariance of the innovations. Note that no sparsity is assumed here for $\Sigmaeps$ and that the entry-wise consistency of the sample covariance of the innovations  follows immediately from the number of finite moments and $\gpeps$ is the price paid for the increasing dimension. More specifically,  if only $q$ moments of the innovations are finite, then $\gpeps=\log(p)+(np)^{2/q}$ and this assumption follows by the Nagaev inequality for the independent case while  for the dependent case we refer to,  among others,   \cite{liu2013probability}. In the case of  sub-Gaussian innovations we have $\gpeps=\log(p)$. 
Assumption \ref{ass.Est}\eqref{ass.Est.5} ensures that, given the innovations,  the raw shocks can be identified. Note that in a low-dimensional case, let us say  for a system of dimension $\cpu$, it is usually implicitly assumed that $\mathcal{I}=1,\dots,\cpu$. This  means that  this assumption generalizes in a direct way a corresponding condition  in  the low-dimensional case.  Note that the rate at which  the sparsity and the dimension of the SVAR system is allowed to  grow with the sample size can be derived from the bounds presented  in the theorems stated   in this section. 



The regularized estimated SVAR slope parameters $\hat A_1^{(re)},\dots,\hat A_d^{(re)}$ with corresponding stacked form $\hat \A^{(re)}$ can be used to obtain a (regularized) estimators of the moving average matrices $\Psi_h, h=1,\dots,H$.  For this, let $\hat \Psi_h^{(re)}=\E^\top (\hat \A^{(re)})^h. \E$ 
Under Assumption~\ref{ass.Est} we then  have 
\[\|\hat \Psi_h^{(re)} -\Psi_h\|_\infty=O_P\Big(\kp^3 (\gp/n)^{(1-\zeta)/2}\Big),\ \  h=1,\dots, H;\]
see Lemma~\ref{lem.ass} of the Appendix. Note that the stability Assumption \ref{ass.Est}(ii) ensures that $\|\Psi_h\|_\infty$ does not increases  too fast while  the sparsity assumptions are only imposed on the SVAR slope parameters and not on the corresponding moving average matrices $\Psi_h$. Furthermore, since the relationship between $\Psi_h$ and $\A$ is nonlinear, sparsity of the SVAR slope parameters $ A_s$ implies  only  in very special situations  sparsity of the corresponding moving average matrices $\Psi_h$. For this reason, we do not  use  regularized estimators  to directly estimate  the coefficient matrices $\Psi_h$.

The innovations  $\{\eps_t\}$ can be estimated in the standard way as residuals   $\hat \eps_t=X_t-\sum_{j=1}^d \hat A_j^{(re)} X_{t-j}$, $t=d+1, d+2, \ldots, n$.\footnote{We omit centering of the residuals in order to not overload the notation and the proofs. However, in practice we recommend to use centering.} As in the low-dimensional case, the matrix $B$ describing the contemporaneous effect of the shock is identified only up to a rotation ${\Rot}\in \R^{\cpu\times \cpu}$, i.e.,  the matrix $B$ equals $B=\cov(\eps_t,\eps_{t;\mathcal{I}}){\Rot}$, where the set $\mathcal{I}$ refers to the set described  in Assumption \ref{ass.Est}\eqref{ass.Est.5}. We denote by $\Theta_h{\Rot}^{-1}=\Psi_h \cov(\eps_t,\eps_{t;\mathcal{I}})$  the ``raw'' impulses responses where  the term ``raw'' is used here to describe  the situation where the true rotation is not yet identified.\footnote{Note that with an additional normalization the ``raw'' impulse responses coincide with the ``generalized'' impulse responses.} 
Furthermore, the matrix  $B{\Rot}^{-1}=\cov(\eps_t,\eps_{t;\mathcal{I}})$ can be consistently estimated entry-wise by using  the sample covariance of the residuals $\hat{\eps}_t$. This, together with the estimated moving average matrices, lead to  the following estimator of  the  raw impulses responses 
\begin{align} \label{eq.EstTheta}
    \widetilde \Theta_h^{(re)}=\hat \Psi_j^{(re)} \big(\frac{1}{n-d}\sum_{t=d+1}^n \hat \eps_{t} \hat \eps_{t;\mathcal{I}}^\top \big)=\hat \Psi_j^{(re)}\widetilde B,
\end{align}
with an obvious notation for $ \widetilde{B}$. Consistency of the above estimators is established   in the following theorem.

\begin{thm} \label{thm.estimation}
Under Assumption~\ref{ass.Est} the estimator (\ref{eq.EstTheta}) of the raw impulse responses  $e_j^\top \Psi_h \cov(\eps_t,\eps_{t;\mathcal{I}})e_r$, $j=1,\dots,p$, $r=1,\dots,\cpu$, and  $h=0,\dots,H$, satisfies,
\begin{align}
    e_j^\top (\widetilde \Theta_h^{(re)}-\Psi_h \cov(\eps_t,\eps_{t;\mathcal{I}})) e_r=O_P\Big(\kp^3 \Big(\frac{\gp}{n}\Big)^{(1-\zeta)/2}+\kp\Big(\frac{\gpeps}{n}\Big)^{1/2}\Big).
    \end{align}
\end{thm}

We mention here  that the above  consistency result  for the raw impulse responses  is obtained without imposing any   sparsity assumptions on  the contemporaneous part $B$ or on the covariance matrix $\Sigmaeps$. Furthermore,  $\cpu=p$ is also allowed. Notice that  the stability assumptions ensure that  long-run effects also can be consistently estimated. Hence, one can follow  the existing literature and  impose  identification restrictions, like  short-run, long-run, or sign restrictions, see  \cite{kilian2017structural}. Note that for partial identification purposes only a column of the matrix ${\Rot}$ needs to be identified. 

As regards  identification, the raw impulse responses are restricted such that the rotation $\Rot$ can be identified. Replacing the raw impulse response with the above estimator leads to an estimator of $\Rot$. To elaborate, let $\mathcal{R}=\{(i,h) : i \in \{1,\dots,p\}, h \in \{0,\dots,H\}\}=\{(i_1,h_1),\dots,(i_{\cpu},h_{\cpu})\}$ be a set of indices
corresponding to the variables and time horizons for which  identification restrictions are imposed. Let the restrictions be formalized in a function $\grest : \R^{d\times d\times(H+1)}\to \R^{\cpu\times \cpu}$ such that $\grest(\Psi_{0} \cov(\eps_t,\eps_{t;\mathcal{I}}),\dots,\Psi_{H} \cov(\eps_t,\eps_{t;\mathcal{I}}))$ gives the restrictions one wants to impose. The set $\mathcal{R}$ needs to be set up in such a way that the following matrix of raw impulse responses 
\[\Theta_\mathcal{R}=\begin{pmatrix}
e_{i_1}^\top \Psi_{h_1} \cov(\eps_t,\eps_{t;\mathcal{I}}) \\
\vdots\\
e_{i_{\cpu}}^\top \Psi_{h_{\cpu}} \cov(\eps_t,\eps_{t;\mathcal{I}})
\end{pmatrix}\]
 has full rank $\cpu$.\footnote{If the Cholesky decomposition is used for identification, we set $\mathcal{R}=\{(i,0) : i \in \mathcal{I}\}$. Then, $\Theta_\mathcal{R}=\var(\eps_{t,\mathcal{I}})$. Let $\var(\eps_{t,\mathcal{I}})=PP^\top$, where $P$ is the Cholesky factor. Then $\grest(\Psi_{0} \cov(\eps_t,\eps_{t;\mathcal{I}}),\dots,\Psi_{H} \cov(\eps_t,\eps_{t;\mathcal{I}}))=\grest(\cov(\eps_t,\eps_{t;\mathcal{I}}),\cdot,\dots,\cdot)=P$ and $\Rot=\var(\eps_{t,\mathcal{I}})^{-1} P=(P^{-1})^{\top}$. Consequently, $\hat R$ is the inverse transposed Choleksy factor of $\fracd{1}{(n-d)}\sum_{t=d+1}^n \hat \eps_{t;\mathcal{I}} \hat \eps_{t;\mathcal{I}}^\top$}  We then have  ${\Rot}= \Theta_\mathcal{R}^{-1}\grest(\Psi_{0} \cov(\eps_t,\eps_{t;\mathcal{I}}),\dots,\Psi_{H} \cov(\eps_t,\eps_{t;\mathcal{I}}))$ and
  $\hat {\Rot} $  is an estimator of ${\Rot}$  obtained by  replacing the raw impulse response $\Psi_{h} \cov(\eps_t,\eps_{t;\mathcal{I}})$ by its estimator $\widetilde \Theta_h^{(re)}$. Let $\hat \Theta_{h;jr}^{(re)}=e_j^\top \widetilde \Theta_{h}^{(re)} \hat {\Rot} e_r$. The following theorem specifies the  error rates in estimating  the impulse responses of the $r$th shock following the procedure discussed so far. 

\begin{thm} \label{thm.estimation2}
Let ${\Rot}\in \R^{\cpu\times \cpu}$ be a rotation matrix such that $B=\cov(\eps_t,\eps_{t;\mathcal{I}}){\Rot}$. Furthermore, let $\mathcal{R}=\{(i,h) : i \in \{1,\dots,p\}, h \in \{0,\dots,H\}\}$ and $\grest(\Psi_{0} \cov(\eps_t,\eps_{t;\mathcal{I}}),\dots,\Psi_{H} \cov(\eps_t,\eps_{t;\mathcal{I}}))$ be such that 
$ {\Rot}= \Theta_\mathcal{R}^{-1}\grest(\Psi_{0} \cov(\eps_t,\eps_{t;\mathcal{I}}),\dots,\Psi_{H} \cov(\eps_t,\eps_{t;\mathcal{I}}))$
 and $\rho(\Theta_\mathcal{R}^{-1})<1/\alpha,\alpha>0$. Then, under Assumption~\ref{ass.Est} the error in estimating the  impulse responses  $e_j^\top \Theta_h e_r,j=1,\dots,p$ and $h=0,\dots,H$,  satisfies
\begin{align}
    e_j^\top \big(\hat\Theta_h^{(re)} - \Theta_h\big) e_r=O_P\Big(\kp\cpu^{1/2}\|\cov(\eps_t,\eps_{t;\mathcal{I}})\|_2
    \Big[\kp^2\Big(\frac{\gp}{n}\Big)^{(1-\zeta)/2}+\Big(\frac{\gpeps}{n}\Big)^{1/2}\Big]\Big). \label{eq.thm.est.all}
    \end{align}
If  only short-run restrictions are used, i.e., when  $\mathcal{R}=\{(i,h) : i=1,\dots,p, h =0\}$, then 
\begin{align}
    e_j^\top \big(\hat\Theta_h^{(re)} - \Theta_h\big) e_r=
    O_P\Big(\kp^3\Big(\frac{\gp}{n}\Big)^{(1-\zeta)/2} +
    \kp\cpu^{1/2}\|\cov(\eps_t,\eps_{t;\mathcal{I}})\|_2
    \Big(\frac{\gpeps}{n}\Big)^{1/2}\Big). \label{eq.thm.est.short}
\end{align}
\end{thm}
The identified shocks are given by $\hat u_{t}=  {\hat \Rot}^\top \hat \eps_{t;\mathcal{I}}$. 

\section{De-Sparsified Estimators of Impulse Responses}
\label{section.inferece}

Despite their consistency property, regularized estimators of impulse responses estimators are of limited use if one is interested in  inferring  properties of the corresponding theoretical  coefficients. The reason for this lies in the fact that the (limiting) distribution of regularized estimators is unknown and difficult to investigate. In fact, findings in the  much simpler i.i.d.~case  suggest that this unknown  distribution would neither be Gaussian nor it is clear how    can it be  approximated using alternative approaches, like  for instance,  the  bootstrap, see \cite{knight2000asymptotics} and \cite{chatterjee2010asymptotic}. For this reason, we follow the alternative approach of developing de-sparsified estimators as has been proposed in the i.i.d.~case by  \cite{zhang2014}, see also \cite{deGeer2014}. In this section, we introduce de-sparsified  or de-biased estimators of the impulse response coefficients  $\Theta_{h;jr}=e_j^\top\Psi_h Be_r$ 
 which possesses  a manageable limiting distribution and can, therefore,  be used for statistical inference.

The general idea is to obtain  a de-sparsified estimator for $\Theta_{h;jr}$ by   combining   a de-sparsified estimator of the regularized estimator of the moving average matrix $\Psi_h$ with a non-regularized estimator of the contemporaneous part $B$. We first introduce   de-sparsified estimators for the moving average parameter matrices. Then, we use these estimators to construct  de-sparsified  estimators of the impulse responses of interest. We derive the limiting distribution of these estimators and, finally,  we present  a valid bootstrap procedure for  estimating  this distribution.

\subsection{De-Sparsified Moving-Average Parameter Matrices}

In order to adapt the basic idea of de-sparsifying  to the estimation of the parameter matrices $\{\Psi_h,h=1,\dots,H\}$ we first reformulate the estimation of the coefficient matrix $\Psi_h$ as a regression problem. This  reformulation  leads to a more direct estimator of $\Psi_h$ compared to an estimator of the form $\hat \Psi_h=\E^\top (\hat \A)^h \E$ as applies for $ \widehat\Psi_h^{(re)}$. 
First, recall the stacked form representation $W_t = \A W_{t-1} + \E \eps_t$, where $W_t=( X_t^\top,X_{t-1}^\top,\dots,X_{t-d+1}^\top)^\top$. 
By recursive substitution, we obtain $W_{t+h} = \A^h W_{t} +\sum_{j=0}^{h-1} \A^j \E \eps_{t+h-j}$. Using  $X_{t+h}=\E^\top W_{t+h}$ and $\Psi_j=\E^\top \A^j \E$ eventually leads to the representation
\begin{align} \label{eq.x_t_k}
X_{t+h}=\E^\top \A^h W_t + \sum_{j=0}^{h-1} \Psi_j \eps_{t+h-j},
\end{align}
where $\Psi_h$ is the coefficient matrix of the regressor $X_t$ in \eqref{eq.x_t_k}. Notice that  expression  (\ref{eq.x_t_k}) has also been  used in the local projection approach considered by \cite{jorda2005estimation}. Our aim here, however,  is not to transfer 
the idea of \cite{jorda2005estimation} to the high-dimensional setting by applying  some regularized type estimators to representation (\ref{eq.x_t_k}).
Moreover, since $\Psi_h$ is not necessarily sparse, the expression in equation (\ref{eq.x_t_k}) may not even  be helpful for deriving a direct (sparse) estimator of $\Psi_h$. Instead, we use representation \eqref{eq.x_t_k} as a starting point to construct a de-sparsified estimator of $\Psi_h$.

Recall that  the advantage of de-sparsifying is that it leads to a manageable  limiting distribution of the  estimator obtained and that this is achieved  by  introducing  a bias-correction to an initial, regularized estimator. The name is motivated by the fact that the initial estimator is usually sparse, whereas the applied bias-correction leads to an estimator which is not sparse anymore. As will be seen later on, the initial estimator uses the relation $\E^\top \A^h \E$ with regularized VAR slope estimators $\hat A_1^{(re)},\dots,\hat A_d^{(re)}$ as input. Furthermore, the obtained estimator of $\Psi_h$ may not be sparse even if the VAR slope estimators $\widehat{A}_{j}^{(re)}$ are sparse. Nevertheless, we still call  the  estimator obtained a de-sparsified estimator,  since the initial estimators of the slope parameters  used are  regularized and  the term ``de-sparsified'' is  commonly used  for estimators obtained by  the procedure  discussed in this section. 

To proceed with our construction of the de-sparsified estimator of $\Psi_h$, let $U_{t+h}=\sum_{j=0}^{h-1} \Psi_j \eps_{t+h-j}$. Note 
 that $\{U_{t}, t \in \Z\}$ is a  $h$-dependent process, i.e., $U_{t_1}$ and $U_{t_2}$ are independent if $|t_1-t_2|>h$. Furthermore,   $U_{t+h}$ is independent from $W_t$.
Let 
$ \Xi_h=\E^\top \A^h\in \R^{p\times(dp)}.$ 
Then, \eqref{eq.x_t_k} can be written in regression form as  
\[X_{t+h}=\Xi_h W_t + U_{t+h}.\]
Now, recall  the  basic idea of  de-sparsifying: Rotate the regressor $W_t$ in such a way that  orthogonality of the regressors is achieved and, as a consequence,   ordinary  least squares  can  be applied to estimate  the components of the parameter matrix $\Xi_h=(\xi_{h,j}, j=1,2, \ldots, p)$. Here $ \xi_{h,j}$ denotes the $j$th row of the matrix $ \Xi_h$ and note that $\Xi_{h;jr}=\Psi_{h;jr},j,r=1,\dots,p$. Due to the high-dimension of the regression problem, however, only an approximate orthogonal rotation of the regressors can be achieved. To elaborate, define first the rotated regressors as the projections   $\hat Z_{t;r}=\hat \beta_{r}^\top W_{t} $, where the coefficient vector $\hat \beta_{r}$ is given by 
\begin{align} \label{eq.beta-hat}
    \hat \beta_{r}=(e_{r}^\top \Gammah^{-1} e_r )^{-1} \Gammah^{-1}  e_{r},
\end{align}
and $\Gammah$ is some estimator of the lag-zero autocovariance matrix $\Gamma^{(st)}(0)$ of $\{W_t\}$, which will  be discussed later on.  
Notice that      $\Gammas=\var(W_t)=\var((X_1^\top,\dots,X_d^\top)^\top)$  is  the lag-zero autocovariance matrix of the stacked VAR$(d)$  process. Using the rotated regressors $\widehat{Z}_{t;r}$ and  motivated by a least squares estimator with orthogonal regressors, an estimator $ \widetilde \Psi_{h;jr}$ of the coefficient $\Psi_{h;jr}=\Xi_{h;jr}$,  $ j,r\in \{1,\dots,p\}$,  of the matrix $ \Psi_h$ is then  given by 
\begin{align} \label{eq.PsiTilde}
    \widetilde \Psi_{h;jr}&=\Big(\sum_{t=d}^{n-h} \hat Z_{t;r} X_{t+h;j}\Big)/\Big(\sum_{t=d}^{n-h} \hat Z_{t;r} W_{t;r}\Big) \nonumber \\
    &=\Psi_{h;jr}+\Big(\sum_{t=d}^{n-h} \hat Z_{t;r} W_{t;r}\Big)^{-1}\Big(\xi_{h,j} I_{dp;-r}  I_{dp;-r}^\top\sum_{t=d}^{n-h} \hat Z_{t;r}  W_t+ \sum_{t=d}^{n-h} \hat Z_{t;r} U_{t+h;j} \Big).
\end{align}
Observe  that in the low-dimensional case $(p<n)$, 
the ``rotated regressor" $\hat Z_{t;r}$ can be constructed  so that the condition $\sum_{t=d}^{n-h} \hat Z_{t;r} W_t \equiv 0$ for all $j \not=r$  is satisfied. This implies that  the first  term within the  last parentheses on the right hand side of equation (\ref{eq.PsiTilde}) would disappear in this case.   
Since   such a construction is not possible in the high-dimensional case $(p>n)$,   the aforementioned term in (\ref{eq.PsiTilde}) does not disappear. Therefore, this  term introduces  a bias of  the  estimator $ \widetilde{\Psi}_{h;jr}$ which, however, can be estimated   using 
 some (regularized) estimator $\hat \Xi^{(re)}_h$ of $\Xi_h$. As a consequence, this bias term can   be   removed from  $ \widetilde{\Psi}_{h;jr}$. This procedure leads to a new  estimator which 
 is  called de-biased  or de-sparsified  estimator of $\Psi_h$.
This estimator, say $\hat \Psi_h^{(de)}$, is given by 
\begin{align}
    \hat \Psi_{h;jr}^{(de)}&=\widetilde \Psi_{h;jr}-\Big(\sum_{t=d}^{n-h} \hat Z_{t;r} W_{t;r}\Big)^{-1}\Big(\sum_{t=d}^{n-h} \hat Z_{t;r} \hat \Xi_{h;j\cdot }^{(re)} I_{dp;-r}  I_{dp;-r}^\top W_t\Big) \nonumber \\
    &=\hat \Psi_{h;jr}^{(re)}+\Big(\sum_{t=d}^{n-h} \hat Z_{t;r} W_{t;r}\Big)^{-1}\Big[\sum_{t=d}^{n-h} \hat Z_{t;r} (X_{t+h;j}-\hat \Xi_{h;j\cdot}^{(re)} W_t)\Big]. \label{eq.de_psi}
\end{align}  

As we have seen,  the derivation of  the de-sparsified estimator $ \hat{\Psi}_{h;jr}^{(de)}$ given above needs  estimators of $\Gammas$  and of  $\Xi_h$.
In general,  the estimation of $\Gammas$ and its inverse is a difficult task in the high-dimensional setting. However, in our setting this problem is more tractable and  can be solved  using  the underlying VAR structure of the system. This VAR structure relates $\Gammas$ to the slope parameters and to the variance matrix of the innovations.  Hence, an estimator 
of $\Gamma^{(st)}(0)$  can be obtained by plugging given estimators  of the parameter matrices  $ A_j$, $j=1,2, \ldots, d$, and of $ \Sigma_\varepsilon$ into the expression
\begin{equation}
    \label{eq.GammaHat}
 \Gamma^{(st)}(0)=\sum_{j=0}^\infty \mathds{A}^{j}\E \Sigmaeps \E^\top (\mathds{A}^\top)^j=\operatorname{vec}^{-1}_{dp,dp} \Big((I_{(dp)^2}-\mathds{A}\otimes\mathds{A})^{-1} \operatorname{vec}(\E \Sigmaeps \E^\top)\Big),
\end{equation}\todo{Do we need these vec representation here?}
where $\operatorname{vec}(\cdot)$ refers to the operator of stacking the columns of a matrix to a vector and $\operatorname{vec}^{-1}_{r,s}(\cdot)$ refers to the inverse operation, i.e., the one transforming the stacked columns back to a $r\times s$ matrix. For more details on  the $\operatorname{vec}(\cdot)$ operator  we refer to  the Appendix A.1.12.1 in  \cite{luetkepohl2007new}. Similarly, estimation of  $\Xi_h$ is based on $\Xi_h=\E \A^h$. Properties of  the estimators of $\Gammas$  and  $\Xi_h$ obtained in this way  are stated  in Lemma~\ref{lem.ass} and Lemma~\ref{lem.est.gamma} of the Appendix.

To derive the asymptotic distribution of  the de-sparsified estimator $\hat \Psi_h^{(de)}$ given in (\ref{eq.de_psi}),  we need to impose some additional conditions   on the underlying SVAR process,   on  its sparsity, including sparsity with respect to $B$ and $\Sigma_w$, as well as on  the consistency properties of the   estimators of  $A_j, j=1,2, \ldots, d$.

\begin{asp} \label{ass1} {~}
\begin{enumerate}[(i)]
\item   $\A^\top\in \mathcal{U}(\kp,\zeta)$, where $\zeta$ is as in Assumption~\ref{ass.Est}\eqref{ass.Est.1}.
 \label{ass1.1}
\item \label{ass1.2} $\|\mathds{A}^k\|_1 =O(\kp \varphi^k),$ where $\varphi$ is as in Assumption~\ref{ass.Est}\eqref{ass.Est.2}.
\item \label{ass1.3} The 
estimators $\hat A_s,s=1,\dots,d$ with the stacked form $\hat \A$ satisfy
    \[\|\hat \A-\A\|_l=O_P\Big(\kp^{3/2} \Big(\frac{\gp}{n}\Big)^{(1-\zeta)/2}\Big),\]
    where  $l \in \{1,\infty\}.$
\item 
$B^\top \in \mathcal{U}(\cpB,\beta)$, $\Sigma_w \in \mathcal{U}(\cpD,\beta)$.
\label{ass1.4}
\item $ \| \Gamma^{(st)}(0)^{-1} \|_1 = O(\kpp)$ and $\| \Gamma^{(st)}(0)^{-1} \|_2 = O(1)$.
\label{ass1.5}
 \item \label{ass1.5b}
 Let $\mathcal{R}=\{i : i \in \{1,\dots,p\}\}$ 
 be a set of indices and $\grest(\cov(\eps_t,\eps_{t;\mathcal{I}}))$ corresponding to short-run restrictions such that ${\Rot}= (\cov(\eps_{t;\mathcal{R}},\eps_{t;\mathcal{I}}))^{-1}\grest(\cov(\eps_{t;\mathcal{R}},\eps_{t;\mathcal{I}}))$. These identification restrictions define a function ${\grest}$ such that ${\Rot} e_r=\grest(\cov(\eps_{t;\mathcal{R}},\eps_{t;\mathcal{I}}))e_r=:\grestr(\cov(\eps_{t;\mathcal{R}},\eps_{t;\mathcal{I}}))$ and ${\grestr}$ is continuously differentiable with derivative $\nabla {\grestr}$ satisfying $ \nabla {\grestr}(\cov(\eps_{t;\mathcal{R}},\eps_{t;\mathcal{I}}))\not=0$ and for  $x$ in a neighborhood of $\cov(\eps_{t;\mathcal{R}},\eps_{t;\mathcal{I}})$ and some vector $u, \|u\|_2=1$ it holds true that $u^\top \nabla {\grestr}(x)=O(\|u\|_2 \cpu^2 \|x\|_2)$.  
\item  $\max_{\|v\|_2=1}(E (v^\top \eps_0)^q)^{1/q}\leq C<\infty$ for some $q\geq 8$ such that 
$$
{\gpeps}^{1/2} \kp^{5}\kpp (\gp/n)^{(1-\zeta)/2} (\cpu\cpB+\cpD)=o(1)
$$
and
$$
{\gpeps}^{1/2} \kp^{5/2}\kpp\big[\cpD\cpu^{1-\beta}+\cpB\cpu^{(3-\beta)/2}\big]\Big[\kp^2(\gp/n)^{1-\zeta}+\sqrt{\gpeps/n}\Big]^{1-\beta}=o(1).
$$

\label{ass1.6}
\end{enumerate}
\end{asp}



\cite{krampe2020Est} showed that  the column-wise consistency required in  Assumption \ref{ass1}(\ref{ass1.3}) can be achieved by thresholding  initial (regularized) estimators of $A_s$, $s=1,2, \ldots, d$,  like those introduced in the discussion of 
Assumption~\ref{ass.Est}(iii) in Section~\ref{sec.estimation}. To elaborate, let $ \widehat{A}_s^{(re)}$ be such a regularized estimator, for instance, the adaptive lasso estimator \eqref{eq.adlasso}. Then, suitable candidates  for $ \widehat{A}_s$ satisfying Assumption \ref{ass1}(\ref{ass1.3}) are thresholded estimators denoted by $\hat A^{(thr)}$ which are  given by
\begin{equation} \label{eq.init-thr}
    \hat A_s^{(thr)}=\THR(A^{(re)}_{s}):=\Big(\THR(\hat A^{(re)}_{s;jr})\Big)_
    {j,r=1,\dots,p},s=1,\dots,d.
    \end{equation}
Here, $\widehat{A}^{(re)}_{s;jr}$ denotes the $ (j,r)$th element of $ \widehat{A}_{s}$ and $\THR(\cdot)$ is a thresholding function with threshold parameter $\lambda$  which acts by thresholding every element $ \widehat{A}^{(re)}_{s;jr}$ of the matrix $A^{(re)}_{s} $. Such a thresholding  function can, for instance, be the adaptive lasso thresholding function given by $\THR^{al}(z)=z(1-|{\lambda}/z|^\nu)_+$ with $\nu\geq1$. Soft thresholding ($\nu=1$) and hard thresholding ($\nu=\infty$) are boundary cases of this function, see also \cite{rothman2009generalized} and \cite{cai2011adaptive} for alternative choices of $ \THR$. Notice   that the additional thresholding step discussed above leads to sparse estimators of $A_s$, $s=1,2, \ldots, d$. For details, we refer here to the proof of Theorem~\ref{thm.boot.valid} given in the Appendix.


Regarding Assumption~\ref{ass1}(iv)   the  following is mentioned. In order to estimate the covariance matrix $\Sigmaeps$ of the innovations, we propose the following procedure. Based on the estimated residuals $\hat \eps_t=X_t-\sum_{s=1}^d \hat A_s^{(re)} X_{t-j}$, we can estimate $\widetilde  B=1/(n-d)\sum_{t=d+1}^n \hat \eps_t \hat \eps_{t,\mathcal{I}}^\top$ and $\hat \Sigma_w=1/(n-d)\sum_{d+1}^n (\hat \eps_t- \widetilde B \hat \eps_{t;\mathcal{I}})(\hat \eps_t- \widetilde B \hat \eps_{t;\mathcal{I}})^\top$. Assumption~\ref{ass1}\eqref{ass1.4} implies sparsity of the matrices $B$ and $\Sigma_w$. This means that ${\Sigma}_w$ can be estimated by thresholding $\hat \Sigma_w$. Since only $B$ and not necessarily a rotated version of $B$ is sparse,  $B$ is estimated by thresholding $\hat B:=\tilde B \hat {\Rot}=1/(n-d)\sum_{t=d+1}^n \hat \eps_t \hat \eps_{t,\mathcal{I}}^\top \hat {\Rot}$. Note that $\hat {\Rot}$ in Section~\ref{sec.estimation} was obtained without any form of regularization on $B$.   Hence, the following regularized estimators of  $ B$ and $\Sigma_w$, and consequently of $ \Sigmaeps$, can be obtained,
\begin{align}
    \hat B^{(re)}=\THRarg{\lambda_B}(\hat B), \ \  \hat{\Sigma}_w^{(re)}=\THRarg{\lambda_w}(\hat \Sigma_w) \ \text{and, }  \ \hat{\Sigma}_{\varepsilon}^{(re)}=\hat B^{(re)}(\hat B^{(re)})^\top +\hat\Sigma_w^{(re)}. \label{eq.est.sigma.eps}
\end{align}
Under these assumptions  we can establish   the following rates for the estimators introduced above,
$$\|\hat B^{(re)}-B\|_1=O_P(\cpB\cpu^{(1-\beta)/2}[\kp^2(\gp/n)^{(1-\zeta)}+\sqrt{\gpeps/n}]^{1-\beta})$$
and 
$$
\|\Sigmah^{(re)}-\Sigmaeps\|_l=O_P\big([\cpD\cpu^{1-\beta}+\cpB\cpu^{(3-\beta)/2}][\kp^2(\gp/n)^{1-\zeta}+\sqrt{\gpeps/n}]^{1-\beta}\big), l \in [1,\infty]
;$$ see Lemma~\ref{lem.est.B} of the Appendix  for details.

Regarding Assumption 2\eqref{ass1.5b}, first note that we focus on short-run identifying restrictions in order to avoid any conflict with necessary sparsity constraints. Furthermore, if $\mathcal{R}=\mathcal{I}$, $\var(\eps_{t,\mathcal{I}})=PP^\top$, where $P$ is the Cholesky factor, and the Cholesky decomposition is used for identification, we have ${\Rot}e_r=(P^\top)^{-1} e_r$, i.e., ${\grestr}(PP^\top)=(P^\top)^{-1} e_r$. The derivative of the Cholesky factor and the matrix inverse can be found in  \cite{luetkepohl2007new}, p. 668-669; see also Remark~\ref{remark.cholesky} of the next section. 




The next theorem is the main result of this subsection and establishes asymptotic normality of the de-sparsified estimator $ \widehat{\Psi}_h^{(de)}$ proposed.

\begin{thm}\label{thm.clt.psi}
If Assumption~\ref{ass.Est} and~\ref{ass1} hold true, then  for $h \in \{1,\dots,H\}$, $ j \in \{1,\dots,p\}$, and $v \in \R^p$ such that  $s.e._\Psi(j,h,v)\not=0$ and $\|v\|_1=O(\cpB)$,   we have that,  
\begin{align*}
\sqrt{n}& \,\frac{\displaystyle e_j^\top(\hat \Psi_{h}^{(de)}-\Psi_{h}) v}{\displaystyle \widehat{s.e.}_\Psi(j,h,v)}\overset{d}{\to}\mathcal{N}(0,1),
\end{align*}
where 
\[\widehat{s.e.}_\Psi(j,h,v)^2=\sum_{t_1,t_2=0}^{h-1} \Big(1-\frac{h+d+|t_2-t_1|}{n}\Big) e_j^\top \hat \Psi_{t_1}\hat\Sigmaeps\hat \Psi_{t_2}^\top e_j v^\top \E^\top \Gammah^{-1} \hat \Gamma^{(st)}(t_2-t_1)\Gammah^{-1}\E v\] 
is an estimator of 
\[s.e._\Psi (j,h,v)^2=\sum_{t_1,t_2=0}^{h-1} e_j^\top \Psi_{t_1}\Sigmaeps \Psi_{t_2}^\top e_j v^\top \E^\top \Gammas^{-1} \Gamma^{(st)}(t_2-t_1) \Gammas^{-1} \E v.\]
Furthermore, it holds true that 
\begin{align*} 
n\cov\Big(e_j^\top(\hat \Psi_{h_1}^{(de)} & -\Psi_{h_1}) v,e_j^\top(\hat \Psi_{h_2}^{(de)}-\Psi_{h_2}) v\Big) \to\\
&  \sum_{t_1=0}^{h_1-1} \sum_{t_2=0}^{h_2-1} e_j^\top \Psi_{t_1}\Sigmaeps \Psi_{t_2}^\top e_j v^\top \E^\top \Gammas^{-1} \Gamma^{(st)}(h_2-h_1+t_2-t_1) \Gammas^{-1} \E v
\end{align*}
and 
$\|\sqrt{n} (\hat \Psi_h^{(de)}-\Psi_h) e_r/\widehat{s.e.}_\Psi(j,h,e_r)\|_{\max}=O_P(\gpeps^{1/2}).$
\end{thm}

As it is seen from the above theorem,  the asymptotic variance   $s.e._\Psi(j,h,v)^2$ of the de-sparsified estimator of the components of $ \Psi^{(de)}_h$ only depends  on the second-order moments of the process $\{X_t\}$ and of the  innovations $\{\eps_t\}$. Furthermore, this variance tends to increase with 
horizon $h$ and  since the underlying SVAR system  is assumed to be stable, this variance  converges to a finite limit as $h\to \infty$. 

\subsection{De-Sparsified Impulse Responses}

In the previous subsection we have constructed de-sparsified estimators of the moving average parameter matrices $\Psi_h$, $h=1,2, \ldots, H$. However, in structural impulse response analysis the  parameters of interest are the elements  $ e_j ^\top \Theta_h e_r =e_j^\top\Psi_h Be_r$
of  the matrices $\Theta_h=\Psi_h B, h=0,\dots,H$, for some index $ j \in\{1,\dots,p\}$,  which refers to the variable of interest and for some index  $r$ which refers to the shock of interest.    Notice that in order  to construct an estimator of $\Theta_h$, $h=0,\dots,H$, an estimator of  the matrix $ B$  is also  needed. Here  we consider again the estimator $ \widehat{B}=\sum_{t=d+1}^n \hat \eps_t \hat \eps_{t,\mathcal{I}}^\top \hat {\Rot}\big/(n-d)$ and establish its asymptotic normality in Theorem~\ref{thm.clt.B}  below. 

\begin{thm} \label{thm.clt.B} 
Under Assumptions~\ref{ass.Est} and~\ref{ass1}, we have
for $r \in \mathcal{I}$ and for any vector $v \in \R^{p}$ with $s.e._{B}(v,r)\not=0$ and $\|v\|_1=O(\kp)$,  that, as $n\to \infty$, 

$$
\frac{\sqrt{n} v^\top (\hat B-B) e_r}{\widehat{s.e.}_{B}(v,r)}\overset{d}{\to} \mathcal{N}(0,1).
$$
Here
\begin{align*}
\widehat{s.e.}_{B}(v,r)^2=&\frac{1}{n-d}\sum_{t=d+1}^n \Big(v^\top (\hat \eps_t \hat \eps_{t}-\Sigmah) I_{d;\mathcal{I}} \hat {\Rot}e_r+v^\top \tilde B \nabla {\grestr}(I_{d;\mathcal{R}}^\top \Sigmah I_{d;\mathcal{I}}) \veco(\hat \eps_{t;\mathcal{R}} \hat \eps_{t;\mathcal{I}}^\top-I_{d;\mathcal{R}}^\top \Sigmah I_{d;\mathcal{I}})\Big)^2, 
\end{align*}
is an estimator of
\begin{align*}
{s.e.}_{B}(v,r)^2=&
\var\Big(v^\top \eps_t u_{t;r}+\eps_{t;\mathcal{R}}^\top\veco_{\cpu,\cpu}^{-1}(v^\top B {\Rot}^{-1} \nabla {\grestr}(I_{d;\mathcal{R}}^\top \Sigmaeps I_{d;\mathcal{I}})){\Rot}^{-1} u_t\Big),
\end{align*}
where $\nabla {\grestr}(x)=\partial \veco {\grestr}(x)/\partial \veco(x)^\top$  and if $x$ is symmetric, then  $\nabla {\grestr}(x)=\partial \veco {\grestr}(x)/\partial \vech(x)^\top L_{\cpu}$. Notice that  $L_{\cpu}$ is the elimination matrix,  see \citet[Ch.~A.12.2]{luetkepohl2007new} for details. 
Furthermore, $$\|\fracd{\sqrt{n} (\hat B-B) e_r}{\widehat{s.e.}_{B}(v,r)}\|_{\max}=O_P(\sqrt{\gpeps}).$$ \end{thm}

We note here  that the asymptotic variance of $\hat Be_r$ also depends  on  the  fourth-order moments of $\{\eps_{t}\}$ and
on  the derivative of the function ${\grestr}$. Observe that if $\{u_t\}$ and $\{w_t\}$ are not only uncorrelated but  also mutually independent, then the fourth-order moments involved in the above expressions for $s.e._{B}(v,r)^2 $, respectively,   $\widehat{s.e.}_{B}(v,r)^2 $, are those of the process  $\{u_t\}$ only. 

\begin{rmk} \label{remark.cholesky}
If $\mathcal{R}=\mathcal{I}$, $\var(\eps_{t,\mathcal{I}})=PP^\top$, where $P$ is the Cholesky factor and the Cholesky decomposition is used for identification, we then have ${s.e.}_{B}(v,r)^2=\var(v^\top D w_t u_{t;r}+v^\top BP^\top (e_r^\top \otimes (PP^\top)^{-1}) L_{\cpu}^\top (L_{\cpu} (I_{{\cpu}^2}+K_{{\cpu}{\cpu}}) (P\otimes I_{\cpu}) L_{\cpu}^\top)^{-1} \vech (\eps_{t;\mathcal{I}}\eps_{t;\mathcal{I}}^\top))$, where $K_{{\cpu}{\cpu}}$ is a commutation matrix, see \citet[Ch.~A.12.2]{luetkepohl2007new}. If additionally $p=\cpu$, i.e., $\eps_t=B u_t$, we then have $ {s.e.}_{B}(v,r)^2=\var((e_r^\top \otimes v^\top) L_{\cpu}^\top (L_{\cpu} (I_{{\cpu}^2}+K_{{\cpu}{\cpu}}) (P\otimes I_{\cpu}) L_{\cpu}^\top)^{-1} \vech (\eps_{t;\mathcal{I}}\eps_{t;\mathcal{I}}^\top))$. A proof of these assertions is  given in the Appendix. Note further that the latter expression for the  variance coincides with  the variance of the contemporaneous effect in the low-dimensional case, see Proposition 3.6 and equation (3.7.8)  in \cite{luetkepohl2007new}. Note the misprint in the aforementioned  equation. 
\end{rmk}

Now, given asymptotically normal  estimators of $B$ and  $\Psi_h$,  we can 
construct
a suitable estimator of the impulse response coefficient  $\Theta_{h;jr}=e_j ^\top \Theta_h e_r=e_j^\top\Psi_h B e_r,$ of interest. Observe  that in the high-dimensional setting considered here,  the statistic  $e_j^\top (\hat \Psi_h^{(de)} \hat B)e_r$ is not a suitable candidate for estimating $\Theta_{h;jr}$. This is due to the fact that the matrix estimators $\hat \Psi_h^{(de)}$ and $\hat B$ 
are not necessarily  consistent with respect to  some matrix norm. As a consequence,  the error term $\sqrt{n}e_j^\top (\hat \Psi_h^{(de)}-\Psi_h)(\hat B-B) e_r$, respectively, its variance,  may  grow with the dimension $p$. In other words,  the variance of  $\sqrt{n} (e_j^\top (\hat \Psi_h^{(de)} \hat B)e_r-e_j^\top \Psi_h Be_r)$ may  diverge. To overcome such  problems, 
the following estimator of  $\Theta_{h;jr} $ is introduced
\begin{align}
    \hat\Theta_{h;jr}^{(de)}=e_j^\top (\hat \Psi_h^{(de)} \hat B)e_r - e_j^\top (\hat \Psi_h^{(de)}-\hat \Psi_h^{(re)})(\hat B-\hat B^{(re)}) e_r. \label{eq.theta.de}
\end{align}
The second term in the above expression is included in order  to correct and, therefore, to control for the estimation error $\sqrt{n}e_j^\top (\hat \Psi_h^{(de)}-\Psi_h)(\hat B-B) e_r$. 
Notice that the estimator $ \hat{\Psi}_h^{(re)}$ 
used   in the above expression is the one discussed in Section 3 while $ \hat{B}^{(re)}$ is the regularized estimator of $B$ given in (\ref{eq.est.sigma.eps}). 
Combining the results of Theorem~\ref{thm.clt.psi} and Theorem~\ref{thm.clt.B}, we can establish  a Gaussian limit for  the estimator $\hat\Theta_{h;jr}^{(de)}$ given in \eqref{eq.theta.de}. This result is stated in the following theorem.

\begin{thm}\label{thm.clt.theta}
Let $r \in \{1,\dots,\cpu\}$ be the shock of interest and $j \in \{1,\dots,p\}, h \in\{0,\dots,H\}$. If Assumptions~~\ref{ass.Est} and \ref{ass1} hold true, then, as $n\to \infty$,
$$
\frac{\sqrt{n}}{\widehat{s.e.}_\Theta(h,j,r)}(\hat \Theta_{h;jr}^{(de)}-\Theta_{h;jr})\overset{d}{\to}\mathcal{N}(0,1),
$$
where
$\widehat{s.e.}_\Theta(h,j,r)^2=\widehat{s.e.}_\Psi(j,h,\hat B^{(re)}e_r)^2+\widehat{s.e.}_B(\hat \Psi_h^{(re)} e_j,r)^2.$
\end{thm}

Similar to the  low-dimensional case, for which the estimator of the variance of the innovations  is asymptotically independent of the estimator of the slope parameters, we get here the result that the asymptotic variance of $\hat\Theta_{h;jr}^{(de)}$ is just  the sum of the  variances  of the moving average estimator $\hat \Psi_h^{(de)}$ and the estimator of contemporaneous impact matrix $\hat B$.


\subsection{Bootstrapping De-Sparsified Impulse Responses}
\label{section.bootstrap}
In addition to the asymptotic Gaussian approximation of the distribution of the de-sparsified estimator of $ \Theta_{h;jr}$ given in   \eqref{eq.theta.de},  a   bootstrap procedure is also proposed in this section to estimate this distribution. The limiting Gaussian distribution in Theorem~\ref{thm.clt.theta} can be used to construct confidence intervals or to implement   tests for hypotheses about the impulse responses $ \Theta_{h;jr}$ of interest. However, the bootstrap can be an useful alternative to the limiting distribution for reasons beyond possible finite sample advantages. First, the bootstrap  avoids  a direct estimation of the standard deviation of the estimator  $\hat \Theta_h^{(de)}$, what may be difficult depending on the particular approach used to identify the structural shocks. Second, if one wants to relax  the i.i.d.~assumption  of  the structural errors, like for instance by allowing for   conditional heteroscedasticity in the innovations,   the bootstrap can easily be adapted  to take care of such a situation; see our comments after Theorem~\ref{thm.boot.valid} below. 

To appropriately implement  a valid bootstrap procedure, we need to additional assume  that the processes  $\{u_t\}$ and $\{w_t\}$ are not only uncorrelated but  also mutually independent. To elaborate on the importance of this assumption, consider  $\hat \Theta^{(de)}_h$ and recall that  this estimator  is based on the  estimators  $\hat \Psi_h^{(de)}$ and $\hat B$. As we have seen in Theorem~\ref{thm.clt.B}, the asymptotic variance of $\hat B$ also depends on the fourth-order moments   of $\{\eps_t\}$. This  means that a valid bootstrap procedure also has to correctly  imitate  
  the fourth-order moments of $\{\eps_{t}\}$. 
However, in the high-dimensional case considered here,  generating the pseudo innovations by drawing with replacement from the estimated residuals $\{\hat \eps_t\}$ does not lead  to a valid procedure. This is due to the fact  that, in our set-up,  the sample covariance matrix $\sum_{t=d+1}^n \hat \eps_t \hat \eps_t^\top\big/(n-d)$  is not a consistent estimator of  $\Sigma_\eps$. Now,  the assumption that  $\{u_t\}$ and $\{w_t\}$ are mutually independent simplifies the problem,  since only the fourth-order moments of the low-dimensional process  $\{u_t\}$ 
affect the distribution of  $\hat \Theta^{(de)}_h$ in this case. Consequently,  only these fourth order moments  have to be correctly imitated  by the bootstrap and not those  of the entire  vector $\eps_t$. This is   achieved   
in the following algorithm by  drawing with replacement from the corresponding set of  estimated residuals $\widehat{u}_t$.

%
  
Now, given estimators $\hat A_1^{(thr)}, \dots,\hat A_d^{(thr)}$, $\hat B^{(re)}$, and  $\hat\Sigma_w^{(re)}$ of the structural autoregressive model, the following  bootstrap procedure  can be used to consistently estimate the distribution of $  \sqrt n (\hat \Theta_{h;jr}^{(de)}- \Theta_{h;jr})$. 

\begin{enumerate}
\item[] {\it Step 1:} \ Generate pseudo innovations
$\{\eps_t^*=\hat B^{(re)} u_t^*+w_t^*, t \in \Z\}$, 
where $u_{t}^*$ is drawn with replacement from the set of estimated and centered residuals $\{\hat u_{t}, t =d+1,\dots,n\}$ and $w_{t}^*$ are i.i.d. with  $w_{t}^* \sim \mathcal{N}(0,\hat\Sigma_w^{(re)})$. 
\item[] {\it Step 2:} \  Generate a pseudo time series  $ X_1^*, X_2^*, \dots,X_n^* $  using the model equation 
$$X_t^*=\sum_{s=1}^d \hat A_s^{(thr)} X^*_{t-s} + \eps_t^*, \ \ t=1,2, \ldots, n$$
and some starting values $ X_0^*, X_{-1}^*, \ldots, X^*_{1-d}$,  where a   burn-in procedure can be used to eliminate the effects of starting values.
 \item[] {\it Step 3:} \ Let $\hat \Theta_{h;jr}^{*(de)}$ be the same de-sparsified estimator of $ \Theta_{h;jr}$ as  the estimator $\hat \Theta_{h;jr}^{(de)}$ given in $\eqref{eq.theta.de}$, but based on the pseudo time series $X_1^*, X_2^*, \ldots, X_n^*$. 
 \item[] {\it Step 4:} \ Approximate the distribution of $  \sqrt n (\hat \Theta_{h;jr}^{(de)}- \Theta_{h;jr})$ by the distribution  of the bootstrap analogue $ \sqrt n (\hat \Theta_{h;jr}^{*(de)}- \hat \Theta_{h;jr}^{(boot)})$, where $\hat \Theta_{h;jr}^{(boot)}=e_{j}^\top \E^\top(\hat\A^{(thr)})^h \E \hat B^{(re)} e_r$.
\end{enumerate}

Notice that thresholded estimators $ \hat A_s^{(thr)}, s=1,\dots,d $,  are used in Step 2 of the bootstrap algorithm. This ensures that  $\hat A_s^{(thr)},s=1,\dots,d$, are   with high probability approximately sparse matrices, i.e., they fulfill Assumptions~\ref{ass.Est} and \ref{ass1}. 
In this way,  the generated pseudo time series $ X_1^\ast, X_2^\ast, \ldots, X_n^\ast$      (asymptotically) stems  from an approximately sparse SVAR$(d)$ model which appropriately imitates the  properties of the underlying SVAR($d$) model. $\hat \Theta^{(boot)}$ is introduced in Step 4 in order to center the bootstrap distribution properly. Note that $\hat \Theta^{(boot)}$ is based on the thresholded estimators, that is,  $\hat \Theta^{(boot)}\not=\hat \Theta^{(Re)}$, in general.


The following theorem  establishes validity of  the bootstrap procedure  in consistently estimating  the distribution of interest. In this theorem,   Mallow's $d_2$ metric is used to measure the distance between two distributions.  For two random variables $X$ and $Y$ with  
cumulative distribution functions  $F_X$ and $F_Y$, respectively,  Mallow's distance between $F_X$ and $F_Y$ is defined as $d_2(X,Y)=\{ \int_0^1 \left(F_X^{-1} (x)-F_Y^{-1}(x)\right)^2 dx\}^{1/2}$, see \cite{bickel1981}. 

\begin{thm}\label{thm.boot.valid}
Let $r \in \{1,\dots,\cpu\}$ be the shock of interest and let $j \in \{1,\dots,p\}$  and  $ h \in\{0,\dots,H\}$. If Assumptions~\ref{ass.Est} and~\ref{ass1} hold, and $\{u_t\}$ and $\{w_t\}$ are mutually independent, then, as $n\to \infty$,
\begin{align*}
d_2\Big(&  \frac{\sqrt{n}}{\widehat{s.e.}_\Theta(h,j,r)}(\hat \Theta_{h;jr}^{(de)}-\Theta_{h;jr}),
\frac{\sqrt{n}}{\widehat{s.e.}^*_\Theta(h,j,r)}(\hat \Theta_{h;jr}^{*(de)}-\hat \Theta_{h;jr}^{(boot)})\Big)=o_P(1).
\end{align*}
\end{thm}

The above theorem,   enables  the use 
of the  bootstrap in order to construct  confidence intervals
 or to perform statistical tests for the impulse responses $ \Theta_{h;jr}$. 
Adaption of  the  bootstrap  procedure proposed to the case of  conditional heteroscedasticity  of the structural shocks $u_t$ can easily be done following  \cite{bruggemann2016inference}. In particular,  in this case,  the i.i.d.~bootstrap applied to generate the pseudo  innovations $u^\ast_t$ in Step 1 of the above algorithm  can be replaced by, for instance,  a version of the block bootstrap applied to the time series of estimated  residuals $\widehat{u}_t$, $t=d+1, \ldots, n$. 

\section{Inference procedures for Impulse Responses and FEVDs} 
\subsection{Confidence Intervals for Impulse Responses} \label{section.CI}
Recall, Theorem~\ref{thm.clt.theta} establishes a Gaussian limit for the  distribution  of $\sqrt{n}(\hat \Theta_{h;jr}^{(de)}-\Theta_{h;jr})$.
By Theorem~\ref{thm.boot.valid} we further have that   the distribution of $\sqrt{n}(\hat \Theta_{h;jr}^{(de)}-\Theta_{h;jr})$ is consistently estimated  by the  distribution of the bootstrap random variable  $\sqrt{n}(\hat \Theta_{h;jr}^{*(de)}-\hat \Theta_{h;jr}^{(boot)})$. Denote by $q^*(\alpha)$ the $\alpha$-quantile of this bootstrap distribution. Using the de-sparsified estimator, an asymptotically $1-\alpha$ confidence intervals for $\Theta_{h;jr}$ can be constructed in the usual way as 
\begin{align}
    \label{eq.conf.interval.de}\big[\hat \Theta_{h;jr}^{(de)}-q^*(1-\alpha/2)/\sqrt{n},
\hat \Theta_{h;jr}^{(de)}-q^*(\alpha/2)/\sqrt{n}\big].
\end{align}
Another possibility is to center the confidence intervals around the regularized estimator $\hat \Theta_{h;jr}^{(re)} $, that is 
\begin{align}
    \label{eq.conf.interval}
     \big[\ \hat \Theta_{h;jr}^{(re)}-q^*(1-\alpha/2)/\sqrt{n},\
\hat \Theta_{h;jr}^{(re)}-q^*(\alpha/2)/\sqrt{n}\ \big].
\end{align}
The regularized estimator  as well as the de-sparsified estimator are both point-wise consistent. In contrast to the regularized estimator and  by construction, the variance of the de-sparsified estimator, however, does not decrease as the response horizon $h$ increases. Notice that a  decrease of the variance of $ \hat \Theta_{h;jr}^{(re)}$ is  expected since  the corresponding true coefficients  $\Theta_{h,;jr} $  decrease  exponentially fast to zero as the horizon $h$ increases. Furthermore,  our simulations also show that de-sparsified estimators  have a  larger variability than  regularized estimators. For these reasons, the confidence interval (\ref{eq.conf.interval})  may overshoot the nominal level  of $1-\alpha$ in finite samples. 
Notice, however,  that both   confidence intervals  have the same  length $\big(q^*(1-\alpha/2)-q^*(\alpha/2)\big)/\sqrt{n}$. That means the gain in coverage of \eqref{eq.conf.interval} is not accompanied  with a loss in power since the length of the intervals  is not affected by the particular centering used. Therefore, we recommend  \eqref{eq.conf.interval}  as a  confidence interval for $\Theta_{h;jr}$ in the high-dimensional setting.

The same arguments also apply if the confidence interval is constructed by using the asymptotic normality established in  Theorem~\ref{thm.clt.theta}. This leads  to the following confidence interval for $ \Theta_{h;jr}$  
\begin{align}
\label{eq.conf.interval.norm}
\big[\ \hat \Theta_{h;jr}^{(re)}-\widehat{s.e.}_\Theta(h,j,r)q(1-\alpha/2)/\sqrt{n},\
\hat \Theta_{h;jr}^{(re)}+\widehat{s.e.}_\Theta(h,j,r)q(1-\alpha/2)/\sqrt{n}\ \big],    
\end{align}
where $q(1-\alpha/2)$ denotes  the $1-\alpha/2$ quantile of the standard normal distribution.

Using  the adaptive lasso as a regularized estimator, the following algorithm summarizes the  steps needed to construct confidence intervals for the impulse response coefficients of interest. 
\begin{enumerate}
\item[] {\it Step 1:} Estimate $\hat A^{(re)}_s,s=1,\dots,d$, by using the row-wise adaptive lasso \eqref{eq.adlasso} with tuning parameter $\lambda_A$.  The threshold parameter $ \lambda_A$ is selected using BIC.  Obtain $\hat A_s^{(thr)}$  by hard thresholding the adaptive lasso estimator with the same  threshold parameter $\lambda_A$. Compute $\hat \Psi_h^{(re)}=\E^\top \hat (\A^{(re)})^h \E,h=1,\dots,H$. 
\item[] {\it Step 2:} Estimate the residuals by $\hat \eps_t=X_t-\sum_{s=1}^d \hat A_s^{(re)} X_{t-j}, t=d+1,\dots,n$. Denote by 
$\widetilde{\eps}_t$ the centered residuals $\widetilde{\eps}_t=\hat{\eps}_t -\sum_{t=d+1}^n\hat{\eps}_t\big/(n-d)$, and compute $\widetilde  B=1/(n-d)\sum_{t=d+1}^n \widetilde \eps_t \widetilde \eps_{t,\mathcal{I}}^\top$ and $\hat \Sigma_w=1/(n-d)\sum_{d+1}^n (\widetilde \eps_t- \widetilde B \widetilde \eps_{t;\mathcal{I}})(\widetilde \eps_t- \widetilde B \widetilde \eps_{t;\mathcal{I}})^\top$. 
\item[] {\it Step 3:} Set up restrictions $\mathcal{R},j=1,\dots,\cpu,$ such that ${\Rot}= \Theta_{\mathcal{R}}^{-1}\grest(\Psi_{0} \cov(\eps_t,\eps_{t;\mathcal{I}}),\dots,\Psi_{H} \cov(\eps_t,\eps_{t;\mathcal{I}}))$ and obtain an estimated rotation matrix $\hat {\Rot}$ with the estimated raw impulse responses $\tilde \Theta_{h}= \hat \Psi_h^{(re)} \tilde B,h=0,\dots,H$.  Obtain then, $\hat B=\tilde B \hat {\Rot}$ and the regularized impulse responses $\hat \Theta_h^{(re)}=\hat \Psi^{(re)}_h \hat B$. 
\item[] {\it Step 4:} Compute  $\hat B^{(re)}=\THRarg{\lambda_B}(\hat B), \hat{\Sigma}_w^{(re)}=\THRarg{\lambda_w}(\hat \Sigma_w)$, and  $\hat\Sigmaeps^{(re)}=\hat B^{(re)}(\hat B^{(re)})^\top +\hat\Sigma_w^{(re)}$. Use soft thresholding and  cross-validation to select the threshold tuning parameters $\lambda_B$ and $ \lambda_w$. 
Given $\hat\Sigma_w^{(re)}$ and $\hat A^{(thr)}_s,s=1,\dots,d$, compute the estimator $\Gammah$ using expression \eqref{eq.GammaHat} for $\Gamma^{(st)}(0)$. 
\item[] {\it Step 5:} Compute for $h=0,\dots,H,$ $\hat \Psi_h^{(de)}$ using \eqref{eq.de_psi} and $\hat \Theta_h^{(de)}$ using \eqref{eq.theta.de}. 
\item[] {\it Step 6:} Use the bootstrap algorithm of Section~\ref{section.bootstrap} to estimate the distribution of $\sqrt{n}(\widehat{\Theta}^{\ast^{(de)}}_{h;jr}- \widehat{\Theta}^{(boot)}_{h;jr}) $ and use \eqref{eq.conf.interval}  as a confidence interval for $ \Theta_{h;jr}$. Alternatively,  the normal approximation  \eqref{eq.conf.interval.norm} can be used to construct a confidence interval for the  same parameter. 
\end{enumerate}
\subsection{Forecast Error Variance Decompositions}
The forecast error variance decomposition of variable $j$ at horizon $h$ and of shock $i$ is defined as 
\begin{align}
    w_{i,j}^h=\frac{\sum_{k=0}^{h-1} \Theta_{k;ij}^2}{\sum_{k=0}^{h-1} e_i^\top \Psi_k \Sigmaeps \Psi_k^\top e_i}, \label{eq.FEVD}
\end{align}
see Section 4.2 in \cite{kilian2017structural}. Note that by   definition,   $w_{i,j}^h$ gives the portion  of explained variance of the linear $h$-step ahead prediction for  variable $i$ caused by the $j$th shock of $\{u_t\}$. Replacing the unknown quantities in \eqref{eq.FEVD} by  the regularized estimators developed in the previous sections leads to the estimator 
$$\hat w_{i,j}^{h,(re)}={\sum_{k=0}^{h-1} (\hat\Theta_{k;ij}^{(re)})^2}/({\sum_{k=0}^{h-1} e_i^\top \hat \Psi_k^{(re)} \Sigmah (\hat \Psi_k^{(re)})^\top e_i}),$$
of $ w^{h}_{i,j}$, where $\Sigmah$ is the sample covariance of $\{\hat \eps_t\}$. Notice that the above  estimator  is consistent under Assumption~\ref{ass.Est} and no sparsity assumption on $B$ or $\Sigmaeps$, respectively, are required. 
Furthermore,  the de-sparsified estimator $\hat\Theta_{k;ij}^{(de)}$ can be used to construct a valid test of  hypotheses about the  $ w^{h}_{ij}$'s under Assumptions~\ref{ass.Est} and \ref{ass1}. In particular, 
 the following testing problem can be considered:
\begin{align}\label{eq.test.problem}
H_0: w_{i,j}^h \leq \delta  \text{ versus } H_1: w_{i,j}^h > \delta,    \end{align}
where we  distinguish the cases $\delta=0$ and $\delta>0$ in the following. 

If $\delta=0$, a test can be easily constructed  by using Theorem~\ref{thm.clt.theta}.
With the covariance expression given in Theorem~\ref{thm.clt.psi}, 
the asymptotic normality of Theorem~\ref{thm.clt.theta} can  be extended to establish asymptotic normality of the  vector $(\hat\Theta_{k;ij}^{(de)})_{k=0,\dots,h-1}$. An asymptotically valid test at level $\alpha$ for the hypotheses given in \eqref{eq.test.problem} with $ \delta=0$ is then obtained by rejecting  the null hypothesis   if 
$$
\begin{pmatrix}
\hat\Theta_{0;ij}^{(de)} & \dots & \hat\Theta_{h-1;ij}^{(de)}
\end{pmatrix}
\hat \Sigma_T^{-1}
\begin{pmatrix}
\hat\Theta_{0;ij}^{(de)} & \dots & \hat\Theta_{h-1;ij}^{(de)}
\end{pmatrix}^\top
> q_{\chi^2_h}(1-\alpha)/n.
$$
Here $q_{\chi^2_h}(1-\alpha)$ is the upper $\alpha$ quantile of the $\chi^2$-distribution with $h$ degrees of freedom and the covariance matrix $\widehat{\Sigma}_T$ is obtained  from  Theorem~\ref{thm.clt.psi} and Theorem~\ref{thm.clt.theta} as  
$$\hat \Sigma_T=\Big(\hat \cov_\Psi(j,r,h_1,h_2)+\widehat{s.e.}_B(\hat \Psi_h^{(re)} e_j,r)^2\Big)_{h_1,h_2=0,\dots,h-1},$$ 
where
\begin{align*}
\hat \cov_\Psi(j,r,h_1,h_2)=&  \sum_{t_1=0}^{h_1-1} \sum_{t_2=0}^{h_2-1} e_j^\top \hat \Psi_{t_1}^{(re)}\hat \Sigmaeps (\hat \Psi_{t_2}^{(re)})^\top e_j (\hat B^{(re)} e_r)^\top \\
& \ \ \ \ \times\E^\top (\Gammah)^{-1} \hat\Gamma^{(st)}(h_2-h_1+t_2-t_1) (\Gammah)^{-1} \E \hat B^{(re)} e_r.
\end{align*}

The case  $\delta>0$ is more involved. To construct an asymptotically $\alpha$ level test in this case,  the randomness of the  denominator in (\ref{eq.FEVD}) should also be taken into account. This can be done as follows. Observe first that 
 $\sum_{k=0}^{h-1} e_i^\top \Psi_k \Sigmaeps \Psi_k^\top e_i=\var(U_{t+h;j})$, where $U_{t+h}=\sum_{j=0}^{h-1}\Psi_j\eps_{t+h-j}$; see  the errors in expression (\ref{eq.x_t_k}). Hence, the denominator of $w_{i,j}^h$ can be estimated  by the sample variance of $\hat U_{t+h;i}=e_i^\top(X_{t+h}-\hat \Xi_h^{(re)} W_t)$. This estimator 
 is asymptotically normal by the same arguments as those used in Lemma~\ref{lem.Sigma1.clt} of the Appendix. We can then write $w_{i,j}^h=f_w(\Theta_{0;ij},\dots,\Theta_{h-1;ij},\var(U_{t+h;i})),$ where the function $f_w$ is defined by $f_w(x_1,\dots,x_h,y)=\sum_{k=1}^h x_k^2/y$. Let  $\nabla f_w$ be the vector of partial derivatives of $ f_w$, that is, $\nabla f_w=(2x_1,\dots,2 x_h,-\sum_{k=1}^h x_k^2/y^2)$. We define  $$\hat w_{i,j}^{h,(de)}=\sum_{k=0}^{h-1} (\hat \Theta_{k;ij}^{(de)})^2/\hat \var(\hat U_{t+h;i})$$ and obtain, by using the delta-method, that  
$\sqrt{n}(\hat w_{i,j}^{h,(de)}- w_{i,j}^{h})/\hat \sigma_{T,\delta}\stackrel{d}{\rightarrow}  \mathcal{N}(0,1)
$, where  straightforward calculations show that\footnote{Note that $\hat \kappa$ is the asymptotic covariance of $n\cov(\hat \var(\hat U_{t+h;i}),\hat \Theta_{h_1;ij}^{(de)}),h_1=0,\dots,h$ and we have $n\cov(\hat \var(\hat U_{t+h;i}),\hat \Theta_{h_1;ij}^{(de)})=
n\cov(\hat \var(\hat U_{t+h;i}),e_i^\top \hat \Psi_{h_1}^{(de)}B e_j)+
n\cov(\hat \var(\hat U_{t+h;i}),e_i^\top \Psi_{h_1} (\tilde B \hat {\Rot} e_j+B \nabla {\grestr}(I_{d;\mathcal{R}}^\top (\Sigmah-\Sigmaeps) I_{I;\mathcal{I}}))) 
+o_P(1)$. With similar arguments as in the proofs of Theorems~\ref{thm.clt.psi} and \ref{thm.clt.B} the used expression for  $\hat \kappa$ follows.} 
$$\hat \sigma_{T,\delta}^2=\nabla \widehat{f}_w^\top \begin{pmatrix}
    \hat \Sigma_T & \hat\kappa \\
    \hat\kappa^\top & \hat f_{U^2}(0)
    \end{pmatrix} \nabla \widehat{f}_w.$$
    Here, $\hat f_{U^2}(0)=  \sum_{k=0}^{h-1} \hat \Gamma_{U^2}(k)(1+\ind(k>0))$, where
    $\hat \Gamma_{U^2}(k)=\hat \cov(\hat U_{t;i}^2,\hat U_{t+k;i}^2)$
    is the sample covariance of $\hat U_{t;i}^2$ and $\hat U_{t+k;i}^2$, and
    $$ \nabla \widehat{f}_w=
\Big(2 \hat \Theta_{0;ij}^{(de)},\dots,2 \hat \Theta_{h-1;ij}^{(de)},-\sum_{k=1}^h (\hat \Theta_{k;ij}^{(de)})^2/(\hat \var(\hat U_{t+h;i}))^2\Big)^\top.$$ 
The estimator $ \hat{\kappa}$ appearing in the expression for $\hat \sigma_{T,\delta}^2 $ is given by  
\begin{align*}
\hat \kappa & =\Big(\sum_{k_1,k_2,k_3=0}^{h-1} e_i^\top \hat \Psi_{k_1}^{(re)} \Sigmah \E^\top (\hat \A^{(re)})^{k_2} \times (\Gammah)^{-1} \E \hat B^{(re)} e_i\\
& \ \ \ \  \times e_j^\top \hat \Psi_{h_1}^{(re)} \Sigmah (\hat \Psi_{k_3}^{(re)})^\top e_i \ind(k_1-h+k_3-k_2\geq 0) \\
& \ \  \ \ +\sum_{k=0}^{h-1} \hat \cov\Big((e_j^\top \hat \Psi_{k}^{(re)} \hat \eps_t)^2, (e_i^\top \hat \Psi_{h_1}^{(re)} \hat \eps_t)(\hat u_{t;j}\\
& \ \ \ \ +e_i^\top \hat \Psi_{h_1}^{(re)}\widetilde B \nabla {\grestr}(I_{d;\mathcal{R}}^\top \Sigmah I_{d;\mathcal{I}}) \veco(\hat \eps_{t;\mathcal{R}} \hat \eps_{t;\mathcal{I}}^\top-I_{d;\mathcal{R}}^\top \Sigmah I_{d;\mathcal{I}}))\Big)\Big)_{h_1=0,\dots,h}.
\end{align*}

An asymptotically valid test at level $\alpha$ 
is then given by rejecting the null hypothesis   $w_{i,j}^h\leq \delta$ if
$$
\sum_{k=0}^{h-1} (\hat \Theta_{k;ij}^{(de)})^2/\hat \var(\hat U_{t+h;i})>\delta+q(1-\alpha)\hat \sigma_{T,\delta}/\sqrt{n}.
$$
Here $q(1-\alpha)$ denotes  the upper $\alpha$-quantile of the standard normal distribution.


\cite{diebold2014network} popularized the method of using forecast error variance decompositions to construct networks and for computing connectedness measures. They analyzed the connectedness of fifteen major US financial institutions. That is, they considered a low-dimensional framework and estimated the forecast error variance decompositions using  (standard) vector autoregressive models. In a high-dimensional framework and based on sparse SVAR models, \cite{demirer2018estimating} and \cite{barigozzi2017network} analyzed the connectedness of many firms. \cite{barigozzi2017network} estimated $w_{i,j}^h$ by using a similar regularized estimator as $\hat w_{i,j}^{h,(re)}$ and obtained a network by defining an edge from component $i$ to $j$ if $\hat w_{i,j}^{h,(re)}$ exceeds   some threshold. This means that a network of relevant connections between actors can be constructed by setting an edge from $i$ to $j$,  if $w_{i,j}^h>\tau$, where $\tau$ is a threshold of  relevance. The test proposed above can  be used to test, at some predefined level of statistical uncertainty,  whether an edge is present or not. Controlling the false discovery rate as in  \cite{benjamini2001control} and \cite{fan2017estimation}, this testing  approach can also be suitable to test multiple edges or even the entire network.

\section{Numerical Results}
\label{section.numerical}
In this section we investigate by means of simulations the finite sample performance of the procedures proposed to construct confidence intervals. The intervals are produced following the steps of the algorithm described at the end of Section~\ref{section.CI}, where the lag length is treated as known. All results presented in this section are based on implementations in  \emph{R} \citep{R}. 
To implement the adaptive lasso with BIC tuning parameter selection, we have modified an estimator of the \emph{HDeconometrics} package \citep{garcia2017real} that itself relies on the \emph{glmnet} package \citep{glmnet}. For $\hat \Sigma_w^{(re)}$ we used the packages \emph{PDCSE} \citep{PDSCE} and for $\hat B^{(re)}$ we modified an estimator of  \emph{FinCovRegularization} \citep{FinCovRegularization}.

For our numerical examples we consider two classes of structural VAR processes the specifications of which are defined below. The parameters of the VARs are generated at random for the $1,000$ Monte Carlo replications that we run for each specification. 
If not denoted otherwise, sparsity of a matrix is obtained by setting entries -- beginning with the absolute smallest values -- to zero such that the specified amount of sparsity is obtained. This results in exact sparsity. 
\begin{enumerate}
      \item[\emph{Class 1}:] VAR$(2)$, $p=100, n=100$, the slope matrix is sparse with $k_A=5$ and the maximum eigenvalue of the slope (stacked) matrix is $0.9$ but otherwise unstructured.  $(u_t^\top,w_t^\top)^\top$ are i.i.d.~standard Gaussian and $\cpu=4, \mathcal{I}=\{1,\dots,\cpu\}$, sparsity parameters $\cpB=5,\cpD=5$, and the eigenvalues of $\Sigma$ are in a range of $0.5$ to $5$ but otherwise unstructured.  The shock of interest is $r=4$. 
      For this class, we consider the following modifications:
      \begin{enumerate}[A)]
    \item $\cpu=8, \cpB=10,\cpD=10, r=6$,
    \item $n=200$,
    \item $\kp=10$,
    \item $t$-distributed innovations with $10$ degrees of freedom.
\end{enumerate}
      \item[\emph{Class 2}:] VAR$(3)$, $p=100, n=100$, the slope matrix is sparse with $k_A=5$ and the maximum eigenvalue of the slope (stacked) matrix is $0.95$ but otherwise unstructured.  $(u_t^\top,w_t^\top)^\top$ are i.i.d.~standard Gaussian and $\cpu=4, \mathcal{I}=\{1,\dots,\cpu\}$, sparsity  parameters $\cpB=5,\cpD=5$, and the eigenvalues of $\Sigma$ are in the range of $0.5$ to $5$ but otherwise unstructured. The shock of interest is $r=4$. 
      For this class, we consider the following modifications:
      \begin{enumerate}[A)]
    \item $p=200$,
    \item $n=200$.
\end{enumerate}

  \end{enumerate}
For instance, \emph{Class 1 A+B} refers to DGPs generated with specification $d=2,p=100,n=200,\cp=5,\rho=0.9,\cpu=8,\mathcal{I}=\{1,\dots,8\},r=6,\cpB=10,\cpD=10$. 
In all cases considered, the Cholesky decomposition is used for identification with ordering $\{1,\dots,\cpu\}$. As outlined in Section~\ref{section.CI}, the Gaussian approximation as well as the bootstrap distribution is used to construct confidence intervals 
at the level $1-\alpha=0.95$. The Gaussian approximation is denoted by  {\OLS} and the bootstrap approximation  by {\DESP}. The quantiles of the bootstrap distribution are based on $1,000$ replicates. As mentioned in Section~\ref{section.CI}, the confidence intervals can be centered around the de-sparsified estimator as well as around the regularized estimator. We present results using  both centering methods. Centering with the regularized estimator is indicated by  \emph{Re} while  centering with the de-sparsified estimator is indicated  by {\emph{De}}.

\begin{figure}[t]
    \centering
    \resizebox{\textwidth}{!}{\input{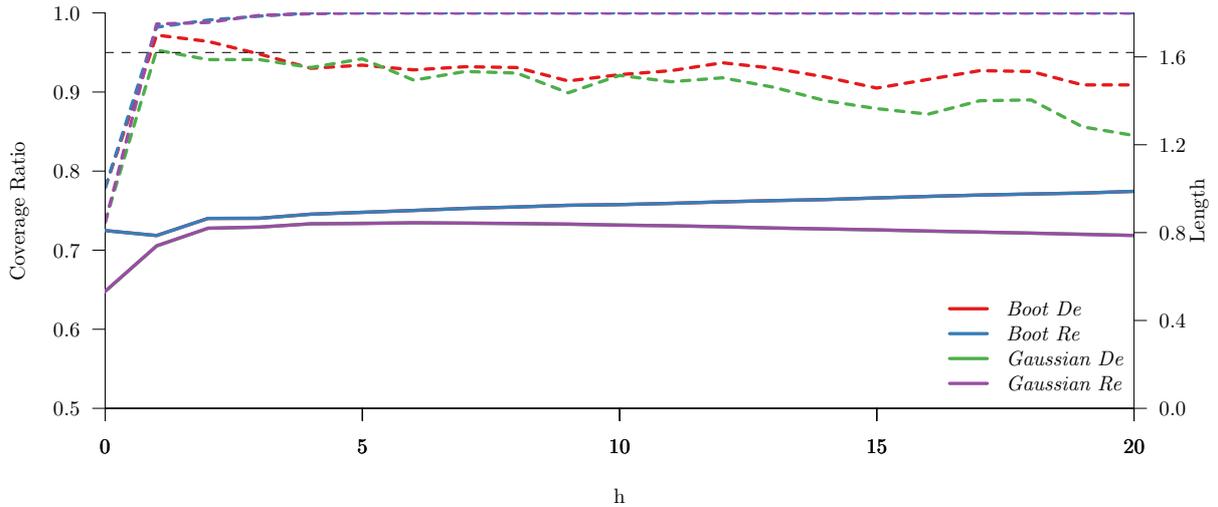}}
    \caption{Average coverage ratios (dashed lines and left vertical axis)  and lengths (solid lines and right vertical axis) of the confidence intervals for variable $j=r$, horizon $h=0,1,\dots,20$ and SVARs of \emph{Class 1}($p=100, n=100, d=2, \rho=0.9, \kp=5,\cpu=4,\cpB=\cpD=5, r=4$). The confidence intervals are constructed at the nominal level  level $0.95$ which indicated  by the horizontal dashed line in black. {\DESP} \emph{De} and {\DESP} \emph{Re} as well as {\OLS} \emph{De} and {\OLS} \emph{Re} have the same length by construction, respectively.}
    \label{fig:base}
\end{figure}

Figure~\ref{fig:base} provides a comparison of the different approaches used to construct confidence intervals for the DGPs of \emph{Class 1}. As it can be seen, both approaches, {\OLS} and {\DESP},  give overall very similar results regarding the interval lengths and the coverage ratios (CR) of the confidence intervals. Variance estimation, i.e., the effect at impact, seems to be  a difficult task as severe  undercoverage (CR $\approx 0.75\%$) can occur for all methods considered. At impact {\DESP} is wider which leads to slightly less undercoverage. Apart from $h=0$, however, both approaches are close to or overshoot the nominal level.  If the confidence intervals are centered around the de-sparsified estimator (\emph{De}), then they  are close to the nominal level (CR: $0.9 - 0.94$). Thus, they are slightly too liberal. This is different for the confidence intervals centered around the regularized estimator (\emph{Re}).  Mostly, the \emph{Re}-intervals have coverage ratios of $1$ or close to $1$ which is partly due to the construction used; see also the discussion in Section~\ref{section.CI}. However, the overcoverage is not associated with a loss in power since \emph{De} and \emph{Re} have the same interval lengths.

\begin{figure}[t]
    \centering
    \resizebox{\textwidth}{!}{\input{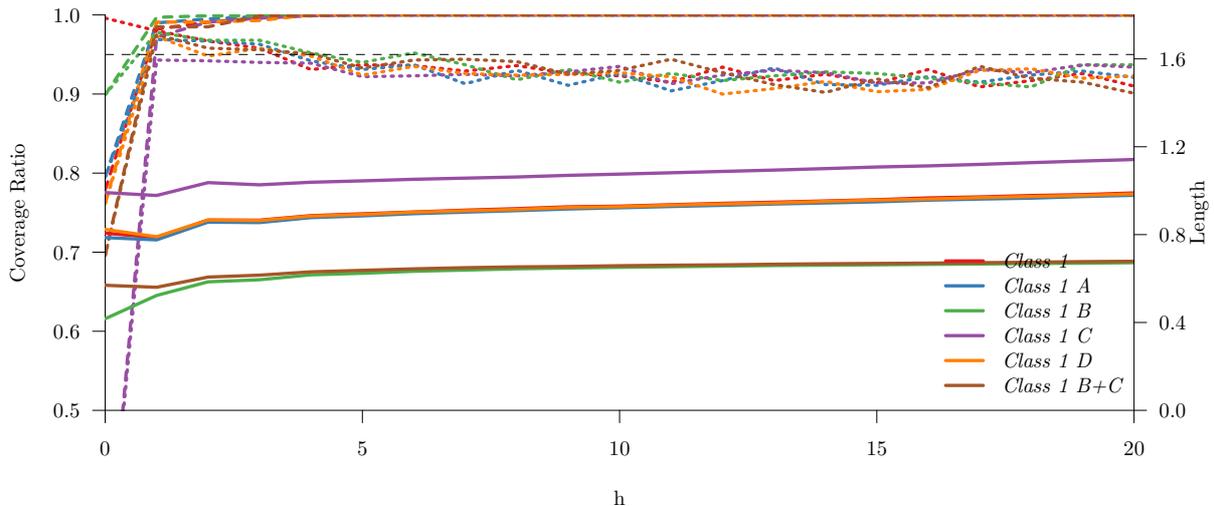}}
    \caption{Average coverage ratios (dashed lines: {\DESP} \emph{Re}, dotted lines: {\DESP} \emph{De}, and left vertical axis) and lengths (solid lines and right vertical axis) of the confidence intervals for variable $j=r$ and horizon $h=0,1,\dots,20$ constructed by {\DESP} \emph{Re} and {\DESP} \emph{De} for SVARs of \emph{Class 1}($p=100, n=100, d=2, \rho=0.9, \kp=5,\cpu=4,\cpB=\cpD=5, r=4$) with modifications: \emph{Case A:} $\cpu=8, \cpB=10,\cpD=10, r=6$; \emph{Case B:} $n=200$; \emph{Case C:} $\kp=10$; \emph{Case D:} $t$-distributed innovations with $10$ degrees of freedom. The confidence intervals are constructed at level $0.95$ as indicated by the horizontal dashed line in black. {\DESP} \emph{De} and {\DESP} \emph{Re} have the same length by construction.}
    \label{fig:comp}
\end{figure}

In the following, the presentation focuses on confidence intervals for variable $j=r$ constructed by {\DESP} \emph{Re} and {\DESP} \emph{De} only, that is, the bootstrap intervals centered around the regularized and de-sparsified estimators.  Figure~\ref{fig:comp} shows a comparison of the different modifications of \emph{Class 1}. Modifications on the innovations, a doubled dimension of the shocks and sparsity parameter of $B$ and $\Sigma_w$ (\emph{Class 1 A}), or a different distribution (\emph{Class 1 D}) do not affect considerably coverage ratios or interval lengths. However, a doubled sparsity parameter in the slope matrix (\emph{Class 1 C}) decreases dramatically the coverage ratio at impact. Furthermore, the confidence intervals are on average $17\%$ wider than in the previous cases. This effect shrinks if  the sample size is also  doubled. A doubled sample size and sparsity parameter of the slope matrix (\emph{Class 1 B+C}) results in less undercoverage at impact and the confidence intervals are only $2\%$ wider than the ones of \emph{Class 1 B}. Compared to the base case, a doubled sample size (\emph{Class 1 B})  improves coverage ratios at impact and decreases the length by the expected factor $1/\sqrt{2}$. 

\begin{figure}[t]
    \centering
    \resizebox{\textwidth}{!}{\input{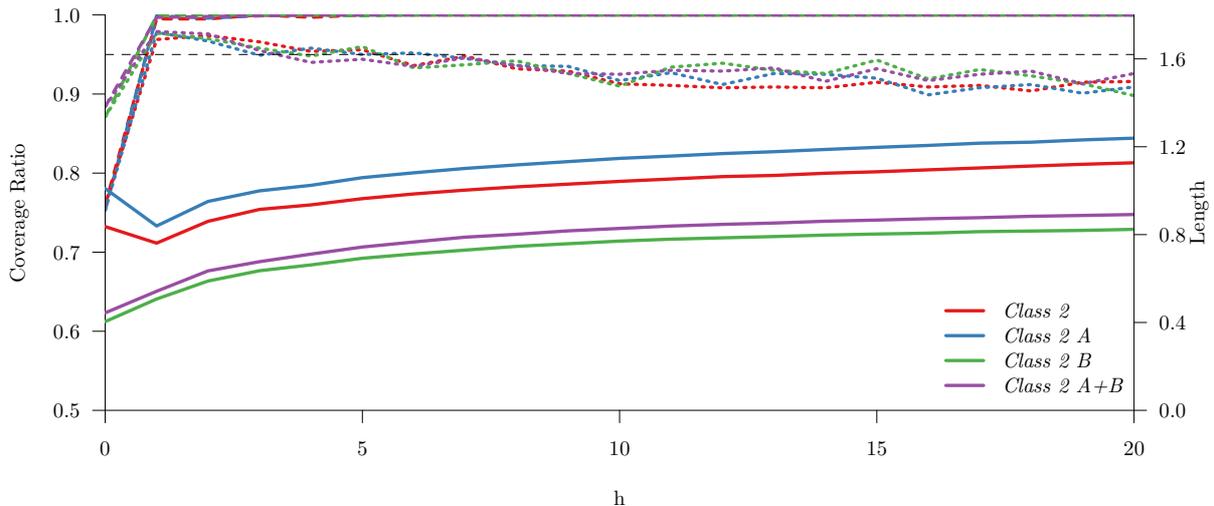}}
    \caption{Average coverage ratios (dashed lines: {\DESP} \emph{Re}, dotted lines: {\DESP} \emph{De}, and left vertical axis) and lengths (solid lines and right vertical axis) of the confidence intervals for variable $j=r$ and horizon $h=0,1,\dots,20$ constructed by {\DESP} \emph{Re} and {\DESP} \emph{De} for SVARs of \emph{Class 2} \ ($p=100, n=100$, $d=3$, $\rho=0.95$, $\kp=5$, $\cpu=4,\ \cpB=\cpD=5,\  r=4$) with modifications: \emph{Case A:} $p=200$ and  \emph{Case B:} $n=200$. The confidence intervals are constructed at level $0.95$ as indicated  by the horizontal dashed line in black. {\DESP} \emph{De} and {\DESP} \emph{Re} have the same length by construction.}
    \label{fig:comp2}
\end{figure}

Figure~\ref{fig:comp2} shows results for different modifications of \emph{Class 2}. It seems that doubling the dimension of the system does not affect the coverage ratios. However,  the lengths of the confidence intervals increase by an average of $12\%$ in case of a sample size of $n=100$ (\emph{Class 2} vs.~\emph{Class 2 A}) and in case of $n=200$ (\emph{Class 2 A+B} vs.~\emph{Class 2 
B}) by  $9\%$. Finally, let us compare the results for \emph{Class 1} and \emph{Class 2} shown in Figures \ref{fig:comp} and \ref{fig:comp2}, respectively. We see that a higher persistence and a larger lag length (\emph{Class 2}) leads to confidence intervals which are wider on average by $22\%$. On the contrary, the coverage ratios are not much affected. 


To sum up, the simulation results have shown that our approaches for obtaining confidence intervals for structural impulse responses performs well in finite sample situations. The intervals centered around the regularized estimator of the impulse response of interest have higher coverage without loosing power in relation to the intervals centered around the de-sparsified estimator. This finding motivates our recommendation to use the regularized estimator for centering the intervals. The differences between the asymptotic Gaussian and the bootstrap approximations seem to be less relevant for the DGPs considered.

\section{Conclusion}
In this paper, we have investigated  how sparse SVAR models can be used to implement  impulse response analysis and to obtained forecast error variance decompositions for large time series systems. We presented a consistent estimator for impulse responses and additionally showed how valid inference can be obtained using a de-sparsified approach.  This approach can be used to construct confidence intervals for impulse responses or  tests for forecast error variance decompositions. In some applications sparsity can be a reasonable assumption. In this case, one can directly apply the inference procedures developed in this paper to analyse the  time series at hand. In other applications,  sparsity  may only be a reasonable assumption after controlling for common factors. Therefore, studying inference in a combined factor plus sparse VAR approach as suggested by \cite{barigozzi2017network} or to include factors as a low-rank matrix within the estimation of the SVAR system are very interesting venues for future research. Further promising future work comprises extensions to time varying coefficients and VAR models with cointegrated time series.\\


{\bf Acknowledgments.}  The authors are grateful to the editor, the associate editor, and three referees for their valuable and insightful comments that led to an improved manuscript. We also thank the participants of the 31th $(EC)^2$ Conference, the seminar of University of Tilburg, and the seminar of Maastricht University for very helpful comments. The research of the first and last  authors was supported by the Research Center (SFB) 884 ``Political Economy of Reforms''(Project B6), funded by the German Research Foundation (DFG). Furthermore, the first author acknowledges support by the state of Baden-W{\"u}rttemberg through bwHPC.

\FloatBarrier
\section*{Appendix}
\subsection*{Auxiliary results and proofs}
\begin{lem} \label{lem.ass}
Let $\hat \Psi_h^{(re)}=\E^\top \hat \A^{(re)} \E$ be the estimator of $\Psi_h, h =1,\dots,H$, where $\hat A_1^{(re)},\dots,\hat A_d^{(re)}$ are regularized estimators of the VAR slope parameters with stacked form $\hat \A^{(re)}$. Furthermore, let $\hat \Xi_h^{(re)}=\E^\top \hat \A^{(re)}$ be the estimator of $\Xi_h=\big(\Psi_h,-\sum_{j=0}^{l+1} \Psi_{h+1+l-j} A_{j}, \ l=0,\dots,d-2\big)=\E^\top \A, h=1,\dots,H$. Let Assumption~\ref{ass1} (i),(ii),(iii) a), and (vi) hold true. Then, 
$$
    \|\hat \Psi_h^{(re)} - \Psi_h\|_\infty=O_P\left( \kp^3 (\frac{\gp}{n})^{(1-\zeta)/2}\right)=\|\hat \Xi_h^{(re)} -\Xi_h\|_\infty.
$$
\end{lem}
\begin{proof}[Proof of Lemma~\ref{lem.ass}]
Using the stacked form representation of the SVAR$(d)$ 
system,   $W_t= \mathds{A} W_{t-1} + \mathds{U}_t$ and because the process $\{W_t\}$ is stable due to Assumption~\ref{ass.Est}\eqref{ass.Est.2}, we have the representation $W_t=\sum_{j=0}^\infty \mathds{A}^j \mathds{U}_{t-j}$. Note further that, $X_t= \E^\top W_t$ and $\mathds{U}_t = \E B u_t$.
Since $X_t=\sum_{j=0}^\infty \Psi_j B u_{t-j}= \sum_{j=0}^\infty  \E^\top  \mathds{A}^j \E B u_t$, we have $\Psi_k=\E^\top \mathds{A}^k \E$. 
Note that for $j \in \N$ we have 
$
(\hat \A^{(re)})^j -\A^j=(\hat \A^{(re)}-\A+\A)((\hat \A^{(re)})^{j-1}-\A^{j-1})+(\A-\hat \A^{(re)})\A^{j-1}$. Using this recursive formula, we obtain
$
(\hat \A^{(re)})^j -\A^j=\sum_{s=0}^{j-1}[(\hat \A^{(re)}-\A)+\A]^s(\hat \A^{(re)}-\A)\A^{j-1-s}.
$ Note further that Assumption~\ref{ass.Est}\eqref{ass.Est.2} implies $\sum_{s=0}^\infty \|\A^s\|_\infty=O(\kp/(1-\varphi))=O(\kp)$ and we have $\|((\hat \A^{(re)}-\A)-\A)^s\|_\infty\leq \sum_{l=0}^s \| \hat \A^{(re)}-\A\|_\infty^l \|\A^{s-l}\|_\infty$. Since $\|\A- \hat \A^{(re)}\|_\infty=O_P(\kp (\gp/n)^{(1-\zeta)/2})$, we have
$
\|\Psi_h - \hat \Psi_h\|=    \|\E^\top (\mathds{A}^h-\hat{\mathds{A}}^h) \E\|_\infty\leq \|\mathds{A}^h-(\hat{\mathds{A}}^{(re)})^h\|_\infty=O_P(\kp ^3 (\gp/n)^{(1-\zeta)/2})
$
and similarly $\|\hat \Xi_h^{(re)} -\Xi_h\|_\infty\leq \|\mathds{A}^h-(\hat{\mathds{A}}^{(re)})^h\|_\infty$. 
\end{proof}

\begin{proof}[Proof of Theorem \ref{thm.estimation}]
We have $\Theta_h {\Rot}^{-1}=\Psi_h BH^{-1}=\Psi_h \cov(\eps_t,\eps_{t;\mathcal{I}})$. By Lemma~\ref{lem.ass} we have $\|\hat \Psi_h^{(re)} - \Psi_h\|_\infty=O_P(\kp^3 (\gp/n)^{(1-\zeta)/2})$. For $BH^{-1}$ consider the following.  The residuals $\{\eps_t\}$ can be estimated by $\hat \eps_t=X_t-\sum_{j=1}^d \hat A_j^{(re)} X_{t-j}=\eps_t+\sum_{j=1}^d (A_j-\hat A_j^{(re)}) X_{t-j}$. Furthermore, note that under Assumption~\ref{ass.Est}~(\ref{ass.Est.3}) the estimation error is of order 
$\|1/(n-d)\sum_{t=d+1}^n \hat \eps_t \hat \eps_t^\top-1/(n-d)\sum_{t=d+1}^n \eps_t \eps_t^\top\|_{\max}=O_P(\kp^2((\gp/n)^{(1-\zeta)}+(\gp/n)^{(1-\zeta)/2}\sqrt{\gpeps/n}))$, see for instance equation (32) in \cite{krampe2020Est}. Consequently, we obtain  $\|1/(n-d)\sum_{t=d+1}^n \hat \eps_t \hat \eps_t^\top-\Sigmaeps\|_{\max}=O_P(\sqrt{\gpeps/n}+\kp^2(\gp/n)^{(1-\zeta)})$. With the estimated residuals, we can obtain the raw shocks, i.e., those shocks which are identified up to a rotation matrix ${\Rot}\in \R^{\cpu\times \cpu}$, and the raw impulse response. 
Since $e_j^\top \Big(\Theta_h {\Rot}^{-1} - \tilde \Theta_h^{(re)}\Big) e_r=\|\hat \Psi_h^{(re)} - \Psi_h\|_\infty \| BH^{-1}\|_{\max}+\|\Psi_h\|_\infty\|\tilde B- BH^{-1}\|_{\max}+\|\hat \Psi_h^{(re)}-\Psi_h\|_\infty\|\tilde B- BH^{-1}\|_{\max}$, the assertion follows. 
\end{proof}

\begin{proof}[Proof of Theorem~\ref{thm.estimation2}]
First note that we have $e_j^\top (\hat\Theta_h^{(re)} -  \Theta_h) e_r=e_j^\top(\hat \Psi_h^{(re)}-\Psi_h) Be_r+e_j^\top \Psi_h (\hat B^\dagger - B) e_r+e_j^\top(\hat \Psi_h^{(re)}-\Psi_h)(\hat B^\dagger - B) e_r+e_j^\top \Psi_h \cov(\eps_t,\eps_{t;\mathcal{I}}) (\hat {\Rot} - {\Rot}) e_r+e_j^\top(\tilde \Theta_h^{(re)}-\Theta_h {\Rot}^{-1}) (\hat {\Rot}-{\Rot})e_r$, where $\hat B^\dagger=1/(n-d)\sum_{t=d+1}^n \hat \eps_t \hat\eps_{t;\mathcal{I}} {\Rot}$. Then, since $1/(n-d) \sum_{t=d+1} (\hat\eps_{t;\mathcal{I}} {\Rot} e_s)^2\overset{P}{\to} \var(u_{t,s})=1, s=1,\dots,\cpu$, as $n\to \infty$ and Assumption~\ref{ass.Est}~(\ref{ass.Est.3}), we have $\|\hat B^\dagger-B\|_{\max}=O_P(\kp^2(\gp/n)^{(1-\zeta)}+\sqrt{\gpeps/n})$.

Let $\tilde \Theta_\mathcal{R}=\begin{pmatrix}
e_{i_1}^\top \tilde \Theta_{h_1}^{(Re)} \\
\vdots\\
e_{i_{\cpu}}^\top \tilde\Theta_{h_{\cpu}}^{(Re)}
\end{pmatrix}$ and let $H_\mathcal{R}$ denote the maximal horizon present in $\mathcal{R}$. We have $\|\tilde \Theta_\mathcal{R}-\Theta_\mathcal{R}\|_2\leq \sqrt{\cpu} \max_{0\leq h\leq H_\mathcal{R}}\|\tilde \Theta_h^{(Re)}-\Theta_h\|_{\max}$ and the same for the vectors of restrictions.
Since $\rho(\Theta_\mathcal{R}^{-1})<\infty$, the same rate holds for $\|\tilde \Theta_\mathcal{R}^{-1}-\Theta_\mathcal{R}^{-1}\|_2$. If $H_\mathcal{R}\geq 1$, i.e., in the case of \eqref{eq.thm.est.all}, we have $\max_{0\leq h\leq H_\mathcal{R}}\|\tilde \Theta_h^{(Re)}-\Theta_h\|_{\max}$ is order of $O_P(\sqrt{\cpu}(\kp^3 (\gp/n)^{(1-\zeta)/2}+\kp\sqrt{\gpeps/n}))$ whereas if $H_\mathcal{R}=0$, i.e., in the case of \eqref{eq.thm.est.short}, we have $\max_{0\leq h\leq H_\mathcal{R}}\|\tilde \Theta_h^{(Re)}-\Theta_h\|_{\max}=\|\tilde B-\cov(\eps_t,\eps_{t;\mathcal{I}})\|_{\max}=O_P(\sqrt{\cpu}(\kp^2(\gp/n)^{(1-\zeta)}+\sqrt{\gpeps/n}))$  and the assertion follows.



\end{proof}

\begin{rmk}
To quantify the dependence structure of the stochastic processes,
we use the concept of functional/physical dependence, see  \cite{wu2005nonlinear,wu2011asymptotic}. To elaborate, 
we write for a random variable $X$, we write $ \|X\|_{E,q}$ for $\big(E|X|^q\big)^{1/q}$, where   $q\in \N$ and let $Y_{t;i}=G_i(\eps_t,\eps_{t_1},\dots,), i=1,\dots,p, t \in \Z,$ be a stochastic  process generated causally by the i.i.d. processes $\{\eps_t\}$ for some function $G=(G_1,\dots,G_p)$. We denote by $Y_{t;i}^{\prime(k)}=G_i(\eps_t,\eps_{t-1},\dots,\eps_{t-k+1},\eps_{t-k}^\prime,\eps_{t-k-1},\eps_{t-k-2},\dots)$ the process where $\eps_{t-k}$ is replaced by an i.i.d.~copy of it.
Furthermore, define the functional dependence coefficients in the following way. Let $\delta_{k,q,i}=\|Y_{0;i}-Y_{0;i}^{\prime(k)}\|_{E,q}, k \geq 0$, $\Delta_{m,q;i}=\sum_{k=m}^\infty \delta_{k,q;i}$, $\|Y_{;i}\|_{q,\alpha}=\sup_{m\geq 0} (m+1)^\alpha \Delta_{m,q;i}$, and $\nu_{q;i}=\sum_{j=1}^\infty (j^{q/2-1} \delta_{k,q;i})^{1/(q+1)}$. For one-dimensional processes, e.g. $\{Z_t\}$, we drop the index $i$ in the subscript, i.e, we write $\delta_{k,q}, \Delta_{m,q}, \|Z\|_{q,\alpha}$ and $ \nu_q$.

\noindent The processes $\{X_t\}$ possesses under Assumption~\ref{ass1} the following causal representation $X_t=\sum_{j=0}^\infty \Psi_k \eps_{t-j}$. Hence, for some vector $v$ with $\|v\|_2=1$ and if $\|v^\top \Psi_k\|_2>0$, we have $\|v^\top (X_k-X_k^\prime)\|_{E,q}=\| v^\top\Psi_k(\eps_k-\eps_k^\prime)\|_{E,q} \leq \|v^\top \Psi_k\|_2  \| v^\top  \Psi_k/(\|v^\top \Psi_k\|_2) \eps_1\|_{E,q}=O(\lambda^k), \lambda\in(0,1)$. Such a geometrical decay implies $\nu_q=O(1)$ for $\{v^\top X_t\}$ and some transformations of it, see also {\Lemmaelf} in \cite{krampe2018bootstrap}.
\end{rmk}

\begin{lem} \label{lem.filtering.eps}
Let $\{\Phi_j^{(k)}, j=0,1,\dots\},k=1,2$ be linear filters with $\sum_{j=0}^\infty \|\Phi_j^{(k)}\|_2=O(1),k=1,2$. Then under Assumption~\ref{ass.Est}\eqref{ass.Est.4}
$$
\|1/\sqrt{n} \sum_{t=1}^n \sum_{j,k=0}^\infty \Phi_j^{(1)} (\eps_{t-j} \eps_{t-k}^\top - \ind(j=k)\Sigmaeps) (\Phi_k^{(2)})^\top \|_{\max}=O(\sqrt{\gpeps}).
$$
\end{lem}
\begin{proof}[Proof of Lemma~\ref{lem.filtering.eps}]
Since $
    \|1/n \sum_{t=1}^n \sum_{j,k=0}^\infty \Phi_j^{(1)} \eps_{t-j} \eps_{t-k}^\top (\Phi_k^{(2)})^\top\|_{\max}\leq \sum_{j,k=0}^\infty \|\Phi_j^{(1)}\|_2 \|\Phi_j^{(2)}\|_2 \times \linebreak \|1/n \sum_{t=1}^n \Phi_j^{(1)}/\|\Phi_j^{(1)}\|_2 (\eps_{t-j}\eps_{t-k}-\ind(j=k)\Sigmaeps)(\Phi_j^{(2)})^\top/\|\Phi_j^{(2)}\|_2\|_{\max}
$ and $\{\eps_t\}$ is i.i.d., the assertion follows by Assumption~\ref{ass.Est}\eqref{ass.Est.4}, see also Section 2.2. in \cite{wu2016performance}.
\end{proof}

\begin{lem} \label{lem.est.B}
Let $\THR$ be a threshold function with threshold value $\lambda$ fulfilling the conditions $(i)$ to $(iii)$ in Section 2 in \cite{cai2011adaptive}, see also \cite{rothman2009generalized}. Then, let 
\begin{align*}
    \hat B^{(re)}=\THR(\hat B) \ \text{ and }  \ \hat\Sigmaeps^{(re)}=\hat B^{(re)}(\hat B^{(re)})^\top +\hat\Sigma_w^{(re)},
\end{align*}
where $\hat B=1/(n-d)\sum_{t=d+1}^n \hat \eps_t \hat \eps_{t;\mathcal{I}}^\top \hat {\Rot}$ and $\hat{\Sigma}_w^{(re)}=\THR(\hat \Sigma_w),$ where $\hat \Sigma_w=1/(n-d)\sum_{d+1}^n (\hat \eps_t- \widetilde B \hat \eps_{t;\mathcal{I}})(\hat \eps_t- \widetilde B \hat \eps_{t;\mathcal{I}})^\top$ and $\widetilde  B=1/(n-d)\sum_{t=d+1}^n \hat \eps_t \hat \eps_{t,\mathcal{I}}^\top$ with   $\hat \eps_t=X_t-\sum_{s=1}^d \hat A_s^{(re)} X_{t-j}$ the estimated residuals.

Then, 
under 
Assumption~\ref{ass.Est} and Assumption~\ref{ass1} \eqref{ass1.4}, \eqref{ass1.5}, 
the following assertions are true:
$$
\|\hat B - B \|_{\max}=O_P\Big(\sqrt{\cpu}[\kp^2(\gp/n)^{(1-\zeta)}+\sqrt{\gpeps/n}]\Big)
$$
$\|(\hat B^{(re)}-B)\|_\infty=O_P(\cpu^{3/2}[\kp^2(\gp/n)^{(1-\zeta)}+\sqrt{\gpeps/n}])$, \\
$\|(\hat B^{(re)}-B)\|_1=O_P(\cpB\cpu^{(1-\beta)/2}[\kp^2(\gp/n)^{(1-\zeta)}+\sqrt{\gpeps/n}]^{1-\beta})$,\\
$\|(\hat B^{(re)}-B)\|_2=O_P(\cpu^{(5-\beta)/4}\sqrt{\cpB}[\kp^2(\gp/n)^{(1-\zeta)}+\gpeps/n]^{1-\beta/2})$,\\ $\|\Sigma\|_\infty=O(\cpu\cpB+\cpD),\,\|\Sigma\|_2=O(1)$, and 
\begin{align*}
\|\Sigmah^{(re)}-\Sigmaeps\|_l&=O_P\bigg(\big[\cpD\cpu^{1-\beta}+\cpB\cpu^{(3-\beta)/2}\big]\Big[\kp^2(\gp/n)^{1-\zeta}+\sqrt{\gpeps/n}\Big]^{1-\beta}\bigg)\\
&=:O_P(\cp (\gpeps/n)^{(1-\beta)/2}), \ \ l \in [1,\infty].    
\end{align*}
\end{lem}


\begin{proof}[Proof of Lemma~\ref{lem.est.B}]
First note that $\|\hat B-B\|_{\max}\leq \|\hat B -\hat B^\dagger\|_{\max}+\|\hat B^\dagger-B\|_{\max}$, where $\hat B^\dagger=1/(n-d)\sum_{t=d+1}^n \hat \eps_t \hat\eps_{t;\mathcal{I}} {\Rot}$ as in the proof of Theorem~\ref{thm.estimation2} and we have $\|\hat B^\dagger-B\|_{\max}=O_P(\kp^2(\gp/n)^{(1-\zeta)}+\sqrt{\gpeps/n})$. Furthermore, $\|\hat B -\hat B^\dagger\|_{\max}\leq \|\cov(\eps_t,\eps_{t;\mathcal{I}})\|_2 \|\hat {\Rot} - {\Rot}\|_2+\|1/(n-d)\sum_{t=d+1}^n \hat \eps_{t} \hat \eps_{t;\mathcal{I}}^\top -\cov(\eps_t,\eps_{t;\mathcal{I}})\|_{\max} \|{\Rot}-\hat {\Rot}\|_1$. Hence, since $\|\cov(\eps_t,\eps_{t;\mathcal{I}})\|_2<M$ due to  Assumption~\ref{ass1}\eqref{ass1.4} and by the arguments of the proof of Theorem~\ref{thm.estimation2}, we have $\|\hat B -B\|_{\max}=O_P(\sqrt{\cpu}(\kp^2(\gp/n)^{(1-\zeta)}+\sqrt{\gpeps/n}))$. Since $B\in \R^{ \cpB \times \cpu},$ the rates $\|\hat B^{(re)}-B\|_l, l\in \{1,2,\infty\}$ follows by the same arguments as in the proof of Theorem~1 in \cite{cai2011adaptive}. 

Furthermore, we have $\Sigmaeps=BB^\top+DD^\top$ and $B^\top \in \mathcal{U}(\cpB,\beta)$, $DD^\top \in \mathcal{U}(\cpD,\beta)$. That implies $\|BB^\top\|_\infty\leq \|B\|_\infty \|B^\top \|_\infty\leq \cpu M \|B^\top \|_\infty=O(\cpu\cpB)$. Note that for some vector $p$-dimensional $v \in \mathcal{U}(k,q)$, we have $\|v\|_1=\sum_{i=1}^p |v_i| \ind(|v_i|>\nu_q)+\sum_{i=1}^p |v_i| \ind(|v_i|\leq\nu_q)\leq M |S_{\nu_q}|+\nu_q k=O(k),$ where $S_{\nu_q}=\{i=1,\dots,p : |v_i|>\nu_q\}$ and $|S_{\nu_q}|\leq k/(\nu_q)^q$ and $\nu_q=q M/(1-q)$. 

For $DD^\top$, let $V=I_{p}-BH^{-1}I_{p;\mathcal{I}}$ and $\hat V=I_{p}-\tilde B I_{p;\mathcal{I}}$ such that $V\eps_t=w_t$ and $(\hat \eps_t- \tilde B \hat \eps_{t;\mathcal{I}})=\hat V \hat \eps_t$. Furthermore, we have $\|V-\hat V\|_\infty\leq \|BH^{-1}-\tilde B\|_\infty\leq\cpu\|BH^{-1}-\tilde B\|_{\max}=O_P(\cpu[\sqrt{\gpeps/n}+\kp^2(\gp/n)^{(1-\zeta)}])$, $\|V\|_2=O(1)$, and $\|V\|_\infty=O(\cpu)$. Then, since $\|1/(n-d)\sum_{t=d+1}^n V(\hat \eps_t-\eps_t) (\hat \eps_t-\eps_t)^\top V^\top\|_{\max}\leq \|V\|_\infty^2 \|\hat \A-\A\|_\infty^2 \|1/(n-d)\sum_{t=d+1}^n W_t W_t^\top\|_{\max}=O_P(\cpu^2\kp^2 (\gp/n)^{1-\zeta})$, where $W_t=( X_t^\top,X_{t-1}^\top,\dots,X_{t-d+1}^\top)^\top$, we have $\|1/(n-d)\sum_{d+1}^n (\hat \eps_t- \tilde B \hat \eps_{t;\mathcal{I}})(\hat \eps_t- \tilde B \hat \eps_{t;\mathcal{I}})^\top-DD^\top\|_{\max}=\|1/(n-d)\sum_{d+1}^n \hat V (\hat \eps_t \hat \eps_t^\top-\eps_t \eps_t^\top) \hat V^\top+ \hat V \eps_t \eps_t^\top \hat V^\top-V \eps_t \eps_t^\top V^\top+D (w_t w_t^\top-I_p) D^\top\|_{\max}=O_P\Big(\cpu\times \\\Big(\kp^2(\gp/n)^{1-\zeta}+\sqrt{\gpeps/n}\Big)\Big)$. Since $DD^\top \in \mathcal{U}(\cpD,\beta)$,  the assertion follows with the results above and again by the arguments as in the proof of Theorem~1 in \cite{cai2011adaptive}.

\end{proof}

\begin{lem} \label{lem.est.gamma}
Let $\E=(e_1\otimes I_p) \in \R^{dp\times p}$ as in Lemma~\ref{lem.ass}.
If Assumption~\ref{ass.Est} and Assumption~\ref{ass1} hold true, then the estimator of $\Gamma^{(st)}(h)$, given  for $h\geq0$ by  
\begin{align}
\label{eq.Ghat-thr}
    \hat \Gamma^{(st)}(h)&=\sum_{j=0}^\infty (\hat {\mathds{A}})^{j+h}\E \Sigmah^{(re)} \E^\top (\hat{\mathds{A}}^{\top})^j=(\hat {\A})^{h}\operatorname{vec}^{-1}_{dp} \Big((I_{(dp)^2}-\hat {\mathds{A}}\otimes\hat {\mathds{A}})^{-1} \operatorname{vec}(\E\hat \Sigma_{\varepsilon})\Big),
\end{align}
and for $h <0$ by $\hat \Gamma^{(st)}(h)=\hat \Gamma^{(st)}(-h)^\top$,   satisfies 

\begin{align*}
\|\hat \Gamma^{(st)}(h)-\Gamma^{(st)}(h)\|_\infty=&O_P\bigg(\kp^2\Big(\kp^{2.5} (\cpu\cpB+\cpD) (\gp/n)^{(1-\zeta)/2} \\&+ \big[\cpD\cpu^{1-\beta}+\cpB\cpu^{(3-\beta)/2}\big]\Big[\kp^2(\gp/n)^{1-\zeta}+\sqrt{\gpeps/n}\Big]^{1-\beta}\Big)\bigg)    
\end{align*}
and 
$$
\|\hat \Gamma^{(st)}(h)-\Gamma^{(st)}(h)\|_2=O_P(\kp^{1.5} (\gp/n)^{(1-\zeta)/2}+\big[\cpD\cpu^{1-\beta}+\cpB\cpu^{(3-\beta)/2}\big]\Big[\kp^2(\gp/n)^{1-\zeta}+\sqrt{\gpeps/n}\Big]^{1-\beta}).
$$
Furthermore, we also have 
\begin{align*}
\|\hat \Gamma^{(st)}(0)^{-1}-\Gamma^{(st)}(0)^{-1}\|_\infty=&O_P\bigg(\kpp^2\kp^2\Big(\kp^{2.5} (\cpu\cpB+\cpD) (\gp/n)^{(1-\zeta)/2} \\&+ \big[\cpD\cpu^{1-\beta}+\cpB\cpu^{(3-\beta)/2}\big]\Big[\kp^2(\gp/n)^{1-\zeta}+\sqrt{\gpeps/n}\Big]^{1-\beta}\Big)\bigg)    
\end{align*}
and
$$
\|\hat \Gamma^{(st)}(0)^{-1}-\Gamma^{(st)}(0)^{-1}\|_2=O_P(\kp^{1.5} (\gp/n)^{(1-\zeta)/2}+\big[\cpD\cpu^{1-\beta}+\cpB\cpu^{(3-\beta)/2}\big]\Big[\kp^2(\gp/n)^{1-\zeta}+\sqrt{\gpeps/n}\Big]^{1-\beta}).
$$
\end{lem}
\begin{proof}[Proof of Lemma~\ref{lem.est.gamma}]
Note that $\|\hat \A-\A\|_l=O_P(\kp^{1.5} (\gp/n)^{(1-\zeta)/2}).$ Then, we have by the arguments of Lemma~\ref{lem.ass}, that  $\sum_{j=0}^\infty \|(\hat \A)^j -\A^j\|_\infty=O_P( \kp^{3.5} (\gp/n)^{(1-\zeta)/2})$. Furthermore, we have \linebreak
$
\|\Gammas-\Gammah\|_\infty \leq \|\sum_{j=0}^\infty ({\A}^j-(\hat {\A})^j) \E\Sigmaeps(\E^\top ({\A}^\top)^j\|_\infty+\|\sum_{j=0}^\infty ({\hat {\A}})^j \E(\Sigmaeps-\hat \Sigma_\eps)\E^\top ({\A}^\top)^j\|_\infty+\|\sum_{j=0}^\infty ({\hat {\A}})^j\E \hat\Sigma_\eps\E^\top \! (({\A}^\top)^j-(\hat {\A}^{\top})^j)\|_\infty=O_P(\kp^{4.5} (\cpu\cpB+\cpD) (\gp/n)^{(1-\zeta)/2}+ \kp^2\big[\cpD\cpu^{1-\beta}+\cpB\cpu^{(3-\beta)/2}\big]\times \big[\kp^2(\gp/n)^{1-\zeta}+\sqrt{\gpeps/n}\big]^{1-\beta})$, where the last equality follows by Assumption~\ref{ass1} and Lemma~\ref{lem.est.B}. The third assertion follows then due to $A^{-1}-B^{-1}=A^{-1}(B-A)B^{-1}$.
Note that $\|\A^j\|_2=O(\lambda^j)$ implies $\|\Gammas\|_2=O(1/(1-\lambda)^2 \|\Sigmaeps\|_2).$ Furthermore, we have $\|\hat \A-\A\|_2=\|\hat \A^{\top}-\A^\top\|_2\leq \sqrt{\|\hat \A-\A\|_1 \|\hat \A-\A\|_\infty}=O_P(\kp^{1.5} (\gp/n)^{(1-\zeta)/2})$. Following the same arguments above with the norm $\|\cdot\|_2$ leads to the second and fourth assertion.
\end{proof}

\begin{lem} \label{lem.bias}
If Assumption~\ref{ass.Est} and \ref{ass1} hold true, we have for $h \in \{0,\dots,H\}, j \in \{1,\dots,p\},$ and $v \in \R^p$ with $s.e._\Psi(j,h,v)\not=0$ and $\|v\|_1=O(\cpB)$,
\begin{align*}
&\sqrt{n} e_j^\top(\hat \Psi_{h}^{(de)}-\Psi_{h}) v/s.e._\Psi(j,h,v)=\frac{1}{\sqrt{n}} \sum_{t=d}^{n-h} v^\top (I_p \otimes e_1)^\top(\Gammas)^{-1} W_t U_{t+h;j}/s.e._\Psi(j,h,v)\\
&+O_P\Big(\cpB(\gpeps/\sqrt{n}+\|\Gammah-\Gammas\|_1 \kpp \sqrt{\gpeps}\\
&\qquad +\|\Xi_{h}-\hat \Xi_{h}^{(re)}\|_\infty(\sqrt{\gpeps}+\|\Gammas-\Gammah\|_\infty \kpp \sqrt{\gpeps}))\Big),
\end{align*}
where
$s.e._\Psi(j,k,v)^2=\sum_{t_1,t_2=0}^{k-1} e_j^\top \Psi_{t_1}\Sigmaeps \Psi_{t_2}^\top e_j v^\top \E^\top(\Gammas)^{-1} (\Gamma^{(st)}(t_2-t_1)) (\Gammas)^{-1} \E v$. 
\end{lem}
\begin{proof}[Proof of Lemma~\ref{lem.bias}]
Since $\sqrt{n} (e_j^\top (\hat \Psi_h^{(de)}-\Psi_h)v)/s.e._\Psi(j,h,v)=
\sqrt{n} (e_j^\top (\hat \Psi_h^{(de)}-\Psi_h)v/\|v\|_1)/s.e._\Psi(j,h,v/\|v\|_1)$ and $1/s.e._\Psi(j,h,v/\|v\|_1)=O(\|(\Gammas)\|_2\|(\Gammas)^{-1}\|_2 \|v\|_1/\|v\|_2)=O(\cp)$, we assume in the following that the vector is normalized, that is,  $\|v\|_1=1$. 
Let $DN_{h,r,j}=\Big(1/n\sum_{t=d}^{n-h} e_r^\top (\Gammah)^{-1} W_t W_{t}^\top e_j \Big)$ and note that $\hat Z_{t;r}=e_r (\Gammah)^{-1} W_t (e_r (\Gammah)^{-1} e_r)^{-1}$. $DN_{h,r,j}$ is the denominator and we  show that it converges to one.
By Lemma~\ref{lem.filtering.eps}, we have $\max_r |DN_{h,r,r}-1|=\max_r |1/n\sum_{t=d}^{n-h} e_r^\top (\Gammas)^{-1} W_t W_{t}^\top e_r-1| + \max_r |1/n\sum_{t=d}^{n-h} e_r^\top\big[(\Gammah)^{-1} \Gammas - I_{dp}\big] (\Gammas)^{-1} W_t W_{t}^\top e_r|=O_P(\sqrt{\gpeps/n}+\| \Gammas-\Gammah\|_\infty\|(\Gammas)^{-1}\|_\infty)$. Since $\Gamma>0$,  $(\Gammah)^{-1}>0$, we have $\max_r |DN_{h,r,r}^{-1}-1|=O_P(\sqrt{\gpeps/n}+\| \Gammas-\Gammah\|_\infty \kpp)$. Then, 
\begin{align*}
\sqrt{n}& e_j^\top(\hat \Psi_{k}^{(de)}-\Psi_{k}) v
    =
    \sqrt{n}\sum_{r=1}^p v_r\Big(\sum_{t=d}^{n-k} \hat Z_{t;r} W_{t;r}\Big)^{-1}\Big[\sum_{t=d}^{n-k} \hat Z_{t;r} (U_{t+k;j}+(\Xi_{k;j,-r}-\hat \Xi_{k;j,-r}^{(re)}) W_{t;-r})\Big]\\
    =&\frac{1}{\sqrt{n}} \sum_{t=d}^{n-k} \sum_{r=1}^p v_r e_r^\top (\Gammas)^{-1} W_t U_{t+k;j}+\frac{1}{\sqrt{n}} \sum_{t=d}^{n-k} \sum_{r=1}^p v_r (DN_{k,r,r}^{-1}-1) e_r^\top (\Gammas)^{-1} W_t U_{t+k;j}\\&+
    \frac{1}{\sqrt{n}} \sum_{t=d}^{n-k} \sum_{r=1}^p v_r e_r^\top ((\Gammah)^{-1}-(\Gammas)^{-1}) W_t U_{t+k;j}\\&+
    \frac{1}{\sqrt{n}} \sum_{t=d}^{n-k} \sum_{r=1}^p v_r (DN_{k,r,r}^{-1}-1) e_r^\top ((\Gammah)^{-1}-(\Gammas)^{-1}) W_t U_{t+k;j}\\&
    +\frac{1}{\sqrt{n}} \sum_{t=d}^{n-k} \sum_{r=1}^p v_r DN_{k,r,r}^{-1}e_j^\top(\Xi_{k}-\hat \Xi_{k}^{(re)}) I_{dp;-r} I_{dp;-r}^\top  W_t W_t^\top  (\Gammah)^{-1} e_r=I+II+III+IV+V
\end{align*}

Now, we show that $II$ to $V$ are of the specified order such that the assertion follows. 
By the results above and Lemma~\ref{lem.filtering.eps}, we have \\
$|II|\leq \|v\|_1   \max_r |DN_{h,r,r}^{-1}-1| \max_r| 1/\sqrt{n}\sum_{t=d}^{n-h}  e_r^\top (\Gammas)^{-1} W_t U_{t+h;j}|=O_P(\gpeps/\sqrt{n}+\|\Gammas-\Gammah\|_\infty\kpp \sqrt{\gpeps}))$,\\
$|III|\leq \|v\|_1 \|\Gammas-\Gammah\|_\infty\|(\Gammah)^{-1}\|_\infty \max_r|1/\sqrt{n}\sum_{t=d}^{n-h} e_r^\top (\Gammas)^{-1} W_t U_{t+h;j}| = \linebreak O_P(\|\Gammas-\Gammah\|_\infty \kpp \sqrt{\gpeps}),$ and \\
$|IV|=O_P(\|\Gammas-\Gammah\|_\infty\kpp\Big(\gpeps/\sqrt{n}+\|\Gammas-\Gammah\|_\infty\kpp \sqrt{\gpeps}\Big))$. Furthermore, we have
$
    |V|\leq \|v\|_1 \|\Xi_{h}-\hat \Xi_{h}^{(re)}\|_\infty \Bigg[\max_{s,r} |e_s^\top \fracd{1}{\sqrt{n}} \sum_{t=d}^{n-h} I_{dp;-r}^\top  W_t W_t^\top(\Gammas)^{-1} e_r| +
    \max_{s,r} |e_s^\top \fracd{1}{\sqrt{n}} \sum_{t=d}^{n-h} \linebreak I_{dp;-r}^\top  W_t W_t^\top[(\Gammah)^{-1}-(\Gammas)^{-1}] e_r|+
    \max_{s,r} |DN_{h,s,s}^{-1}-1| |\fracd{1}{\sqrt{n}} e_s^\top \sum_{t=d}^{n-h} I_{dp;-r}^\top  W_t W_t^\top(\Gammas)^{-1} e_r| +
    \max_{s,r} |DN_{h,s,s}^{-1}-1|\max_{s,r} |e_s^\top \fracd{1}{\sqrt{n}} \sum_{t=d}^{n-h} I_{dp;-r}^\top  W_t W_t^\top[(\Gammah)^{-1}-(\Gammas)^{-1}] e_r|\Bigg]
    =\|\Xi_{h}-\hat \Xi_{h}^{(re)}\|_\infty [VI+VII+VIII+IX]
$

Since $\max_s e_s^\top I_{dp;-r}^\top=\max_{s\not=r} e_s^\top $ and $Ee_s^\top W_t W_t \Gammas^{-1} e_r=0$ for $s\not=r$, we have \\
$|VI|=\max_{s\not=r}  \fracd{1}{\sqrt{n}} \sum_{t=d}^{n-h} e_s^\top W_t W_t^\top(\Gammas)^{-1} e_r=O_P(\sqrt{\gpeps}),$ \\
$    |VII|=\| \Gammas (\Gammah)^{-1}-I_{dp}\|_1 \times  \|I_{dp;-r}^\top \fracd{1}{\sqrt{n}} \sum_{t=d}^{n-h} W_t W_t^\top (\Gammas)^{-1} \|_{\max}=O_P(\sqrt{\gp} \kpp \|\Gammah-\Gammas\|_1),$
\\
$|VIII|=O_P(\gpeps/\sqrt{n}+\|\Gammas-\Gammah\|_\infty \kpp \sqrt{\gpeps}),$\\
and $|IX|=O_P(\|\Gammas-\Gammah\|_\infty\kpp  (\gpeps/\sqrt{n}+\|\Gammas-\Gammah\|_\infty^2\kpp^2 \sqrt{\gpeps})).$

Hence, we obtain
\begin{align*}
\sqrt{n}& e_j^\top(\hat \Psi_{h}^{(de)}-\Psi_{h}) v/s.e._\Psi(j,h,v)=\frac{1}{\sqrt{n}} \sum_{t=d}^{n-h} v^\top \E^\top(\Gammas)^{-1} W_t U_{t+h;j}/s.e._\Psi(j,h,v)\\
&+O_P\Big(\gpeps/\sqrt{n}+\|\Gammah-\Gammas\|_\infty \kpp \sqrt{\gpeps}\\
&\qquad +\|\Xi_{h}-\hat \Xi_{h}^{(re)}\|_\infty(\sqrt{\gpeps}+\|\Gammas-\Gammah\|_\infty \kpp \sqrt{\gpeps})\Big)
\end{align*}
By plugging in the derived rates and dropping  the terms of higher order, the assertion follows. 
\end{proof}

\begin{proof}[Proof of Theorem~\ref{thm.clt.psi}]
By following the arguments of the proof of Lemma~\ref{lem.est.gamma}, we obtain $|\widehat{s.e.}_\Psi(j,h,v/\|v\|_1)^2-{s.e.}_\Psi(j,h,v/\|v\|_1)^2|\leq O_P(2h(\|(\Gammas)^{-1}-(\Gammah)^{-1}\|_\infty+\|(\Gammas)-(\Gammah)\|_\infty)).$
Let $\tilde v=\E v$. Note $\|v\|=\|\tilde v \|$. With this and Lemma~\ref{lem.bias} we obtain 
$
\sqrt{n} e_j^\top(\hat \Psi_{h}^{(de)}-\Psi_{h}) v/\widehat{s.e.}_\Psi(j,h,v)
=\fracd{1}{\sqrt{n}} \sum_{t=d}^{n-h} \tilde v^\top  \linebreak (\Gammas)^{-1} W_t U_{t+h;j}/s.e._\Psi(j,h,v)+O_P\Big(\|v\|_1/\|v\|_2(\gpeps/\sqrt{n}+\|\Gammah-\Gammas\|_1\kpp(\kpp+ \sqrt{\gpeps})
+\|\Xi_{h}-\hat \Xi_{h}^{(re)}\|_\infty(\sqrt{\gpeps}+\|\Gammas-\Gammah\|_\infty \kpp \sqrt{\gpeps}))\Big).
$

Since $h$ is fixed, $\{U_{t+h}\}$ is an $M$-dependent process, $M=h$, and $U_{t+h}$ is independent to $W_t$ for all $t$, we have that $\{\tilde v^\top (\Gammas)^{-1} W_t U_{t+h;j}/s.e._\Psi(j,h,v),t \in \Z\}$  possesses under Assumption~\ref{ass.Est},~\ref{ass1} also a geometrical decaying functional dependence coefficient and we have $\nu_q<\infty$ for the process $\{\tilde v^\top (\Gammas)^{-1} W_t U_{t+h;j}/s.e._\Psi(j,h,v),t \in \Z\}$. Furthermore, we have
\begin{align*}
    \var(\frac{1}{\sqrt{n}}& \sum_{t=d}^{n-h} \tilde v^\top (\Gammas)^{-1} W_t U_{t+h;j}/s.e._\Psi(j,h,v))\\
    =&E(\frac{1}{n} \sum_{t_1,t_2=d}^{n-h} \tilde v^\top (\Gammas)^{-1} W_{t_1} W_{t_2}^\top (\Gammas)^{-1} \tilde v U_{t_1+h;j} U_{t_2+h;j}/s.e._\Psi(j,h,v)^2)\\
    =&E(\sum_{s_1,s_2=0}^{h-1}\sum_{i=-n+h+d+1}^{n-h-d-1}\!\!\!\!\!\frac{n-h-d-|i|}{s.e._\Psi(j,h,v)^2 n} \tilde v^\top (\Gammas)^{-1} W_{0} W_{i}^\top (\Gammas)^{-1}  \tilde v e_j^\top \Psi_{s_1} \eps_{h-s_1} \eps_{h+i-s_2}^\top \Psi_{s_2}^\top e_j)\\
    =&\sum_{s_1,s_2=0}^{h-1}\!\!\!\!\!\frac{n-h-d-|s_2+s_1|}{s.e._\Psi(j,h,v)^2 n} \tilde v^\top (\Gammas)^{-1} \Gamma^{(st)}(s_2-s_1) (\Gammas)^{-1} \tilde  v e_j^\top \Psi_{s_1}^\top \Sigmaeps \Psi_{s_2} e_j=1+O(1/n).
\end{align*}

Note that $W_t$ and $U_{t+h}$ are independent for all $t$. Furthermore, Lyapounov's condition can be verified  which gives $\fracd{1}{\sqrt{n}}\sum_{t=d}^{n-h} \tilde v^\top (\Gammas)^{-1} W_t U_{t+h;j}/s.e._\Psi(j,h,v) \overset{d}{\to} \mathcal{N}(0,1)$   via an extension  of the central limit theorem for functional dependent random variables,  Theorem 3 of  \cite{wu2011asymptotic},   to triangular arrays;  see also  Theorem 27.3 of 
\cite{billingsley1995probability}. To see that Lyapounov's condition holds, note the following calculation and Assumption~\ref{ass1}\eqref{ass1.1},\eqref{ass1.6}
\begin{align*}
    &\frac{1}{s.e._\Psi(j,h,v/\|v\|_2))^4 n^2} \sum_{t=d}^{n-h} E(  \tilde v^\top /\|v\|_2 (\Gammas)^{-1} W_t U_{t+h;j} )^4\\
    \leq &\frac{1}{s.e._\Psi(j,h,v/\|v\|_2))^4 n} E(  \tilde v^\top /\|v\|_2 (\Gammas)^{-1} \sum_{s=0}^\infty \mathds{A}^s \mathds{U}_{-s})^4 E(\sum_{s=0}^{h-1} e_j \Psi_s \eps_{-s})^4\\
    \leq& \frac{1}{s.e._\Psi(j,h,v/\|v\|_2))^4 n} \Big[\sum_{s=0}^\infty\|\tilde v^\top /\|v\|_2 (\Gammas)^{-1} \mathds{A}^s\|_2^4 \max_{\|w\|_2=1} E (w^\top \eps_0)^4+ 3 (\tilde v^\top (\Gammas)^{-1} \tilde v/\|v\|_2^2)\Big]\\
    &\times \Big[\sum_{s=0}^{h-1} \|e_j^\top \Psi_s\|_2^4 \max_{\|w\|_2=1} E (w^\top \eps_0)^4+ 3(\sum_{s=0}^{h-1} e_j^\top \Psi_s \Sigmaeps \Psi_s e_j)^2\Big]\\
    &=O(1/n \frac{1}{1-\lambda} \|(\Gammas)^{-1}\|_2^4)
\end{align*}

The second assertions follows by Nagaev's inequality for dependent variables, see Theorem 2 in \cite{liu2013probability} and see also {\Lemmaneun} in \cite{krampe2018bootstrap}.
\end{proof}

\begin{lem} \label{lem.Sigma1.clt}
Under Assumption~\ref{ass.Est}, \ref{ass1} we have, for all vectors $v \in \R^{p\cpu}$ and $s.e._{\eps}(v)\not=0$ and \\
$(\|v\|_1/s.e._{\eps}(v) \kp^2((\gp/n)^{(1-\zeta)}+(\gp/n)^{(1-\zeta)/2}\sqrt{\gpeps/n}))=o_p(1)$, 
$$
\sqrt{n} v^\top \Big(\veco(\fracd{1}{(n-d)}\sum_{t=d+1}^{n} \hat \eps_{t} \hat \eps_{t;\mathcal{I}}^\top-(\Sigmaeps I_{d;\mathcal{I}})\Big)/s.e._{\eps}(v)\overset{d}{\to}N(0,1),
$$
where $s.e._{\eps}(v)^2=\var(\sum_{i=1}^{p} \sum_{j=1}^{\cpu} v_{i+(j-1)\cpu} \eps_{1;i} \eps_{1;\mathcal{I}} e_j)$.
\end{lem}

\begin{proof}[Proof of Lemma~\ref{lem.Sigma1.clt}]
Let $\veco^{-1}_{p,\cpu}(v) \in R^{p\times \cpu}$ be a matrix such that $\veco(\veco^{-1}_{p,\cpu}(v))=v$. Then, $v^\top \Big(\veco(\eps_{t} \eps_{t;\mathcal{I}}^\top\Big)=\eps_t^\top \veco^{-1}_{p,\cpu}(v) \eps_{t;\mathcal{I}}$.

We have $\sqrt{n} v^\top (\veco(\fracd{1}{(n-d)}\sum_{t=d+1}^{n} \hat \eps_{t} \hat \eps_{t;\mathcal{I}}^\top)- \Sigmaeps I_{d;\mathcal{I}})/s.e._{\eps}(v)=
\sqrt{n} v^\top \linebreak(\veco(\fracd{1}{(n-d)}\sum_{t=d+1}^{n} \hat \eps_{t} \hat \eps_{t;\mathcal{I}}^\top-\eps_{t} \eps_{t;\mathcal{I}}^\top))/s.e._{\eps}(v))+\sqrt{n} v^\top(\veco(\fracd{1}{(n-d)}\sum_{t=d+1}^{n} \eps_{t} \eps_{t;\mathcal{I}}^\top-\Sigmaeps I_{d;\mathcal{I}}))\times 1/s.e._{\eps}(v))=I+II$. We first show that $I$ is asymptotically negligible and then that $II$ is asymptotically Gaussian.

We have 
\begin{align*}
    |I|\leq& \frac{\|v\|_1}{s.e._{\eps}(v)} \max_r  |\sqrt{n} e_r^\top  (\veco(\frac{1}{(n-d)}\sum_{t=d+1}^{n} \hat \eps_{t} \hat
\eps_{t;\mathcal{I}}^\top-\eps_{t} \eps_{t;\mathcal{I}}^\top))| \\
=&  O_P(\|v\|_1/s.e._{\eps}(v) \times  \kp^2((\gp/n)^{(1-\zeta)}+(\gp/n)^{(1-\zeta)/2}\sqrt{\gpeps/n})).
\end{align*} 
Furthermore, we have
$
II= \fracd{\sqrt{n}}{(n-d)}\sum_{t=d+1}^n \sum_{i=1}^{p} \sum_{j=1}^{\cpu} v_{i+(j-1)p} \times (\eps_{t;i}\eps_{t;\mathcal{I}}^\top e_j-e_i \Sigmaeps I_{d;\mathcal{I}}e_j)/ s.e._{\eps}(v).
$ Since $\{\eps_{t}\}$ is an i.i.d. sequence, we obtain 
$\var(II)=\var(\sum_{i=1}^{p} \sum_{j=1}^{\cpu} v_{i+(j-1)\cpu} \eps_{1;i} \eps_{1;\mathcal{I}} e_j)/s.e._{\eps}(v)^2+O(1/n)=1+O(1/n).$
Let $U D V^\top$ be a singular value decomposition of $\veco^{-1}_{p,\cpu}(v)$. Then, we obtain 
\begin{align*}
        \sum_{t=1}^n& E\Big(\fracd{\sqrt{n}}{((n-d)s.e._{\eps}(v))} \sum_{i=1}^{p} \sum_{j=1}^{\cpu} v_{i+(j-1)p}(\eps_{t;i}\eps_{1;\mathcal{I}} e_j)\Big)^4
    =\fracd{n^3}{((n-d)s.e._{\eps}(v))^4} E( \eps_t^\top \veco^{-1}_{p,\cpu}(v) \eps_{t;\mathcal{I}})^4\\
    &=\fracd{n^3}{((n-d)s.e._{\eps}(v))^4} E( \eps_t^\top U D V^\top \eps_{t;\mathcal{I}})^4
    \leq \fracd{n^3\|v\|_2^4 }{((n-d)s.e._{\eps})^4(v)}  \max_{\|u\|_2=1}E( u^\top \eps_t)^8
    =o(1).
\end{align*}
 Hence, Lyapounov's condition holds and the assertion follows by a central limit theorem for triangular arrays, see among others Theorem 27.3 of 
\cite{billingsley1995probability}.
\end{proof}

\begin{proof}[Proof of Theorem~\ref{thm.clt.B}]
First note that $\sqrt{n} v^\top (\hat B-B)e_r=\sqrt{n}v^\top (\tilde B-BH^{-1}){\Rot}e_r+\sqrt{n}v^\top BH^{-1}(\hat {\Rot}-{\Rot})e_r+\sqrt{n}v^\top(\tilde B-B)(\hat {\Rot}-{\Rot})e_r$. From the proof of Theorem~\ref{thm.estimation},\ref{thm.estimation2}, we have $\sqrt{n}v^\top(\tilde B-B)(\hat {\Rot}-{\Rot})e_r=O_P(\|v\|_1(\sqrt{\gpeps}\cpu(\kp^2(\gp/n)^{1-\zeta}+\sqrt{\gpeps/n})))$. 

Due Assumption~\ref{ass1}\eqref{ass1.5b} and the mean value theorem, we have $$\sqrt{n}v^\top BH^{-1}(\hat {\Rot}-{\Rot})e_r=\sqrt{n}v^\top BH^{-1}\nabla {\grestr}(\tilde \Sigmaeps) (I_{\cpu} \otimes I_{d;\mathcal{R}^\top}) \veco(1/(n-d)\sum_{t=d+1}^n \hat \eps_{t} \hat \eps_{t,\mathcal{I}}-\Sigmaeps I_{d;\mathcal{I}}),$$  where  $\tilde \Sigmaeps=c/(n-d)\sum_{t=d+1}^n \hat \eps_{t;\mathcal{R}} \hat \eps_{t,\mathcal{I}}+(1-c)I_{d;\mathcal{R}}^\top\Sigmaeps I_{d;\mathcal{I}}$ for some $c \in (0,1)$. Furthermore, $\sqrt{n}v^\top (\tilde B-BH{-1}){\Rot}e_r=(e_r^\top {\Rot}^\top \otimes v^\top)\veco(1/(n-d)\sum_{t=d+1}^n \hat \eps_{t} \hat \eps_{t,\mathcal{I}}-\Sigmaeps I_{d;\mathcal{I}})$. Note further that $\|v^\top BH^{-1}\nabla {\grestr}(\tilde \Sigmaeps)-v^\top BH^{-1}\nabla {\grestr}(\Sigmaeps)\|_2=O_P(\|v\|_2 \cpu^{3} \sqrt{\gpeps/n})$. Hence, we have
$
\sqrt{n} v^\top (\hat B-B)e_r=\sqrt{n}[(e_r^\top {\Rot}^\top \otimes v^\top)+v^\top BH^{-1}\nabla {\grestr}(\Sigmaeps) (I_{\cpu} \otimes I_{d;\mathcal{I}^\top})]\veco(1/(n-d)\sum_{t=d+1}^n \hat \eps_{t} \hat \eps_{t,\mathcal{I}}-\Sigmaeps I_{d;\mathcal{I}})+o_p(1)
$. The assertion follows by Lemma~\ref{lem.Sigma1.clt} if $\widehat{s.e.}_{B}(v,r)^2={s.e.}_{B}(v,r)^2+o_P(1)$.
For this note that $\{e_t\}$ is an i.i.d.~sequence with $\max_{ \|u\|_2=1} E(u\top \eps_t)^q)\leq C<\infty, q \geq 8$, see Assumption~\ref{ass1}\eqref{ass1.6}, and for some vectors $u,v$ we have $\var(1/n \sum_{t=1}^n (u^\top \eps_t \eps_t^\top v)^2)\leq 1/n C^2 \|u\|_2 \|v\|_2$. Furthermore, for some vector $v \in \R^{\cpu^2}$ with $\veco^{-1}_{\cpu,\cpu}(v)=UDV^\top$ as its singular value decomposition we obtain the following  $E(1/n \sum_{t=1}^n (v \veco(\eps_{t;\mathcal{R}} \eps_{t;\mathcal{I}}^\top))^4)\leq1/n\sum_{j_1,j_2,j_3,j_4=1}^{\cpu} |\prod_{i=1}^4 \sigma_{j_i}| E \prod_{i=1}^4 |\eps_t^\top U e_{j_i} e_{j_i}^\top V^\top \eps_t| \leq C^2 \|v\|_2^4 \cpu^2/n,$ where $\sigma_{j}$ are the singular values of $\veco^{-1}_{\cpu,\cpu}(v)$ and the last inequality follows by bounding the nuclear norm by the Frobenius norm.

With the previous results, the assertion $\|\sqrt{n} (\hat B-B)e_r\|_{\max}=O_P(\sqrt{\gpeps)}$ follows then directly by Assumption~\ref{ass.Est}\eqref{ass.Est.4}.
\end{proof}

\begin{proof}[Proof of Theorem~\ref{thm.clt.theta}] 
We have 
$
    (\hat\Theta_{h;jr}^{(de)}-\Theta_{h;jr})=
    e_j^\top (\hat \Psi_h^{(de)} \hat B)e_r- e_j^\top \Psi_h B e_r- e_j^\top (\hat \Psi_h^{(de)}-\hat \Psi_h^{(re)})(\hat B-\hat B^{(re)}) e_r
    = e_j^\top\Big[(\hat \Psi_h^{(de)}-\Psi_h) B +\Psi_h (\hat B-B)+
    (\hat \Psi_h^{(de)}-\Psi_h)(\hat B-B)
    -(\hat \Psi_h^{(de)}-\hat \Psi_h^{(re)})(\hat B-\hat B^{(re)})\Big] e_r
    =e_j^\top\Big[(\hat \Psi_h^{(de)}-\Psi_h) B +\Psi_h (\hat B-B)\Big]e_r+e_j^\top\Big[(\hat \Psi_h^{(de)}-\Psi_h)(\hat B^{(re)}-B)
    +(\hat \Psi_h^{(re)}-\Psi_h)(\hat B-B)-(\hat \Psi_h^{(re)}-\Psi_h)(\hat B^{(re)}-B)\Big]e_r=:I+II.
$
Furthermore, we have by H{\"o}lder's inequality
$
    \sqrt{n}|II|\leq \|\sqrt{n} e_j^\top(\hat \Psi_h^{(de)}-\Psi_h)\|_\infty \|(\hat B^{(re)}-B)e_r\|_1+\|e_j(\hat \Psi_h^{(re)}-\Psi_h)\|_1\|\sqrt{n}(\hat B-B)e_r\|_\infty+\|e_j(\hat \Psi_h^{(re)}-\Psi_h)\|_1\|\sqrt{n}(\hat B^{(re)}-B)\Big]e_r\|_\infty
    o_P(1),
$
where the last equality is due to Lemma~\ref{lem.ass},~\ref{lem.est.B}, Theorem\ref{thm.clt.psi},~\ref{thm.clt.B} and Assumption~\ref{ass1}\eqref{ass1.6}. By the proofs of Theorem~\ref{thm.clt.psi} and \ref{thm.clt.B} we have $\cov(\sqrt{n}e_j^\top (\hat \Psi_h^{(de)}-\Psi_h)B e_r,\sqrt{n} e_j^\top \Psi_h (\hat B-B) e_r=1/n\cov(\sum_{t} W_t U_{t+h},\sum_{t} \eps_t \eps_t^\top)+o_P(1)=o_P(1)$ since $\{\eps_t\}$ is i.i.d., and $W_t$ and $U_{t+h}$ are mutually independent. Hence, 
$\var(\sqrt{n}(\hat\Theta_{h;jr}^{(de)}-\Theta_{h;jr}))=n\var(e_j^\top\Big[(\hat \Psi_h^{(de)}-\Psi_h) B +\Psi_h (\hat B-B)\Big]e_r)+o_P(1)=s.e._\Psi(j,h,Be_r)^2+s.e._B(\Psi_h e_j,r)^2$ and the assertion follows by Theorem~\ref{thm.clt.psi} and \ref{thm.clt.B}.
\end{proof}

\begin{proof}[Proof of Theorem~\ref{thm.boot.valid}]
We show  that,  as $n\to \infty$, 
 $\fracd{\sqrt{n}}{\widehat{s.e.}^*_\Theta(h,j,r)}(\hat \Theta_{h;j,r}^{*(de)}-\hat \Theta_{h;j,r}^{(boot)}) \overset{d}{\to} \mathcal{N}\big(0,1\big)\text{ in probability},$
from which  the assertion  follows by the triangular inequality and Theorem~\ref{thm.clt.theta}. To show this, we can mainly follow the arguments of the proofs of Theorem~\ref{thm.clt.psi}, \ref{thm.clt.B}, and \ref{thm.clt.theta}.

For this note first that for some approximate sparse $p$-dimensional vector $u \in  \mathcal{U}(k,q), q \in [0,1)$ with regularized estimate $\hat u, \|\hat u-u\|_1=O_P(k\lambda)$, we have for a thresholded version $\sum_{i=1}^p | \hat u_i|^q \ind(|u_i|\geq \lambda)=\sum_{i=1}^p | \hat u_i|^q \ind(|u_i|\geq \lambda) (\ind(|\hat u_i|\geq|u_i|)+\ind(|\hat u_i|>|u_i|))\leq \sum_{i=1}^p | u_i|^q \ind(|\hat u_i|>|u_i|)+\sum_{i=1}^p (|u_i|+|\hat u_i-u_i|)/|\hat u_i|^{1-q}\ind(|u_i|\geq \lambda) (\ind(|\hat u_i|\geq|u_i|)\leq \sum_{i=1}^p |u_i|^q+\sum_{i=1}^p |\hat u_i-u_i|/\lambda^{1-q}=O_P(k)$. That means the thresholded version of $\hat u$ is again (with high probability) approximately sparse and its sparsity parameter is of the same order as the original one. Furthermore, note that $\{\eps_t^*\}$ is generated as an i.i.d. sequence and it possesses at least as many finite moments as $\{\eps_t\}$. Furthermore, Assumption~\ref{ass1} ensures that the largest absolute eigenvalue of $\hat \A^{(thr)}$ is for $n$ large enough smaller than one and, consequently, $\hat \A^{(thr)}$ fulfills with high probability Assumptions~\ref{ass.Est},\ref{ass1} and the pseudo time series possesses a geometric decaying functional dependence. 

Furthermore, note that the $s.e._B{v,r}^2=\var\Big(v^\top (B u_t+Dw_t) u_{t;r}+(u_t^\top B+w_t^\top D)I_{d;\mathcal{R}}^\top \\\veco_{\cpu}^{-1}(v^\top B {\Rot}^{-1} \nabla {\grestr}(I_{d;\mathcal{R}}^\top \Sigma I_{d;\mathcal{I}})){\Rot}^{-1} u_t\Big)$. Hence, if $\{u_t\}$ and $\{w_t\}$ are mutually independent, the fourth moments occurring in $s.e._B^2(v,r)$ are limited to $\{u_t\}$ only. We have $\var^*(\eps_t^*)=\var^*(\hat B^{(re)}u_t^*+w_t^*)=\hat B^{(re)}(\hat B^{(re)})^\top+\hat{DD^\top}^{(re)}=\Sigmah$. Since $\|\Sigmah-\Sigmaeps\|_l=O_P(\big[\cpD\cpu^{1-\beta}+\cpB\cpu^{(3-\beta)/2}\big]\Big[\kp^2(\gp/n)^{1-\zeta}+\sqrt{\gpeps/n}\Big]^{1-\beta})$  and $\| \hat \A^{(thr)}-\A\|_l=O_P(\kp^{1.5} (\gp/n)^{(1-\zeta)/2})$, we have that $\{X_t^*\}$ approximates well-enough the autocovariance of $X_t$ which gives  $\widehat{s.e.}_\Psi^*(j,h,\hat B^{(re)} e_r)={s.e.}_\Psi(j,h,\hat B^{(re)} e_r)+o_P(1)$. Note that $\|\hat \A-\A\|_\infty=O_P(\kp^{1.5} (\gp/n)^{(1-\zeta)/2})$ implies $\|\hat \Xi^{(re)*}-\Xi\|_\infty=O_P(\kp^{3.5} (\gp/n)^{(1-\zeta)/2})$ and $\|\hat\Gamma^{*(st)}(0)-\Gammas\|_\infty=O_P(\sqrt{\kp}\|\Gammah-\Gammas\|_\infty)$. Thus by the same arguments used in the proof of Theorem~\ref{thm.clt.psi} and Lemma~\ref{lem.bias} with an additional $\kp^{0.5}$, we obtain in probability  $\sqrt{n} e_j^\top(\hat \Psi_{h}^{*(de)}-\hat\Psi_{h}^{(boot)}) \hat B^{(re)} e_r/\widehat{s.e.}_\Psi^*(j,h,\hat B^{(re)} e_r)\overset{d}{\to}\mathcal{N}(0,1)$.

Since $\{u_t^*\}$ is drawn i.i.d.~from $\{\hat u_{t},t=d+1,\dots,n\}$, we have for $r,s \in \{1,\dots,\cpu\}$ that
$\var^*(u_{t;r}^* u_{t;s}^*)=1/(n-d) \sum_{t=d+1}^n (\hat u_{t;r} \hat u_{t;s})^2-e_r^\top e_s$. Thus, $\sum_{r,s} |\var^*(u_{t;r}^* u_{t;s}^*)-\var(u_{t;r} u_{t;s})|=O_P(\cpu^2/\sqrt{n}).$ Hence, also the fourth moment of $\{u_{t}\}$ is approximated well enough and we can follow the arguments of the proof of Theorem~\ref{thm.clt.B} to obtain $\fracd{\sqrt{n}}{\widehat{s.e.}_{B}(\hat \Psi^{(re)} e_j,r)} \hat \Psi^{(re)} e_j^\top (\hat B-B) e_r\overset{d}{\to} \mathcal{N}(0,1).$ The assertion follows then by the arguments of the proof of Theorem~\ref{thm.clt.theta}.
\end{proof}

\begin{proof}[Proof of Remark~\ref{remark.cholesky}]
The assertion follows by rules for matrix differentiation. We have
\begin{align*}
    {s.e.}_{B}(v,r)^2=&\var(v^\top \eps_t u_{t;r}+v^\top B P^\top \frac{\partial P^{-T}e_r}{\partial \vech( PP^\top)} \vech(\eps_{t;\mathcal{I}}\eps_{t;\mathcal{I}}^\top))\\
    =&\var(v^\top \eps_t u_{t;r}-v^\top B P^\top (e_r^\top \otimes I_{\cpu})(P^{-1} \otimes (P^\top)^{-1}) K_{\cpu\cpu} \frac{\partial P}{\partial \vech( PP^\top)} \vech(\eps_{t;\mathcal{I}}\eps_{t;\mathcal{I}}^\top))\\
    =&\var(v^\top \eps_t u_{t;r}-v^\top B P^\top (e_r^\top \otimes I_{\cpu})(P^{-1} \otimes (P^\top)^{-1}) K_{\cpu\cpu} \\
    &\times L_{\cpu}^\top (L_{\cpu}^\top (I_{\cpu^2}+K_{\cpu\cpu})(P\otimes I_{\cpu}) L_{\cpu})^{-1} \vech(\eps_{t;\mathcal{I}}\eps_{t;\mathcal{I}}^\top))\\
    =&\var(v^\top \eps_t u_{t;r}-v^\top B P^\top (e_r^\top \otimes I_{\cpu})(P^{-1} \otimes (P^\top)^{-1}) (I_{\cpu} \otimes P^{-1})(\mp I_{\cpu^2}+ K_{\cpu\cpu}) \\
    &\times (P\otimes I_{\cpu}) L_{\cpu}^\top (L_{\cpu}^\top (I_{\cpu^2}+K_{\cpu\cpu})(P\otimes I_{\cpu}) L_{\cpu})^{-1} \vech(\eps_{t;\mathcal{I}}\eps_{t;\mathcal{I}}^\top))\\
    =&\var(v^\top Dw_t u_{t;r}+v^\top BP^\top (e_r^\top \otimes (PP^\top)^{-1}) L_{\cpu}^\top \\
    &\times  (L_{\cpu} (I_{{\cpu}^2}+K_{{\cpu}{\cpu}}) (P\otimes I_{\cpu}) L_{\cpu}^\top)^{-1} \vech (\eps_{t;\mathcal{I}}\eps_{t;\mathcal{I}}^\top)).
\end{align*}
If $\cpu=p$, we have $B=P$ and $\eps_t=Pu_t$. That means $I_p-BP^{-1}=0$ and $BP^\top=PP^\top$ and the assertion follows.
\end{proof}

\bibliographystyle{apalike}
\bibliography{bib}

\begin{thebibliography}{}

\bibitem[Bai et~al., 2016]{baililu2016}
Bai, J., Li, K., and Lu, L. (2016).
\newblock {Estimation and Inference of FAVAR Models}.
\newblock {\em Journal of Business \& Economic Statistics}, 34(4):620--641.

\bibitem[Banbura et~al., 2010]{banbura2010SIR}
Banbura, M., Giannone, D., and Reichlin, L. (2010).
\newblock {Large Bayesian vector auto regressions}.
\newblock {\em Journal of Applied Econometrics}, 25(1):71--92.

\bibitem[Barigozzi and Brownlees, 2019]{barigozzi2019nets}
Barigozzi, M. and Brownlees, C. (2019).
\newblock Nets: Network estimation for time series.
\newblock {\em Journal of Applied Econometrics}, 34(3):347--364.

\bibitem[Barigozzi and Hallin, 2017]{barigozzi2017network}
Barigozzi, M. and Hallin, M. (2017).
\newblock A network analysis of the volatility of high dimensional financial
  series.
\newblock {\em Journal of the Royal Statistical Society: Series C (Applied
  Statistics)}, 66(3):581--605.

\bibitem[Basu and Michailidis, 2015]{basu2015}
Basu, S. and Michailidis, G. (2015).
\newblock Regularized estimation in sparse high-dimensional time series models.
\newblock {\em The {A}nnals of {S}tatistics}, 43(4):1535--1567.

\bibitem[Bellec et~al., 2018]{bellec2018slope}
Bellec, P.~C., Lecu{\'e}, G., Tsybakov, A.~B., et~al. (2018).
\newblock Slope meets lasso: improved oracle bounds and optimality.
\newblock {\em Annals of Statistics}, 46(6B):3603--3642.

\bibitem[Benjamini and Yekutieli, 2001]{benjamini2001control}
Benjamini, Y. and Yekutieli, D. (2001).
\newblock The control of the false discovery rate in multiple testing under
  dependency.
\newblock {\em Annals of statistics}, pages 1165--1188.

\bibitem[Bernanke et~al., 2005]{bernanke2005}
Bernanke, B.~S., Boivin, J., and Eliasz, P. (2005).
\newblock Measuring the effects of monetary policy: A factor-augmented vector
  autoregressive {(FAVAR)} approach.
\newblock {\em Quarterly Journal of Economics}, 120(1):387--422.

\bibitem[Bickel and Freedman, 1981]{bickel1981}
Bickel, P.~J. and Freedman, D.~A. (1981).
\newblock Some asymptotic theory for the bootstrap.
\newblock {\em Ann. Statist.}, 9(6):1196--1217.

\bibitem[Bickel and Levina, 2008]{bickel2008}
Bickel, P.~J. and Levina, E. (2008).
\newblock Covariance regularization by thresholding.
\newblock {\em The {A}nnals of {S}tatistics}, 36(6):2577--2604.

\bibitem[Billingsley, 1995]{billingsley1995probability}
Billingsley, P. (1995).
\newblock Probability and measure. wiley series in probability and mathematical
  statistics.

\bibitem[Br{\"u}ggemann et~al., 2016]{bruggemann2016inference}
Br{\"u}ggemann, R., Jentsch, C., and Trenkler, C. (2016).
\newblock Inference in vars with conditional heteroskedasticity of unknown
  form.
\newblock {\em Journal of econometrics}, 191(1):69--85.

\bibitem[Cai and Liu, 2011]{cai2011adaptive}
Cai, T. and Liu, W. (2011).
\newblock Adaptive thresholding for sparse covariance matrix estimation.
\newblock {\em Journal of the American Statistical Association},
  106(494):672--684.

\bibitem[Callot and Kock, 2014]{callot2014forecasting}
Callot, L. and Kock, A. (2014).
\newblock {\em Oracle Efficient Estimation and Forecasting with the Adaptive
  Lasso and the Adaptive Group Lasso in Vector Autoregressions}.
\newblock Oxford University press.

\bibitem[Canova and Ciccarelli, 2013]{canova2013panel}
Canova, F. and Ciccarelli, M. (2013).
\newblock Panel vector autoregressive models: A survey.
\newblock In {\em VAR Models in Macroeconomics--New Developments and
  Applications: Essays in Honor of Christopher A. Sims}, pages 205--246.
  Emerald Group Publishing Limited.

\bibitem[Chatterjee and Lahiri, 2010]{chatterjee2010asymptotic}
Chatterjee, A. and Lahiri, S. (2010).
\newblock Asymptotic properties of the residual bootstrap for lasso estimators.
\newblock {\em Proceedings of the American Mathematical Society},
  138(12):4497--4509.

\bibitem[Chaudhry et~al., 2017]{chaudhry2017}
Chaudhry, A., Xu, P., and Gu, Q. (2017).
\newblock Uncertainty assessment and false discovery rate control in
  high-dimensional {G}ranger causal inference.
\newblock In Precup, D. and Teh, Y.~W., editors, {\em Proceedings of the 34th
  International Conference on Machine Learning}, volume~70 of {\em Proceedings
  of Machine Learning Research}, pages 684--693, International Convention
  Centre, Sydney, Australia. PMLR.

\bibitem[Chudik and Pesaran, 2016]{chudik2016theory}
Chudik, A. and Pesaran, M.~H. (2016).
\newblock Theory and practice of gvar modelling.
\newblock {\em Journal of Economic Surveys}, 30(1):165--197.

\bibitem[Dees et~al., 2007]{dees2007exploring}
Dees, S., Mauro, F.~d., Pesaran, M.~H., and Smith, L.~V. (2007).
\newblock Exploring the international linkages of the euro area: a global var
  analysis.
\newblock {\em Journal of applied econometrics}, 22(1):1--38.

\bibitem[Demirer et~al., 2018]{demirer2018estimating}
Demirer, M., Diebold, F.~X., Liu, L., and Yilmaz, K. (2018).
\newblock Estimating global bank network connectedness.
\newblock {\em Journal of Applied Econometrics}, 33(1):1--15.

\bibitem[Diebold and Y{\i}lmaz, 2014]{diebold2014network}
Diebold, F.~X. and Y{\i}lmaz, K. (2014).
\newblock On the network topology of variance decompositions: Measuring the
  connectedness of financial firms.
\newblock {\em Journal of Econometrics}, 182(1):119--134.

\bibitem[Fan and Han, 2017]{fan2017estimation}
Fan, J. and Han, X. (2017).
\newblock Estimation of the false discovery proportion with unknown dependence.
\newblock {\em Journal of the Royal Statistical Society. Series B, Statistical
  methodology}, 79(4):1143.

\bibitem[Fava and Lopes, 2020]{fava2020illusion}
Fava, B. and Lopes, H.~F. (2020).
\newblock The illusion of the illusion of sparsity: An exercise in prior
  sensitivity.

\bibitem[Forni et~al., 2009]{forni2009}
Forni, M., Giannone, D., Lippi, M., and Reichlin, L. (2009).
\newblock Opening the black box: structural factor models with large crosss
  sections.
\newblock {\em Econometric Theory}, 23(5):1319--1347.

\bibitem[Garcia et~al., 2017]{garcia2017real}
Garcia, M.~G., Medeiros, M.~C., and Vasconcelos, G.~F. (2017).
\newblock Real-time inflation forecasting with high-dimensional models: The
  case of brazil.
\newblock {\em International Journal of Forecasting}, 33(3):679--693.

\bibitem[Giannone et~al., 2018]{giannone2018illusion}
Giannone, D., Lenza, M., and Primiceri, G.~E. (2018).
\newblock Economic predictions with big data: The illusion of sparsity.

\bibitem[Han et~al., 2015]{han2015direct}
Han, F., Lu, H., and Liu, H. (2015).
\newblock A direct estimation of high dimensional stationary vector
  autoregressions.
\newblock {\em The Journal of Machine Learning Research}, 16(1):3115--3150.

\bibitem[Jord{\`a}, 2005]{jorda2005estimation}
Jord{\`a}, {\`O}. (2005).
\newblock Estimation and inference of impulse responses by local projections.
\newblock {\em American economic review}, 95(1):161--182.

\bibitem[Kascha and Trenkler, 2015]{kascha2015forecasting}
Kascha, C. and Trenkler, C. (2015).
\newblock Forecasting vars, model selection, and shrinkage.

\bibitem[Kilian and L{\"u}tkepohl, 2017]{kilian2017structural}
Kilian, L. and L{\"u}tkepohl, H. (2017).
\newblock {\em Structural vector autoregressive analysis}.
\newblock Cambridge University Press.

\bibitem[Knight and Fu, 2000]{knight2000asymptotics}
Knight, K. and Fu, W. (2000).
\newblock Asymptotics for lasso-type estimators.
\newblock {\em The {A}nnals of {S}tatistics}, pages 1356--1378.

\bibitem[Kock and Callot, 2015]{kock2015oracle}
Kock, A.~B. and Callot, L. (2015).
\newblock Oracle inequalities for high dimensional vector autoregressions.
\newblock {\em Journal of Econometrics}, 186(2):325--344.

\bibitem[Krampe et~al., 2021]{krampe2018bootstrap}
Krampe, J., Kreiss, J.-P., and Paparoditis, E. (2021).
\newblock Bootstrap based inference for sparse high-dimensional time series
  models.
\newblock {\em Bernoulli}, 27(3):1441--1466.

\bibitem[Krampe and Paparoditis, 2021]{krampe2020Est}
Krampe, J. and Paparoditis, E. (2021).
\newblock Sparsity concepts and estimation procedures for high dimensional
  vector autoregressive models.
\newblock {\em Journal of Time Series Analysis}.

\bibitem[Liu et~al., 2013]{liu2013probability}
Liu, W., Xiao, H., and Wu, W.~B. (2013).
\newblock Probability and moment inequalities under dependence.
\newblock {\em Statistica sinica}, pages 1257--1272.

\bibitem[L{\"u}tkepohl, 2005]{luetkepohl2007new}
L{\"u}tkepohl, H. (2005).
\newblock {\em New Introduction to Multiple Time Series Analysis}.
\newblock Springer Berlin Heidelberg.

\bibitem[Masini et~al., 2020]{masini2019regularized}
Masini, R.~P., Medeiros, M.~C., and Mendes, E.~F. (2020).
\newblock Regularized estimation of high-dimensional vector autoregressions
  with weakly dependent innovations.
\newblock {\em arXiv preprint arXiv:1912.09002}.

\bibitem[Medeiros and Vasconcelos, 2016]{medeiros2016forecasting}
Medeiros, M.~C. and Vasconcelos, G.~F. (2016).
\newblock Forecasting macroeconomic variables in data-rich environments.
\newblock {\em Economics Letters}, 138:50--52.

\bibitem[Neykov et~al., 2018]{neykov2018}
Neykov, M., Ning, Y., Liu, J., and Liu, H. (2018).
\newblock A unied theory of confidence regions and testing for high-dimensional
  estimating equations.
\newblock {\em Statistical Science}, 33(3):427--443.

\bibitem[{R Core Team}, 2021]{R}
{R Core Team} (2021).
\newblock {\em R: A Language and Environment for Statistical Computing}.
\newblock R Foundation for Statistical Computing, Vienna, Austria.

\bibitem[Ramey, 2016]{ramey2016shocks}
Ramey, V.~A. (2016).
\newblock Macroeconomic shocks and their propagation.
\newblock In {\em Handbook of macroeconomics}, volume~2, pages 71--162.
  Elsevier.

\bibitem[Rothman, 2013]{PDSCE}
Rothman, A.~J. (2013).
\newblock {\em PDSCE: Positive definite sparse covariance estimators}.
\newblock R package version 1.2.

\bibitem[Rothman et~al., 2009]{rothman2009generalized}
Rothman, A.~J., Levina, E., and Zhu, J. (2009).
\newblock Generalized thresholding of large covariance matrices.
\newblock {\em Journal of the American Statistical Association},
  104(485):177--186.

\bibitem[Simon et~al., 2011]{glmnet}
Simon, N., Friedman, J., Hastie, T., and Tibshirani, R. (2011).
\newblock Regularization paths for cox's proportional hazards model via
  coordinate descent.
\newblock {\em Journal of Statistical Software}, 39(5):1--13.

\bibitem[Smeekes and Wijler, 2018]{smeekes2018forecast}
Smeekes, S. and Wijler, E. (2018).
\newblock {Macroeconomic forecasting using penalized regression methods}.
\newblock {\em International Journal of Forecasting}, 34(3):408 -- 430.

\bibitem[Song and Bickel, 2011]{song2011large}
Song, S. and Bickel, P.~J. (2011).
\newblock Large vector auto regressions.
\newblock {\em {P}reprint arXiv:1106.3915}.

\bibitem[Stock and Watson, 2005]{stock2005implications}
Stock, J.~H. and Watson, M.~W. (2005).
\newblock Implications of dynamic factor models for {VAR} analysis.
\newblock Technical report, National Bureau of Economic Research.

\bibitem[Stock and Watson, 2016]{stock2016dynamic}
Stock, J.~H. and Watson, M.~W. (2016).
\newblock Dynamic factor models, factor-augmented vector autoregressions, and
  structural vector autoregressions in macroeconomics.
\newblock In {\em Handbook of macroeconomics}, volume~2, pages 415--525.
  Elsevier.

\bibitem[{v}an~de Geer et~al., 2014]{deGeer2014}
{v}an~de Geer, S., B{\"u}hlmann, P., Ritov, Y., and Dezeure, R. (2014).
\newblock On asymptotically optimal confidence regions and tests for
  high-dimensional models.
\newblock {\em The {A}nnals of {S}tatistics}, 42(3):1166--1202.

\bibitem[Wong et~al., 2020]{Wong2020}
Wong, K.~C., Li, Z., and Tewari, A. (2020).
\newblock {Lasso guarantees for $\beta$-mixing heavy-tailed time series}.
\newblock {\em The Annals of Statistics}, 48(2):1124 -- 1142.

\bibitem[Wu, 2005]{wu2005nonlinear}
Wu, W.~B. (2005).
\newblock Nonlinear system theory: Another look at dependence.
\newblock {\em Proceedings of the National Academy of Sciences},
  102(40):14150--14154.

\bibitem[Wu, 2011]{wu2011asymptotic}
Wu, W.~B. (2011).
\newblock Asymptotic theory for stationary processes.
\newblock {\em Statistics and its Interface}, 4(2):207--226.

\bibitem[Wu et~al., 2016]{wu2016performance}
Wu, W.-B., Wu, Y.~N., et~al. (2016).
\newblock Performance bounds for parameter estimates of high-dimensional linear
  models with correlated errors.
\newblock {\em Electronic Journal of Statistics}, 10(1):352--379.

\bibitem[Yamamoto, 2019]{yamamoto2019}
Yamamoto, Y. (2019).
\newblock {Bootstrap inference for impulse response functions in
  factor‐augmented vector autoregressions}.
\newblock {\em Journal of Applied Econometrics}, 34(2):247--267.

\bibitem[Yan and Lin, 2016]{FinCovRegularization}
Yan, Y. and Lin, F. (2016).
\newblock {\em FinCovRegularization: Covariance Matrix Estimation and
  Regularization for Finance}.
\newblock R package version 1.1.0.

\bibitem[Zhang and Zhang, 2014]{zhang2014}
Zhang, C.-H. and Zhang, S.~S. (2014).
\newblock Confidence intervals for low dimensional parameters in high
  dimensional linear models.
\newblock {\em Journal of the Royal Statistical Society: Series B (Statistical
  Methodology)}, 76(1):217--242.

\bibitem[Zheng and Raskutti, 2019]{zheng2019testing}
Zheng, L. and Raskutti, G. (2019).
\newblock Testing for high-dimensional network parameters in auto-regressive
  models.
\newblock {\em Electronic Journal of Statistics}, 13(2):4977--5043.

\end{thebibliography}



\end{document}


\section*{Supplementary Material}
\begin{table}[H]
\centering
\begin{tabular}{l|rrrrrrrr||rrrrrrrr}
 \hline
 &\multicolumn{8}{c||}{De} &
 \multicolumn{8}{c}{Re} \\
&\multicolumn{2}{c}{0} &
\multicolumn{2}{c}{1} &
\multicolumn{2}{c}{8} &
\multicolumn{2}{c||}{20} &
\multicolumn{2}{c}{0} &
\multicolumn{2}{c}{1} &
\multicolumn{2}{c}{8} &
\multicolumn{2}{c}{20} \\
  \hline
  \hline
\multirow{3}{1.15cm}{\emph{Class 1}} & 1.00 & \textit{\small\!0.00} & 0.97 & \textit{\small\!0.79} & 0.94 & \textit{\small\!0.96} & 0.92 & \textit{\small\!1.02} & 1.00 & \textit{\small\!0.00} & 0.99 & \textit{\small\!0.79} & 1.00 & \textit{\small\!0.96} & 1.00 & \textit{\small\!1.02} \\ 
   & 0.78 & \textit{\small\!0.81} & 0.97 & \textit{\small\!0.79} & 0.93 & \textit{\small\!0.92} & 0.91 & \textit{\small\!0.98} & 0.78 & \textit{\small\!0.81} & 0.98 & \textit{\small\!0.79} & 1.00 & \textit{\small\!0.92} & 1.00 & \textit{\small\!0.98} \\ 
   & 1.00 & \textit{\small\!0.77} & 0.98 & \textit{\small\!0.76} & 0.93 & \textit{\small\!0.94} & 0.92 & \textit{\small\!1.00} & 1.00 & \textit{\small\!0.77} & 0.99 & \textit{\small\!0.76} & 1.00 & \textit{\small\!0.94} & 1.00 & \textit{\small\!1.00} \\ 
   \hline
\multirow{3}{1.15cm}{\emph{Class 1 A}} & 1.00 & \textit{\small\!0.00} & 0.98 & \textit{\small\!0.78} & 0.92 & \textit{\small\!0.96} & 0.93 & \textit{\small\!1.03} & 1.00 & \textit{\small\!0.00} & 0.99 & \textit{\small\!0.78} & 1.00 & \textit{\small\!0.96} & 1.00 & \textit{\small\!1.03} \\ 
   & 0.80 & \textit{\small\!0.79} & 0.97 & \textit{\small\!0.78} & 0.93 & \textit{\small\!0.91} & 0.93 & \textit{\small\!0.97} & 0.80 & \textit{\small\!0.79} & 0.99 & \textit{\small\!0.78} & 1.00 & \textit{\small\!0.91} & 1.00 & \textit{\small\!0.97} \\ 
   & 0.99 & \textit{\small\!0.77} & 0.98 & \textit{\small\!0.75} & 0.93 & \textit{\small\!0.93} & 0.93 & \textit{\small\!0.99} & 0.99 & \textit{\small\!0.77} & 0.99 & \textit{\small\!0.75} & 1.00 & \textit{\small\!0.93} & 1.00 & \textit{\small\!0.99} \\ 
   \hline
\multirow{3}{1.15cm}{\emph{Class 1 B}} & 1.00 & \textit{\small\!0.00} & 0.98 & \textit{\small\!0.52} & 0.93 & \textit{\small\!0.66} & 0.93 & \textit{\small\!0.69} & 1.00 & \textit{\small\!0.00} & 1.00 & \textit{\small\!0.52} & 1.00 & \textit{\small\!0.66} & 1.00 & \textit{\small\!0.69} \\ 
   & 0.90 & \textit{\small\!0.42} & 0.98 & \textit{\small\!0.52} & 0.92 & \textit{\small\!0.64} & 0.94 & \textit{\small\!0.67} & 0.90 & \textit{\small\!0.42} & 1.00 & \textit{\small\!0.52} & 1.00 & \textit{\small\!0.64} & 1.00 & \textit{\small\!0.67} \\ 
   & 1.00 & \textit{\small\!0.50} & 0.98 & \textit{\small\!0.50} & 0.93 & \textit{\small\!0.64} & 0.92 & \textit{\small\!0.67} & 1.00 & \textit{\small\!0.50} & 1.00 & \textit{\small\!0.50} & 1.00 & \textit{\small\!0.64} & 1.00 & \textit{\small\!0.67} \\ 
   \hline
\multirow{3}{1.15cm}{\emph{Class 1 C}} & 1.00 & \textit{\small\!0.00} & 0.98 & \textit{\small\!0.74} & 0.93 & \textit{\small\!0.98} & 0.90 & \textit{\small\!1.07} & 1.00 & \textit{\small\!0.00} & 1.00 & \textit{\small\!0.74} & 1.00 & \textit{\small\!0.98} & 1.00 & \textit{\small\!1.07} \\ 
   & 0.85 & \textit{\small\!0.77} & 0.98 & \textit{\small\!0.74} & 0.92 & \textit{\small\!0.94} & 0.92 & \textit{\small\!1.02} & 0.85 & \textit{\small\!0.77} & 0.99 & \textit{\small\!0.74} & 1.00 & \textit{\small\!0.94} & 1.00 & \textit{\small\!1.02} \\ 
   & 0.99 & \textit{\small\!0.74} & 0.98 & \textit{\small\!0.72} & 0.93 & \textit{\small\!0.96} & 0.91 & \textit{\small\!1.05} & 0.99 & \textit{\small\!0.74} & 0.99 & \textit{\small\!0.72} & 1.00 & \textit{\small\!0.96} & 1.00 & \textit{\small\!1.05} \\ 
   \hline
\multirow{3}{1.15cm}{\emph{Class 1 D}} & 1.00 & \textit{\small\!0.00} & 0.98 & \textit{\small\!0.79} & 0.93 & \textit{\small\!0.96} & 0.93 & \textit{\small\!1.03} & 1.00 & \textit{\small\!0.00} & 0.99 & \textit{\small\!0.79} & 1.00 & \textit{\small\!0.96} & 1.00 & \textit{\small\!1.03} \\ 
   & 0.76 & \textit{\small\!0.82} & 0.97 & \textit{\small\!0.79} & 0.92 & \textit{\small\!0.91} & 0.92 & \textit{\small\!0.98} & 0.76 & \textit{\small\!0.82} & 0.99 & \textit{\small\!0.79} & 1.00 & \textit{\small\!0.91} & 1.00 & \textit{\small\!0.98} \\ 
   & 1.00 & \textit{\small\!0.78} & 0.97 & \textit{\small\!0.76} & 0.93 & \textit{\small\!0.94} & 0.92 & \textit{\small\!1.01} & 1.00 & \textit{\small\!0.78} & 0.99 & \textit{\small\!0.76} & 1.00 & \textit{\small\!0.94} & 1.00 & \textit{\small\!1.01} \\ 
   \hline
\multirow{3}{1.15cm}{\emph{Class 1 B+C}} & 1.00 & \textit{\small\!0.00} & 0.98 & \textit{\small\!0.55} & 0.93 & \textit{\small\!0.67} & 0.93 & \textit{\small\!0.70} & 1.00 & \textit{\small\!0.00} & 0.98 & \textit{\small\!0.55} & 1.00 & \textit{\small\!0.67} & 1.00 & \textit{\small\!0.70} \\ 
   & 0.70 & \textit{\small\!0.57} & 0.98 & \textit{\small\!0.56} & 0.94 & \textit{\small\!0.65} & 0.92 & \textit{\small\!0.67} & 0.70 & \textit{\small\!0.57} & 0.98 & \textit{\small\!0.56} & 1.00 & \textit{\small\!0.65} & 1.00 & \textit{\small\!0.67} \\ 
   & 1.00 & \textit{\small\!0.53} & 0.98 & \textit{\small\!0.53} & 0.93 & \textit{\small\!0.66} & 0.92 & \textit{\small\!0.68} & 1.00 & \textit{\small\!0.53} & 0.99 & \textit{\small\!0.53} & 1.00 & \textit{\small\!0.66} & 1.00 & \textit{\small\!0.68} \\ 
   \hline
\multirow{3}{1.15cm}{\emph{Class 2 }} & 1.00 & \textit{\small\!0.00} & 0.98 & \textit{\small\!0.76} & 0.93 & \textit{\small\!1.05} & 0.90 & \textit{\small\!1.16} & 1.00 & \textit{\small\!0.00} & 1.00 & \textit{\small\!0.76} & 1.00 & \textit{\small\!1.05} & 1.00 & \textit{\small\!1.16} \\ 
   & 0.76 & \textit{\small\!0.84} & 0.97 & \textit{\small\!0.76} & 0.93 & \textit{\small\!1.02} & 0.91 & \textit{\small\!1.12} & 0.76 & \textit{\small\!0.84} & 0.99 & \textit{\small\!0.76} & 1.00 & \textit{\small\!1.02} & 1.00 & \textit{\small\!1.12} \\ 
   & 1.00 & \textit{\small\!0.76} & 0.97 & \textit{\small\!0.73} & 0.93 & \textit{\small\!1.03} & 0.91 & \textit{\small\!1.14} & 1.00 & \textit{\small\!0.76} & 1.00 & \textit{\small\!0.73} & 1.00 & \textit{\small\!1.03} & 1.00 & \textit{\small\!1.14} \\ 
   \hline
\multirow{3}{1.15cm}{\emph{Class 2 A}} & 1.00 & \textit{\small\!0.00} & 0.98 & \textit{\small\!0.84} & 0.93 & \textit{\small\!1.17} & 0.92 & \textit{\small\!1.30} & 1.00 & \textit{\small\!0.00} & 1.00 & \textit{\small\!0.84} & 1.00 & \textit{\small\!1.17} & 1.00 & \textit{\small\!1.30} \\ 
   & 0.75 & \textit{\small\!1.01} & 0.98 & \textit{\small\!0.84} & 0.94 & \textit{\small\!1.12} & 0.90 & \textit{\small\!1.23} & 0.75 & \textit{\small\!1.01} & 1.00 & \textit{\small\!0.84} & 1.00 & \textit{\small\!1.12} & 1.00 & \textit{\small\!1.23} \\ 
   & 1.00 & \textit{\small\!0.85} & 0.97 & \textit{\small\!0.82} & 0.93 & \textit{\small\!1.15} & 0.91 & \textit{\small\!1.27} & 1.00 & \textit{\small\!0.85} & 1.00 & \textit{\small\!0.82} & 1.00 & \textit{\small\!1.15} & 1.00 & \textit{\small\!1.27} \\ 
   \hline
\multirow{3}{1.15cm}{\emph{Class 2 B}} & 1.00 & \textit{\small\!0.00} & 0.97 & \textit{\small\!0.52} & 0.93 & \textit{\small\!0.76} & 0.91 & \textit{\small\!0.84} & 1.00 & \textit{\small\!0.00} & 1.00 & \textit{\small\!0.52} & 1.00 & \textit{\small\!0.76} & 1.00 & \textit{\small\!0.84} \\ 
   & 0.87 & \textit{\small\!0.40} & 0.98 & \textit{\small\!0.51} & 0.94 & \textit{\small\!0.75} & 0.91 & \textit{\small\!0.82} & 0.87 & \textit{\small\!0.40} & 1.00 & \textit{\small\!0.51} & 1.00 & \textit{\small\!0.75} & 1.00 & \textit{\small\!0.82} \\ 
   & 1.00 & \textit{\small\!0.50} & 0.97 & \textit{\small\!0.50} & 0.94 & \textit{\small\!0.75} & 0.92 & \textit{\small\!0.82} & 1.00 & \textit{\small\!0.50} & 1.00 & \textit{\small\!0.50} & 1.00 & \textit{\small\!0.75} & 1.00 & \textit{\small\!0.82} \\ 
   \hline
\multirow{3}{1.15cm}{\emph{Class 2 A+B}} & 1.00 & \textit{\small\!0.00} & 0.97 & \textit{\small\!0.55} & 0.94 & \textit{\small\!0.83} & 0.92 & \textit{\small\!0.92} & 1.00 & \textit{\small\!0.00} & 1.00 & \textit{\small\!0.55} & 1.00 & \textit{\small\!0.83} & 1.00 & \textit{\small\!0.92} \\ 
   & 0.88 & \textit{\small\!0.44} & 0.98 & \textit{\small\!0.54} & 0.94 & \textit{\small\!0.80} & 0.91 & \textit{\small\!0.89} & 0.88 & \textit{\small\!0.44} & 1.00 & \textit{\small\!0.54} & 1.00 & \textit{\small\!0.80} & 1.00 & \textit{\small\!0.89} \\ 
   & 1.00 & \textit{\small\!0.54} & 0.97 & \textit{\small\!0.54} & 0.94 & \textit{\small\!0.82} & 0.92 & \textit{\small\!0.91} & 1.00 & \textit{\small\!0.54} & 1.00 & \textit{\small\!0.54} & 1.00 & \textit{\small\!0.82} & 1.00 & \textit{\small\!0.91} \\ 
   \hline
\end{tabular}
\caption{Coverage ratios and lengths (in italic) of confidence intervals constructed by \DESP \emph{ De} and \DESP \emph{Re}. For each class,
        the first row is the average over the variables 1 to r-1, the second row over r, and the third over r+1 to 20.} 
\end{table}
\begin{table}[H]
\centering
\begin{tabular}{l|rrrrrrrr||rrrrrrrr}
 \hline
 &\multicolumn{8}{c||}{De} &
 \multicolumn{8}{c}{Re} \\
&\multicolumn{2}{c}{0} &
\multicolumn{2}{c}{1} &
\multicolumn{2}{c}{8} &
\multicolumn{2}{c||}{20} &
\multicolumn{2}{c}{0} &
\multicolumn{2}{c}{1} &
\multicolumn{2}{c}{8} &
\multicolumn{2}{c}{20} \\
  \hline
  \hline
\multirow{3}{1.15cm}{\emph{Class 1}} & 1.00 & \textit{\small\!0.14} & 0.95 & \textit{\small\!0.74} & 0.92 & \textit{\small\!0.86} & 0.87 & \textit{\small\!0.81} & 1.00 & \textit{\small\!0.14} & 0.99 & \textit{\small\!0.74} & 1.00 & \textit{\small\!0.86} & 1.00 & \textit{\small\!0.81} \\ 
   & 0.74 & \textit{\small\!0.53} & 0.95 & \textit{\small\!0.74} & 0.92 & \textit{\small\!0.84} & 0.86 & \textit{\small\!0.79} & 0.74 & \textit{\small\!0.53} & 0.99 & \textit{\small\!0.74} & 1.00 & \textit{\small\!0.84} & 1.00 & \textit{\small\!0.79} \\ 
   & 0.93 & \textit{\small\!0.68} & 0.95 & \textit{\small\!0.72} & 0.91 & \textit{\small\!0.84} & 0.86 & \textit{\small\!0.79} & 0.93 & \textit{\small\!0.68} & 0.99 & \textit{\small\!0.72} & 1.00 & \textit{\small\!0.84} & 1.00 & \textit{\small\!0.79} \\ 
   \hline
\multirow{3}{1.15cm}{\emph{Class 1 A}} & 1.00 & \textit{\small\!0.21} & 0.96 & \textit{\small\!0.74} & 0.91 & \textit{\small\!0.87} & 0.87 & \textit{\small\!0.82} & 1.00 & \textit{\small\!0.21} & 0.99 & \textit{\small\!0.74} & 1.00 & \textit{\small\!0.87} & 1.00 & \textit{\small\!0.82} \\ 
   & 0.74 & \textit{\small\!0.58} & 0.96 & \textit{\small\!0.73} & 0.92 & \textit{\small\!0.84} & 0.87 & \textit{\small\!0.79} & 0.74 & \textit{\small\!0.58} & 0.99 & \textit{\small\!0.73} & 1.00 & \textit{\small\!0.84} & 1.00 & \textit{\small\!0.79} \\ 
   & 0.93 & \textit{\small\!0.68} & 0.95 & \textit{\small\!0.71} & 0.92 & \textit{\small\!0.84} & 0.87 & \textit{\small\!0.79} & 0.93 & \textit{\small\!0.68} & 0.99 & \textit{\small\!0.71} & 1.00 & \textit{\small\!0.84} & 1.00 & \textit{\small\!0.79} \\ 
   \hline
\multirow{3}{1.15cm}{\emph{Class 1 B}} & 1.00 & \textit{\small\!0.07} & 0.96 & \textit{\small\!0.54} & 0.93 & \textit{\small\!0.67} & 0.92 & \textit{\small\!0.66} & 1.00 & \textit{\small\!0.07} & 0.99 & \textit{\small\!0.54} & 1.00 & \textit{\small\!0.67} & 1.00 & \textit{\small\!0.66} \\ 
   & 0.88 & \textit{\small\!0.36} & 0.95 & \textit{\small\!0.53} & 0.92 & \textit{\small\!0.65} & 0.93 & \textit{\small\!0.64} & 0.88 & \textit{\small\!0.36} & 1.00 & \textit{\small\!0.53} & 1.00 & \textit{\small\!0.65} & 1.00 & \textit{\small\!0.64} \\ 
   & 0.94 & \textit{\small\!0.48} & 0.95 & \textit{\small\!0.51} & 0.94 & \textit{\small\!0.64} & 0.91 & \textit{\small\!0.63} & 0.94 & \textit{\small\!0.48} & 1.00 & \textit{\small\!0.51} & 1.00 & \textit{\small\!0.64} & 1.00 & \textit{\small\!0.63} \\ 
   \hline
\multirow{3}{1.15cm}{\emph{Class 1 C}} & 1.00 & \textit{\small\!0.21} & 0.95 & \textit{\small\!0.73} & 0.92 & \textit{\small\!0.93} & 0.87 & \textit{\small\!0.90} & 1.00 & \textit{\small\!0.21} & 0.99 & \textit{\small\!0.73} & 1.00 & \textit{\small\!0.93} & 1.00 & \textit{\small\!0.90} \\ 
   & 0.75 & \textit{\small\!0.56} & 0.96 & \textit{\small\!0.72} & 0.92 & \textit{\small\!0.91} & 0.87 & \textit{\small\!0.87} & 0.75 & \textit{\small\!0.56} & 0.99 & \textit{\small\!0.72} & 1.00 & \textit{\small\!0.91} & 1.00 & \textit{\small\!0.87} \\ 
   & 0.93 & \textit{\small\!0.67} & 0.96 & \textit{\small\!0.70} & 0.92 & \textit{\small\!0.91} & 0.86 & \textit{\small\!0.88} & 0.93 & \textit{\small\!0.67} & 0.99 & \textit{\small\!0.70} & 1.00 & \textit{\small\!0.91} & 1.00 & \textit{\small\!0.88} \\ 
   \hline
\multirow{3}{1.15cm}{\emph{Class 1 D}} & 1.00 & \textit{\small\!0.17} & 0.95 & \textit{\small\!0.76} & 0.91 & \textit{\small\!0.86} & 0.87 & \textit{\small\!0.81} & 1.00 & \textit{\small\!0.17} & 0.99 & \textit{\small\!0.76} & 1.00 & \textit{\small\!0.86} & 1.00 & \textit{\small\!0.81} \\ 
   & 0.74 & \textit{\small\!0.58} & 0.96 & \textit{\small\!0.76} & 0.92 & \textit{\small\!0.84} & 0.87 & \textit{\small\!0.79} & 0.74 & \textit{\small\!0.58} & 0.99 & \textit{\small\!0.76} & 1.00 & \textit{\small\!0.84} & 1.00 & \textit{\small\!0.79} \\ 
   & 0.92 & \textit{\small\!0.73} & 0.95 & \textit{\small\!0.73} & 0.91 & \textit{\small\!0.84} & 0.86 & \textit{\small\!0.80} & 0.92 & \textit{\small\!0.73} & 0.99 & \textit{\small\!0.73} & 1.00 & \textit{\small\!0.84} & 1.00 & \textit{\small\!0.80} \\ 
   \hline
\multirow{3}{1.15cm}{\emph{Class 1 B+C}} & 1.00 & \textit{\small\!0.07} & 0.96 & \textit{\small\!0.54} & 0.93 & \textit{\small\!0.66} & 0.91 & \textit{\small\!0.65} & 1.00 & \textit{\small\!0.07} & 0.98 & \textit{\small\!0.54} & 1.00 & \textit{\small\!0.66} & 1.00 & \textit{\small\!0.65} \\ 
   & 0.78 & \textit{\small\!0.37} & 0.96 & \textit{\small\!0.54} & 0.94 & \textit{\small\!0.65} & 0.91 & \textit{\small\!0.63} & 0.78 & \textit{\small\!0.37} & 0.98 & \textit{\small\!0.55} & 1.00 & \textit{\small\!0.65} & 1.00 & \textit{\small\!0.63} \\ 
   & 0.94 & \textit{\small\!0.49} & 0.96 & \textit{\small\!0.53} & 0.94 & \textit{\small\!0.64} & 0.91 & \textit{\small\!0.63} & 0.94 & \textit{\small\!0.49} & 0.98 & \textit{\small\!0.53} & 1.00 & \textit{\small\!0.64} & 1.00 & \textit{\small\!0.63} \\ 
   \hline
\multirow{3}{1.15cm}{\emph{Class 2 }} & 1.00 & \textit{\small\!0.14} & 0.97 & \textit{\small\!0.74} & 0.93 & \textit{\small\!1.00} & 0.85 & \textit{\small\!0.97} & 1.00 & \textit{\small\!0.14} & 1.00 & \textit{\small\!0.74} & 1.00 & \textit{\small\!1.00} & 1.00 & \textit{\small\!0.98} \\ 
   & 0.72 & \textit{\small\!0.52} & 0.97 & \textit{\small\!0.73} & 0.93 & \textit{\small\!0.98} & 0.88 & \textit{\small\!0.96} & 0.72 & \textit{\small\!0.52} & 0.99 & \textit{\small\!0.73} & 1.00 & \textit{\small\!0.98} & 1.00 & \textit{\small\!0.96} \\ 
   & 0.93 & \textit{\small\!0.67} & 0.96 & \textit{\small\!0.71} & 0.93 & \textit{\small\!0.97} & 0.87 & \textit{\small\!0.95} & 0.93 & \textit{\small\!0.67} & 0.99 & \textit{\small\!0.71} & 1.00 & \textit{\small\!0.97} & 1.00 & \textit{\small\!0.95} \\ 
   \hline
\multirow{3}{1.15cm}{\emph{Class 2 A}} & 1.00 & \textit{\small\!0.15} & 0.97 & \textit{\small\!0.80} & 0.92 & \textit{\small\!1.08} & 0.87 & \textit{\small\!1.06} & 1.00 & \textit{\small\!0.15} & 1.00 & \textit{\small\!0.80} & 1.00 & \textit{\small\!1.09} & 1.00 & \textit{\small\!1.06} \\ 
   & 0.70 & \textit{\small\!0.56} & 0.96 & \textit{\small\!0.79} & 0.93 & \textit{\small\!1.05} & 0.86 & \textit{\small\!1.03} & 0.70 & \textit{\small\!0.56} & 1.00 & \textit{\small\!0.79} & 1.00 & \textit{\small\!1.05} & 1.00 & \textit{\small\!1.03} \\ 
   & 0.93 & \textit{\small\!0.73} & 0.97 & \textit{\small\!0.78} & 0.93 & \textit{\small\!1.06} & 0.86 & \textit{\small\!1.04} & 0.93 & \textit{\small\!0.73} & 1.00 & \textit{\small\!0.78} & 1.00 & \textit{\small\!1.06} & 1.00 & \textit{\small\!1.04} \\ 
   \hline
\multirow{3}{1.15cm}{\emph{Class 2 B}} & 1.00 & \textit{\small\!0.07} & 0.96 & \textit{\small\!0.54} & 0.94 & \textit{\small\!0.79} & 0.91 & \textit{\small\!0.83} & 1.00 & \textit{\small\!0.07} & 1.00 & \textit{\small\!0.54} & 1.00 & \textit{\small\!0.79} & 1.00 & \textit{\small\!0.82} \\ 
   & 0.88 & \textit{\small\!0.36} & 0.97 & \textit{\small\!0.54} & 0.94 & \textit{\small\!0.78} & 0.92 & \textit{\small\!0.81} & 0.88 & \textit{\small\!0.36} & 1.00 & \textit{\small\!0.54} & 1.00 & \textit{\small\!0.78} & 1.00 & \textit{\small\!0.81} \\ 
   & 0.93 & \textit{\small\!0.48} & 0.96 & \textit{\small\!0.52} & 0.94 & \textit{\small\!0.77} & 0.91 & \textit{\small\!0.81} & 0.93 & \textit{\small\!0.48} & 1.00 & \textit{\small\!0.52} & 1.00 & \textit{\small\!0.77} & 1.00 & \textit{\small\!0.81} \\ 
   \hline
\multirow{3}{1.15cm}{\emph{Class 2 A+B}} & 1.00 & \textit{\small\!0.07} & 0.96 & \textit{\small\!0.59} & 0.94 & \textit{\small\!0.86} & 0.91 & \textit{\small\!0.90} & 1.00 & \textit{\small\!0.07} & 1.00 & \textit{\small\!0.59} & 1.00 & \textit{\small\!0.86} & 1.00 & \textit{\small\!0.90} \\ 
   & 0.87 & \textit{\small\!0.39} & 0.97 & \textit{\small\!0.57} & 0.94 & \textit{\small\!0.84} & 0.91 & \textit{\small\!0.87} & 0.87 & \textit{\small\!0.39} & 1.00 & \textit{\small\!0.58} & 1.00 & \textit{\small\!0.84} & 1.00 & \textit{\small\!0.87} \\ 
   & 0.94 & \textit{\small\!0.53} & 0.96 & \textit{\small\!0.57} & 0.94 & \textit{\small\!0.85} & 0.91 & \textit{\small\!0.89} & 0.94 & \textit{\small\!0.53} & 1.00 & \textit{\small\!0.57} & 1.00 & \textit{\small\!0.85} & 1.00 & \textit{\small\!0.89} \\ 
   \hline
\end{tabular}
\caption{Coverage ratios and lengths (in italic) of confidence intervals constructed by \OLS \emph{ De} and \OLS \emph{Re}. For each class,
        the first row is the average over the variables 1 to r-1, the second row over r, and the third over r+1 to 20.} 
\end{table}